\documentclass[preprint,usenatbib]{aastex61}
\usepackage[utf8]{inputenc}
\usepackage{amsmath}
\usepackage{mathtools}
\usepackage{amsfonts}
\usepackage{amssymb}
\usepackage{makeidx}
\usepackage{indentfirst}
\usepackage{graphicx}
\usepackage{color}
\usepackage{lmodern}
\usepackage{url}
\usepackage{mathrsfs}
\usepackage{longtable}
\usepackage{multirow}
\usepackage [english]{babel}
\usepackage [autostyle, english = american]{csquotes}
\MakeOuterQuote{"}
\usepackage{chemformula}
\usepackage{textcomp}
\usepackage{aas_macros}
\usepackage{natbib}
\usepackage{tabularx}
\usepackage{comment}
\usepackage{xcolor}
\usepackage{color, colortbl}
\definecolor{voilet}{RGB}{127,0,255}
\definecolor{vikas1}{RGB}{255,198,0}
\definecolor{vikas2}{RGB}{0, 255, 0}
\definecolor{vikas3}{RGB}{0, 0, 255}
\definecolor{vikas}{RGB}{255,0,127}
\definecolor{KA}{rgb}{0.76, 0.13, 0.28}

\shorttitle{Exoplanet}
\shortauthors{Soni \& Acharyya}

\begin{document}
\title{
Using a quench level approximation to estimate the effect of metallicity on N-bearing species abundances in 
\ch{H2}-dominated atmospheres
}
\correspondingauthor{Kinsuk Acharyya}
\email{acharyya@prl.res.in}

\author[0000-0002-0603-8777]{Vikas Soni}
\affil{Planetary Sciences Division, Physical Research Laboratory, Ahmedabad, 380009, India}
\affil{Indian Institute of Technology, Gandhinagar, 382355, India}

\author[0000-0002-0603-8777]{Kinsuk Acharyya}
\affil{Planetary Sciences Division, Physical Research Laboratory, Ahmedabad, 380009, India}

\nocollaboration

\begin{abstract}
Variations in atmospheric elemental nitrogen can considerably affect the abundance of major 
nitrogen-bearing species such as \ch{NH3} and HCN. Also, due to vertical mixing and photochemistry, their abundance 
deviates from the thermochemical equilibrium. The goal of this study is to understand the effect of atmospheric 
metallicity on the composition of \ch{NH3}, \ch{N2}, and HCN over a large parameter space in  the 
presence of vertical mixing which, when combined with the work on CHO-bearing species in \cite{Soni2023} 
can provide a comprehensive understanding of the effect of atmospheric metallicity. We used quenching 
approximations and a full chemical kinetics model for the calculations, and a comparison between these two methods was 
made. For generating thermal profiles, petitRADTRANS code is used. Chemical timescales of \ch{NH3} and \ch{N2} 
are found to be complex functions of metallicity, while HCN is inversely proportional. Using \ch{NH3} and CO quenched 
abundances, the HCN quenched abundance can be constrained since it remains in equilibrium with \ch{NH3}, CO, and \ch{H2O}. Quenched \ch{NH3} increases with increasing $K_{zz}$ untill a particular point, after which it becomes independent of vertical mixing. There is a sweet spot in the $K_{zz}$ parameter space to maximize the quenched HCN for a given $T_{\text{int}}$ and $T_{\text{equi}}$; the parameter space moves towards the lower equilibrium temperature, and HCN abundance increases with metallicity. Finally, we used the dataset of quenched abundances to provide a list of potential candidates in which \ch{HCN} observation can be possible.

\end{abstract}
\section{Introduction}
Nitrogen-bearing species are 
essential to a habitable climate \citep{Vladilo2013}; its accurate characterization, along with 
oxygen species, can enable us to differentiate biological signatures from non-biological ones 
\citep{Schwieterman2018}. Besides, it can help us understand disequilibrium chemistry and can 
provide critical constraints to the planet formation and migration history of the exoplanets 
\citep{Piso2016, Cridland2020, Ohno2021}. 
Major nitrogen-bearing species such as  \ch{HCN}, and \ch{NH3} have been detected in exoplanet 
atmospheres \citep{Cabot2019,Giacobbe2021,Guilluy2022, Carleo2022}. 
With JWST, we are entering the golden era for the 
atmospheric characterization of exoplanets, and conclusive detections of nitrogen chemistry are 
possible \citep{MacDonald2017, Tsai2021, Claringbold2023}. The recent detection of \ch{SO2} 
in WASP-39 b gives the first-ever signature of photochemistry \citep{Rustamkulov2023, Alderson2023}, 
which is very promising.

For the solar elemental abundance, nitrogen is the third most abundant heavy element after oxygen and carbon; 
its bulk elemental abundance is a factor of 7.4 and 3.4 less than that of O and C, respectively \citep{Lodders2009}. 
The absorption cross-section of \ch{NH3} and \ch{HCN} is comparable to that of \ch{H2O} in most of the wavelength 
range except when $\lambda>10\mu$m, where the cross-section of \ch{NH3} and \ch{HCN} can be more than two orders of 
magnitude larger than \ch{H2O}. However, the \ch{H2O} abundance is several orders of magnitude larger than those of 
\ch{NH3} and \ch{HCN}. Thus, the total contribution of \ch{H2O} in the planet spectra is considerably larger compared 
to \ch{NH3} and \ch{HCN}, making the observation of \ch{NH3} and \ch{HCN} quite challenging. \ch{N2} remains the 
dominant species in thermochemical equilibrium in the warm exoplanets, but it does not show any observational signature, 
while \ch{NH3} is dominant in the relatively cool atmosphere. The mixing ratio of \ch{HCN} remains small 
($\approx 10^{-8}-10^{-9}$) in the thermochemical equilibrium, The transit-signature of \ch{NH3}/\ch{HCN} are around 50/100 
to 200/300 ppm for the mixing ratio of $\approx 10^{-6}$ for a solar elemental composition \citep{MacDonald2017, Ohno2022}. 
Despite the low abundance of N-bearing species, recent work shows the potential capability of JWST in observing 
N-bearing species \citep{MacDonald2017, Ohno2022}. It is found that the \ch{HCN} signature becomes negligible for 
\ch{HCN/H2O} $<$ 10$^{-2}$. In thermochemical equilibrium, the \ch{HCN} abundance is four to five orders of magnitude 
less than \ch{H2O} for solar metallicity. This gap increases with increasing metallicity. Quenching and photochemistry 
can increase the disequilibrium abundance of \ch{HCN} by more than two orders of magnitude, which increases the possibility 
of its detection \citep{Venot2012, MacDonald2017}. 

In the weakly irradiated atmosphere, the quenched abundance of 
\ch{NH3} is high, and the photodissociation of \ch{NH3} leads to the formation of \ch{HCN}. However, the production is 
limited by the low availability of the photons. In the strongly irradiated atmosphere, the quenched \ch{NH3} abundance is low, 
and photochemically produced \ch{HCN} is limited by the quenched \ch{NH3} abundance. 
As a result, there is a sweet spot for the photochemically produced HCN between 800 to 1400 K \citep{Baeyens2022}. 
Atmosphere with the low-temperature and high vertical mixing, photochemically produced \ch{HCN} can 
diffuse in the higher pressure region ($ P> 10^{-4} $ bar) and can imprint their signature in the transmission spectra 
\citep{Moses2011, Madhusudhan2016, Ohno2022}. Some studies incorporate the zonal wind (mixing along the latitude) and 
meridional wind (mixing along the longitude) and found that the \ch{NH3} and HCN can be largely affected due to horizontal 
mixing. However, this effect is complex and depends upon several parameters (e.g., day-night temperature constant, rotational 
period, and stellar type) \citep{Agundez2014B, Drummond2020, Baeyens2021, Zamyatina2023}. 

Atmospheric abundances are very often model dependent, and the parameter space for reproducing 
certain compositions is degenerate. The thermal profile decides the relative abundance of the 
molecules, and the elemental abundance changes the overall budget of molecules. Several physical 
processes can alter these abundances from their thermochemical equilibrium. Among the various 
parameters, atmospheric metallicity is one of the crucial parameters that dictates atmospheric 
composition \citep{Moses2013, Rajpurohit2020}. It can vary significantly from one planet to another. Considerable variations in 
atmospheric metallicity can be seen in solar system gas giants. The common trend is that the 
atmospheric metallicity increases with decreasing mass (Jupiter, Saturn, Neptune, and 
Uranus have metallicities that are 3.3-5.5, 9.5-10.3, 71-100, and 67-111 $\times$ solar metallicity, respectively), 
although large uncertainties exist in the abundances of individual elements \citep{Atreya2018}. 
Several studies have been made from high-precision spectral analysis to discern the atmospheric 
metallicity of exoplanets, though large uncertainties exist at the current sensitivity level. Exoplanet 
metallicities vary from subsolar (e.g., HAT-P-7 b; \citealp{Mansfield2018}), to near to solar 
(e.g., WASP-43 b; \citealp{Stevenson2017}), to moderately enriched (e.g., WASP-103 b, 
\citealp{Kreidberg2018}; WASP-127 b, \citealp{Spake2021}; WASP-121 b, \citealp{Mikal-Evans2019}; 
WASP-39 b, \citealp{Ahrer2023}) to greatly enriched (e.g., GJ 436 b; \citealp{Knutson2014}). Thus, 
even though only a few exoplanets have been studied, the metallicity space appears to be diverse and 
can range between 0.1 to more than 1000 $\times$ solar metallicity \citep{Wakeford2020}.

The effect of metallicity on the thermochemical equilibrium abundance of \ch{NH3}, \ch{N2}, and \ch{HCN} 
has been studied \citep{Moses2013, Moses2013a, Drummond2018} and it is found that \ch{NH3} and 
\ch{N2} dominate at low and high temperatures, respectively.
As the metallicity increases, the abundance of \ch{NH3} and \ch{N2} increase, and the equal-abundance 
curve of \ch{NH3}-\ch{N2} shifts towards high-pressure and low-temperature regions, leading to an increase 
in the \ch{N2} dominant region in pressure-temperature space. Although the abundance of \ch{HCN} increases 
with metallicity, it always remains lower than both \ch{N2} and \ch{NH3} for all the temperature and pressure 
regions. \ch{HCN} is affected by the C/O ratio, whereas \ch{N2} and \ch{NH3} remain unaffected. \ch{NH3} 
is highly photoactive, and the large chemical conversion time scale of \ch{NH3}$\rightleftarrows$\ch{N2} makes 
its abundance prone to change due to photochemistry and atmospheric mixing. It is shown that disequilibrium 
processes can increase the \ch{NH3} and \ch{HCN} abundance at the photospheric pressure by several orders 
of magnitude in the infrared photosphere \citep{Zahnle2009, Moses2011, Line2011, Madhusudhan2012, 
Moses2013, Heng2016, Tsai2018, MacDonald2017, Ohno2022, Ohno2023a}. 

\cite{Moses2011} studied nitrogen chemistry for two exoplanets, 
HD 189733 b and HD 209458 b, and compared their model results with the transit and eclipse observations. 
They found the enhancement of \ch{NH3} and \ch{HCN} from their equilibrium abundances for both planets. Whereas, 
\ch{N2} closely follows the equilibrium profile until photochemical processes set in and destroy it. They also 
found that deviation from the equilibrium value for \ch{NH3} and \ch{HCN} will affect the spectral 
signatures of exoplanets, particularly for relatively cool transiting exoplanets such as HD 189733 b. 
Subsequently, \cite{Moses2016} found that for specific "young Jupiters" such as HR 8799 b and 51 Eri b, 
quenching will affect the relative abundances of \ch{N2} and \ch{NH3} and it will favour \ch{N2} over 
\ch{NH3} at the quench-point; therefore, \ch{N2/NH3} ratios can be much greater than the chemical-equilibrium 
predictions. They also found that \ch{HCN} is affected by both quenching and photochemistry; 
when deep atmospheric mixing is strong, quenching increases the \ch{HCN} abundance. However, when 
mixing is weak, strong UV irradiation is essential for \ch{HCN} production. Recently, \cite{Giacobbe2021} 
found the presence of \ch{HCN} and \ch{NH3} in HD 209458 b; they concluded that the planet is carbon-rich 
with a C/O ratio close to or greater than one based on atmospheric models in radiative and chemical equilibrium.

In the present work, we extend our previous work \citep{Soni2023} and study the effect of metallicity on the 
nonequilibrium abundance of the H-dominated atmosphere for assorted N-bearing molecules (\ch{NH3}, HCN, and 
\ch{N2}). We use two sets of models; in one we find the disequilibrium abundances in the presence of transport 
using quenching approximation, and in the second set, we use a 1D chemical kinetics model with transport 
and photochemistry. In Section \ref{sec:model}, the photochemistry-transport model and quenching approximation 
are briefly discussed. In Section \ref{sec:3}, the thermochemical equilibrium result is discussed. Sections 
\ref{sec:nh3-n2} and \ref{sec:hcn} include the results of the quenching approximation for \ch{NH3}-\ch{N2} and 
HCN, respectively. We also compared with chemical timescale calculated using quenching approximation 
with the chemical timescale calculated with the widely used analytical expressions from \cite{Zahnle2014} and 
discussed briefly in these sections and provided more details in Appendix A.2.
In Section \ref{sec:apply}, we compare the abundances derived using the quenching approximation 
with the full chemical kinetics model and the error associated with the quenching approximation. 
In Section \ref{sec:constraint}, we use the quench data to discuss the constraints on metallicity and transport 
strength. In Sections \ref{sec:observability} and \ref{sec:candidates}, we discuss the conditions for observing 
N-bearing species and provide a list of candidate exoplanets for HCN detection. Finally, in Section \ref{sec:conclusion}, 
we make the concluding statements.

\section{Model and parameters} \label{sec:model}

We have solved the mass continuity equation for each species. Appendix~\ref{sec:model_A} provides a brief description 
of the model; furthermore, a detailed description and the benchmarking can be found in \cite{Soni2023}. To study the 
effect of metallicity on the nonequilibrium abundance of the nitrogen-bearing species \ch{N2}, \ch{NH3}, and \ch{HCN} 
in a hydrogen-dominated atmosphere for solar N/O ratio (0.135), we considered a large parameter space; the metallicity 
varied between 0.1 and 1000 $\times$ solar metallicity, temperature between 500 and 2500 K, and pressure range between 
$10^{-4}$ and $10^{3}$ bar. The change in metallicity is relative to the solar photospheric elemental abundance, and it 
corresponds to an increase or decrease in the heavy elemental abundance (elements other than H and He) with respect to 
the solar metallicity by a common factor. The solar photospheric metallicity is taken from \cite{Lodders2009}. The range 
of bulk abundance of elements in the present study are C/H = $2.77\times 10^{-5} - 2.77\times 10^{-1}$, N/H = $8.18 \times 
10^{-6} - 8.18\times 10^{-2}$, and O/H = $6.06 \times 10^{-5} - 6.06\times 10^{-1}$.

We ran two sets of models. In the first set, we found the disequilibrium abundances in the presence of transport using the
quenching approximation. For this, we developed a network analysis tool to find the conversion schemes needed to calculate the
chemical timescales \citep{Soni2023}, and then followed the method given in \cite{Tsai2018}. 
In the quenching approximation, the quench level is defined by the pressure 
level at which the chemical and vertical mixing timescales are equal. The abundance at the quench level is called 
the quenched abundance. 
The quenching approximation is the simplest and computationally efficient method as compared to the chemical 
kinetics models to constrain the atmospheric abundance in 
the presence of transport; however, it should be used cautiously. 


The vertical mixing timescale $\tau_{mix}$ can be computed using the mixing length theory, and is given by the 
following equation:
\begin{equation} 
\tau_{mix} = L^2 / K_{zz},
\end{equation}
where $L$ is the mixing length scale of the atmosphere and $K_{zz}$ is the Eddy diffusion coefficient 
\citep{Visscher2011, Heng2017}. Since the Eddy diffusion 
coefficient has a large uncertainty, it is treated as a free parameter. The mixing length scale cannot be computed from the 
first principle, and a simple approximation is to take the pressure scale height as the mixing length. However, \cite{Smith1998} 
found that the mixing length can be $L \approx 0.1-1 \times \text{pressure scale height}$, which leads to $\tau_{mix} = 
(\eta H)^2 / K_{zz}$, where $\eta \in[0.1,1]$ and the exact value of $\eta$ depends upon the rate of change of chemical 
timescale with height. The pressure scale height $H = \frac{K_b  T}{\mu g}$, where $T$, $g$ and $\mu$ are temperature, 
surface gravity, and mean molecular mass of the atmosphere, respectively. It is to be noted that metallicity changes the 
elemental composition, thereby changing the value of $\mu$. When metallicity increases from 0.1 to 1000 $\times$ 
solar metallicity, $\mu$ changes by one order of magnitude.

The chemical timescale can be calculated by finding the appropriate rate-limiting step. 
The following relation gives the timescale of the 
conversion of species $a$ into $b$:  
\begin{equation}
\tau_{a\rightarrow b} = \frac{[a]}{\text{Rate of RLS}_{a\rightarrow b}}.
\end{equation}
Here, [a] is the abundance of species $a$, and $\text{RLS}_{a\rightarrow b}$ is the rate-limiting step in the conversion 
of $a$ into $b$. In a chemical network, a particular species is involved in several reactions; as a result, there are 
many conversion pathways between two species. The number of these pathways increases exponentially as the number of 
reactions in the network increases. However, in a chemical network, only a few conversion schemes are important, 
as most of the conversion schemes are significantly slower than the fastest conversion scheme.  
  
Besides calculating quench abundance, we also ran the full chemical kinetics model, which includes transport and 
photochemistry. We then compared the quenched abundance of \ch{N2}, \ch{HCN}, and \ch{NH3} with chemical kinetics model with transport for the 
two test exoplanets, GJ 1214 b and HD 189733 b  and discuss the quenching approximation's effectiveness. We discuss 
how the quenching approximation can constrain the metallicity and transport strengths, for which we use the test 
exoplanet HD 209458 b. We also use the chemical kinetics model in \S 9 to compare with the HCN abundances calculated 
using quenching approximation.

\begin{figure}[htb!]
	\centering
	\includegraphics[trim={0cm 0cm 0cm 0cm},clip,width=0.8\textwidth]{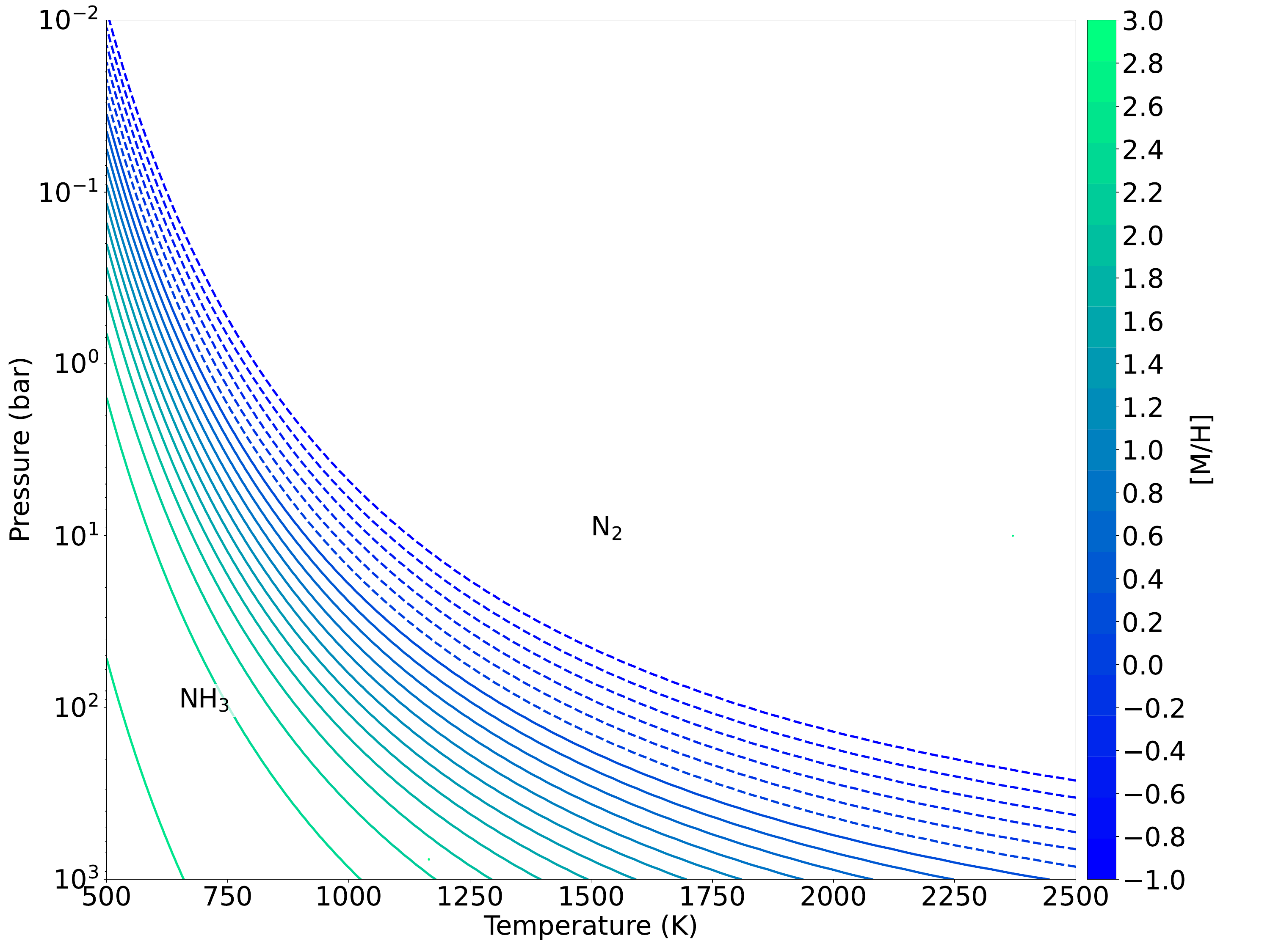}	
	\caption{The equal-abundance curves of \ch{NH3-N2} are shown. The cyan to blue lines delimitate the 
		region where \ch{NH3} (left) or \ch{N2} (right) is the dominant species. The metallicity value 
		in the color bar corresponds to the respective color line in the plot. The dashed lines are 
		for subsolar metallicity, and the solid lines are for supersolar metallicity.}\label{fig:NH3/N2}	
\end{figure}

\begin{figure}[htb!]
	\centering
	\includegraphics[trim={0cm 0cm 3cm 0cm},clip,width=1\textwidth]{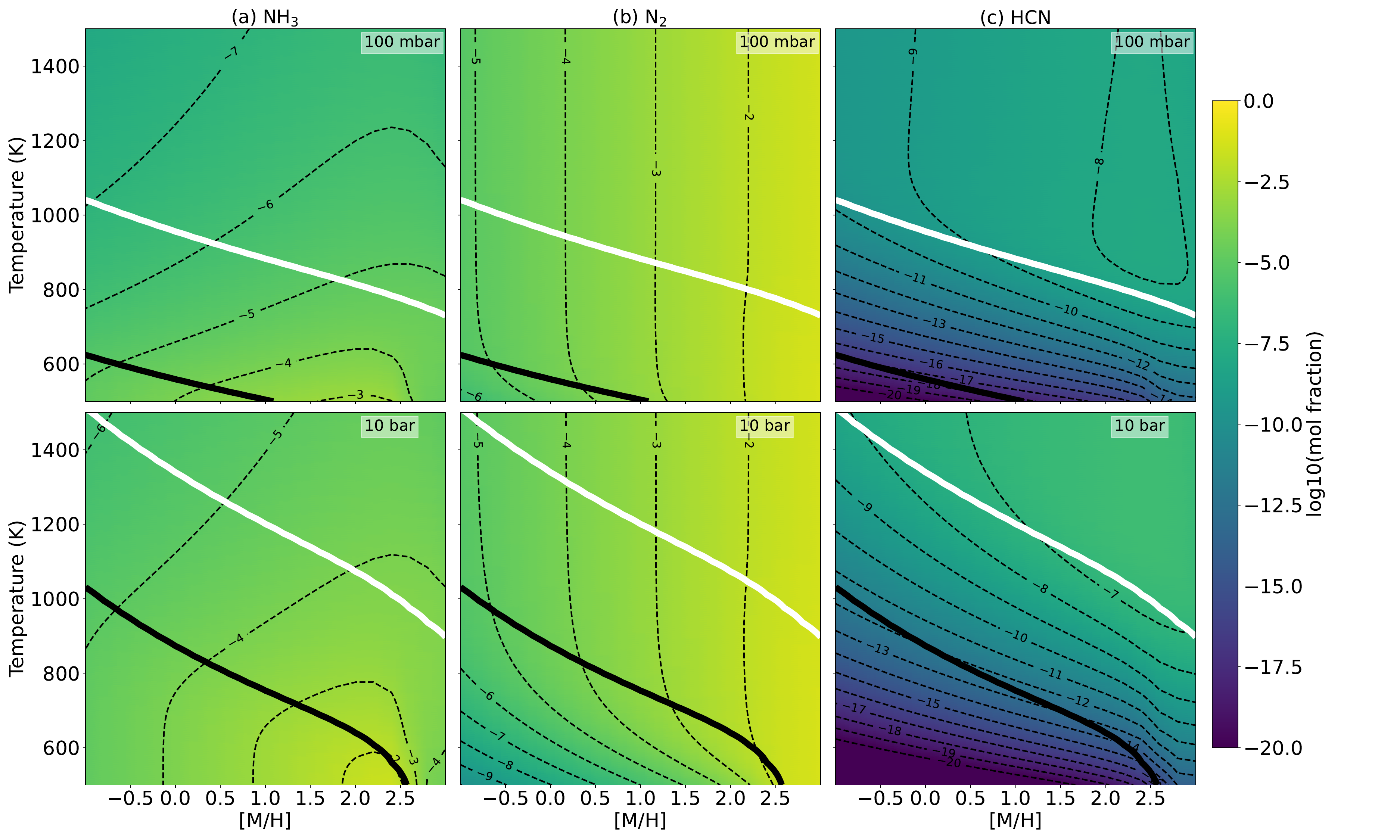}	
	\caption{Variation of the equilibrium mole fraction for (a) \ch{NH3}, (b) \ch{N2} and (c) \ch{HCN} at 100 mbar 
		and at 10 bar pressure with metallicity and temperature is shown. 
		The solid black line is the equal-abundance curve of \ch{NH3-N2}, and the black dashed lines are 
		the constant contours of the mole fraction of the respective gas-phase species.}\label{fig:N2-NH3}	
\end{figure}

\section{\ch{N2-NH3-HCN} equilibrium}\label{sec:3}
In this section, we briefly discuss the effect of metallicity on the equilibrium abundance of \ch{N2-NH3-HCN}, 
which was earlier studied by \cite{Moses2013a}. Figure \ref{fig:NH3/N2} shows the equal-abundance curve of 
\ch{NH3-N2}. It can be seen that the \ch{NH3-N2} curve shifts towards low-temperature and high-pressure 
regions with increasing metallicity, and the rate of shift increases with the metallicity. Thus, in the 
high-temperature and low-pressure regions, \ch{N2} dominates over \ch{NH3}, while in the low-temperature and 
high-pressure regions, \ch{NH3} dominates over \ch{N2}. For most of the parameter space, the \ch{HCN} abundance never 
 exceed the \ch{N2} or \ch{NH3} abundance. Only when the metallicity is very high, the \ch{HCN} mixing ratio 
 exceed the \ch{NH3} mixing ratio in the low-pressure and high-temperature regions.

We show the equilibrium mole-fraction of \ch{NH3} and \ch{N2} in Figure \ref{fig:N2-NH3} for 100 mbar 
(top panel) and 10 bar pressure (bottom panel). The solid black line shows the equal abundance curve of 
\ch{NH3-N2}; \ch{N2} dominates in the regions above this line and \ch{NH3} dominates below the line. \ch{N2} 
and \ch{NH3} abundance both increase linearly with increasing metallicity in the region where they are 
dominant, that is, \ch{N2} above the solid black line and \ch{NH3} below the line. If we compare the \ch{N2} 
and \ch{NH3} profiles with \ch{CO} and \ch{CH4} from \cite{Soni2023}, we see that the behaviors of \ch{N2} 
and \ch{CO} are qualitatively similar. However, the \ch{NH3} equilibrium abundance in the \ch{N2}-dominated 
region first increases with metallicity till [M/H] $\approx$ 2.5, and then starts to decrease due to a 
decrease in the bulk H abundance; in contrast, \ch{CH4} remains constant with metallicity in the CO-dominated 
region for [M/H]$<$2.5, where as, it increases linearly with metallicity in \ch{CH4}-dominated region and this increment 
is limited by the availability of bulk H for [M/H] $>$ 2.5. The equilibrium 
mole fraction of \ch{HCN} for 100 mbar and 10 bar pressure along with the equal-abundance curve of \ch{NH3-N2} and 
\ch{CH4-CO} is plotted in Figure~\ref{fig:N2-NH3}. The \ch{HCN} abundance decreases with metallicity when 
temperature and pressure change from \ch{N2} to \ch{NH3} dominated region, whereas, in a CO-dominated region, 
it becomes a weak function of metallicity. In addition, \ch{HCN} remains in equilibrium with CO, \ch{H2O}, and \ch{NH3}. 
Our result is similar to \cite{Moses2013a}. 

\begin{figure}[htb!]
	\centering
	\includegraphics[trim={0cm 0cm 0cm 0cm},clip, width=0.8\textwidth]{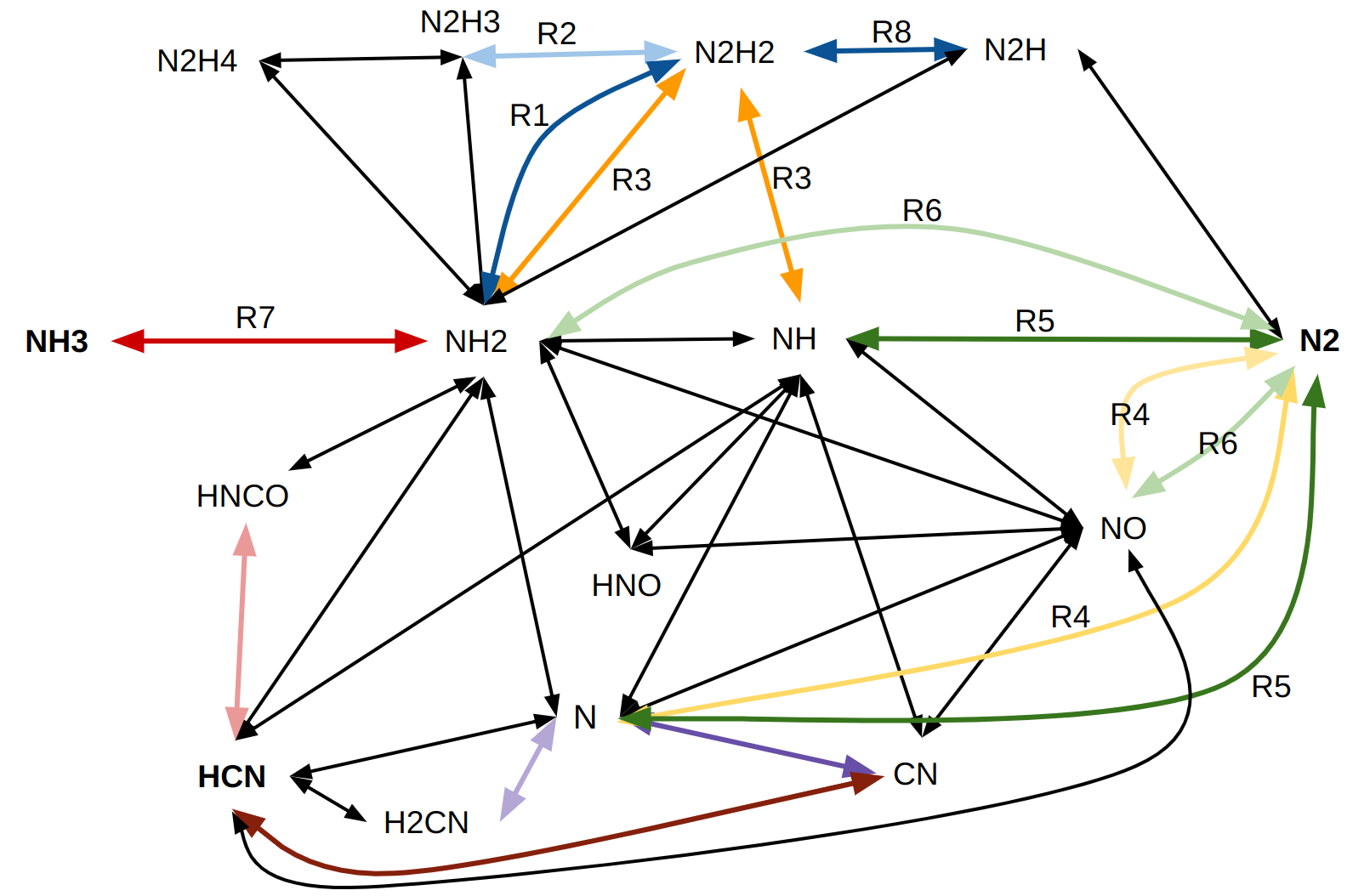}
	\caption{Major chemical pathways between $\ch{HCN}\rightleftarrows\ch{NH3}\rightleftarrows\ch{N2}$ 
		for hydrogen dominated atmosphere. The colored arrows other than black are the rate limiting 
		steps at different pressure-temperature corresponding to the colored regions in Figure \ref{fig:RLS_area_N2} and 
RLS number in Table \ref{HCN_RLS}.}\label{fig:conversion_N2}
\end{figure}

\section{\ch{NH3-N2}}\label{sec:nh3-n2}
\begin{figure}[b!]
	\centering
	\includegraphics[trim={0cm 0cm 8cm 0cm},clip, width=0.8\textwidth]{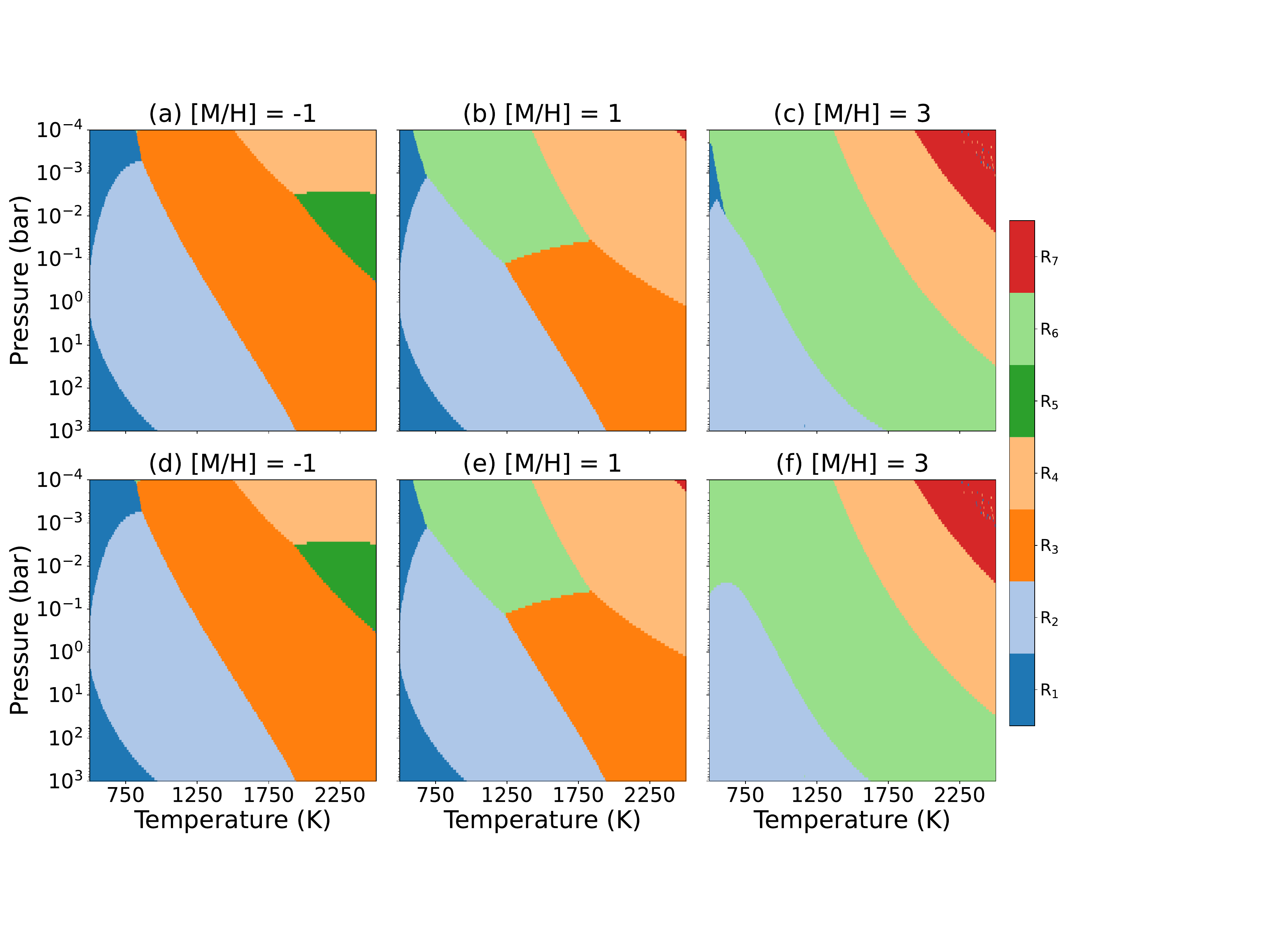}
	\caption{ The pressure-temperature range of different rate-limiting steps is shown for three 
		different metallicities. Each color corresponds to the different RLS in Figure \ref{fig:conversion_N2}. 
		The top and bottom panels represent the RLS parameter space for the conversion of \ch{NH3}$\rightarrow$\ch{N2} 
		($\tau_{ \ch{NH3}}$) and \ch{N2}$\rightarrow$\ch{NH3} ($\tau_{\ch{N2}}$).} \label{fig:RLS_area_N2}	
\end{figure}

Figure \ref{fig:conversion_N2} shows the major chemical pathways in $\ch{HCN}\rightleftarrows\ch{NH3}\rightleftarrows\ch{N2}$ 
conversion. Each arrow (except black) represents a rate-limiting step (RLS) reaction. There are two major schemes in the 
conversion of \ch{NH3} into \ch{N2} \citep{Moses2011, Tsai2018}: (i) the formation of \ch{N2} from \ch{NH3} via progressive 
dehydrogenation of \ch{N2H2}, and (ii) \ch{N2} formed by the deoxidation of \ch{NO} with reacting N or \ch{NH2}. 
Figure \ref{fig:RLS_area_N2} shows the regions of different RLSs (represented with a different color) as a function of 
temperature, pressure, and metallicity. In the low-temperature and high-pressure regions, the first scheme dominates 
(for which the RLS are R1, R2 and R3), whereas in the high-temperature region, the second scheme dominates (R4, R5 and R6).

\begin{table}[ht]\scriptsize
	\caption{ The power law dependence (y2) of the rates of different RLSs on the [M/H] is shown in 
the fourth column (where the reaction rate of RLS $\sim$ [M/H]$^{y2}$) for the different region of the 
temperature and pressure (e.g., \ch{NH3}, \ch{N2}, CH4 or CO dominant region). The First column shows the 
RLS number, the second column shows the respective RLS reaction and the third column shows the power law 
dependence (y1) of the reactant abundance of RLS on the [M/H] (where an abundance of a reactant$\sim$ [M/H]$^{y1}$). 
The value of y1 depends upon the local temperature and pressure, and here, we only use the approximate values 
where the respective RLS dominates.\label{HCN_RLS}.}
	\begin{center}
		\begin{tabular}{|l|l|l|l|}
         \hline
         RLS No & RLS &     (y1)                        & (y2)\\
         \hline
         \hline
                &  For $\ch{NH3}\leftrightarrows\ch{N2}$&  \ch{NH3} or \ch{N2} dominant region & \ch{NH3} or \ch{N2} dominant region \\         
    
		 \hline
         \hline
		         R1      & $\ch{NH2 + NH2} \rightleftarrows \ch{ N2H2 + H2}$      &    \ch{NH2} $\sim$ 1 or $<$ 0.5                            & 2 or $<$ 1 \\
     	\hline
		         R2      & $\ch{N2H3 + M}  \rightleftarrows \ch{ N2H2 + H + M}$   &    \ch{N2H3} $\sim$ 2 or $<$ 1                             & 2 or $<$ 1 \\
		\hline
		         R3      & $\ch{NH + NH2} \rightleftarrows \ch{ N2H2 + H}$        &    \ch{NH} and \ch{NH2} $\sim$ 1 or $<$ 0.5                & 2 or 1 \\
		\hline
		         R4      & $\ch{N + NO}  \rightleftarrows \ch{ N2 + O}$           &    \ch{N} $\sim$ $>$ 1 or $<$ 0.5, \ch{NO} $\sim$ $>$ 2 or 1.5 & $>$ 3 or $<$ 2 \\
		\hline
		         R5      & $\ch{N + NH} \rightleftarrows  \ch{ N2 + H}$           &    \ch{N} $\sim$ $>$ 1 or $<$ 0.5, \ch{NH} $\sim$ 1 or $<$ 0.5 & 2 or 1 \\
		\hline
		         R6      & $\ch{NO + NH2}  \rightleftarrows \ch{ N2 + H2O}$       &    \ch{NO} $\sim$ $>$ 2 or 1.5, \ch{NH2} $\sim$ 1 or $<$ 0.5 & $>$ 3 or $<$ 2 \\
		\hline
				 R7      & $\ch{NH3 + H } \rightleftarrows\ch{ NH2 + H2}$         &    \ch{NH3} $\sim$ $>$ 1 or 0.5, \ch{H} $\sim$ 0           &  $>$ 1 or  0.5 \\
		\hline
				 R8      & $\ch{N2H2 + H}  \rightleftarrows\ch{ N2H + H2}$        &    \ch{N2H2} $\sim$ 2 or $<$ 1, \ch{H} $\sim$ 0           &  2 or  $<$ 1 \\

		\hline
        \hline
                        & For $\ch{CH4}\leftrightarrows\ch{CO}$ \citep{Soni2023}  &    \ch{CH4} or \ch{CO} dominant region                   & \ch{CH4} or \ch{CO} dominant region \\  
		 \hline
				R1      & $\ch{CH3 + H2O} \rightleftarrows \ch{CH3OH + H}$        &    \ch{CH3} $\sim$ 1 or 0, \ch{H2O} 1 or 1               & 2 or 1 \\
		\hline
				R2      & $\ch{OH + CH3}  \rightleftarrows \ch{CH2OH + H}$        &    \ch{OH} $\sim$1 or 1, \ch{CH3} $\sim$ 1 or 0          & 2 or 1 \\
		\hline
				R3      & $\ch{OH + CH3 + M} \rightleftarrows \ch{CH3OH + M}$     &    \ch{OH} $\sim$1 or 1, \ch{CH3} $\sim$ 1 or 0          & 2 or 1 \\
		\hline
				R4      & $\ch{CH2OH + M}  \rightleftarrows \ch{H + H2CO + M}$    &    \ch{CH2OH} $\sim$2 or 1                               & 2 or 1 \\
		\hline
				R7      & $\ch{CH3OH + H}  \rightleftarrows \ch{ CH3O + H2}$      &    \ch{CH2OH} $\sim$2 or 1                               & 2 or 1 \\
       \hline

		\end{tabular}
	\end{center}
\end{table}

The comparison of the different RLS regions in Figure \ref{fig:RLS_area_N2} (\ch{NH3}$\rightleftarrows$\ch{N2}) 
with \cite{Soni2023} (Figure 5; \ch{CH4}$\rightleftarrows$\ch{CO}) shows that the effective region of RLS for 
\ch{NH3}$\rightleftarrows$\ch{N2} exhibits large change with metallicity as compared to \ch{CH4}$\rightleftarrows$\ch{CO}. 
Thus, the reaction rate of RLS in the \ch{NH3}$\rightleftarrows$\ch{N2} conversion shows complex dependence 
on metallicity compared to the \ch{CH4}$\rightleftarrows$\ch{CO} (see fourth column of Table \ref{HCN_RLS}).

For the latter case, the RLS rate has a square dependence on metallicity in \ch{CH4} dominant region and linear 
dependency in CO dominant region. The \ch{CH4} chemical time scale ($\tau_{\ch{CH4}}$ = (abundance of \ch{CH4})/(rate 
of RLS)) decreases linearly with metallicity in most of the parameter range. Where $\tau_{ \ch{CO}}$ remains 
constant with metallicity. In comparison, for the \ch{NH3}$\rightleftarrows$\ch{N2} conversion, the reactants are 
\ch{NO}, \ch{N}, \ch{NH}, \ch{N2H2}, and \ch{N2H3}, and their metallicity-dependence is not always the same 
(see third column in Table \ref{HCN_RLS}). 
In the \ch{N2} dominant region, the rate of R4 and R6 increases as a square with metallicity and are the RLS for the 
\ch{NH3}$\rightleftarrows$\ch{N2} conversion. 
In this region the $\tau_{ \ch{NH3}}$ decrease as a square of 
metallicity and $\tau_{ \ch{N2}}$ decrease linearly with metallicity. The overall  \ch{NH3}$\rightleftarrows$\ch{N2} 
conversion shows the strong dependence on metallicity as compared to the \ch{CH4}$\rightleftarrows$\ch{CO} 
conversion.

The combined effect of metallicity on the rate of RLS (column four in Table \ref{HCN_RLS}) and on the \ch{NH3} and 
\ch{N2} abundance leads to three different types of RLS similar to the \ch{CH4}$\rightleftarrows$\ch{CO} conversion 
\citep{Soni2023}. In the first type, the timescales of the RLS decrease slowly 
with metallicity (R7 in \ch{NH3}$\rightarrow$\ch{N2} and R1-R2-R3-R5 in \ch{N2}$\rightarrow$\ch{NH3}). The second 
type of RLS timescales decrease linearly with metallicity; these contain a reactant with multiple atoms of heavy 
elements or both reactants having one heavy element (R1, R2, R3, and R5 in \ch{NH3}$\rightarrow$\ch{N2}). In the 
third type, timescales decrease as a square of increasing metallicity 
(R4 and R6 in \ch{NH3}$\rightarrow$\ch{N2} conversion), in which case the RLS contains multiple molecules with 
multiple heavy elements. Thus as the number of heavy elements increases in the reactants, the RLS timescale 
decreases faster with increasing metallicity. Also, the effect of metallicity on the timescale of the RLSs is 
much more complex than in the \ch{CH4}$\rightleftarrows$CO conversion due to the presence of a relatively 
large number of reactants for \ch{NH3}$\rightleftarrows$\ch{N2} with different metallicity dependence.

\begin{figure}[htb!]
	\centering
	\includegraphics[width=0.4\textwidth]{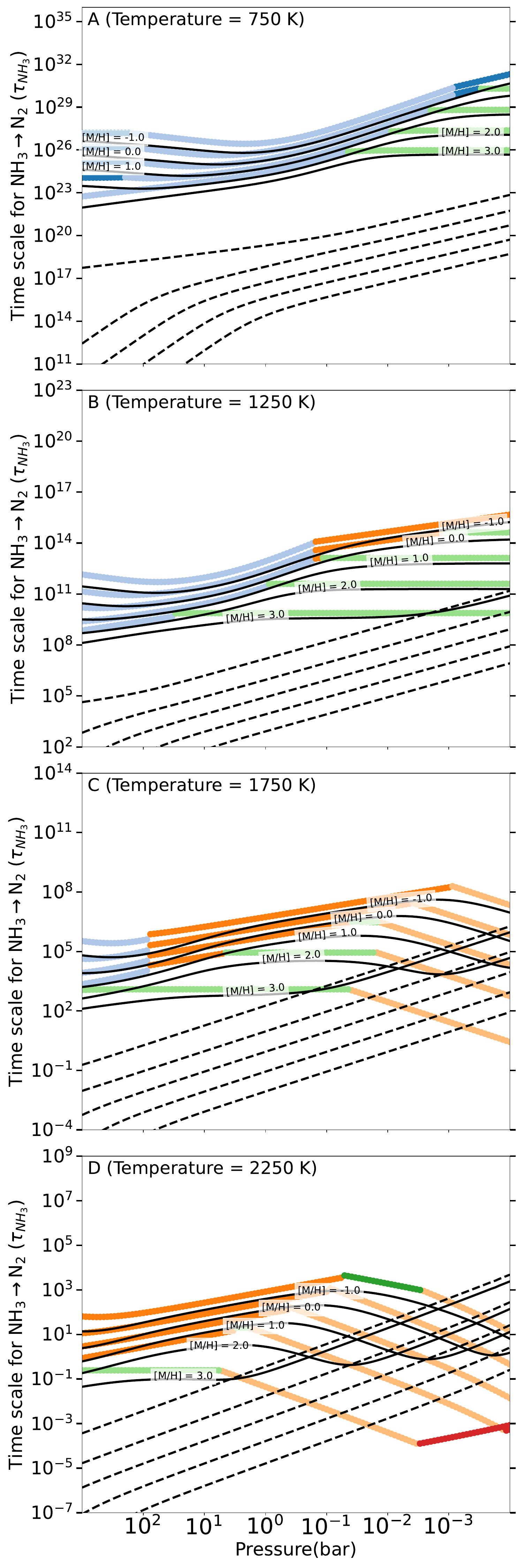}
	\includegraphics[width=0.4\textwidth]{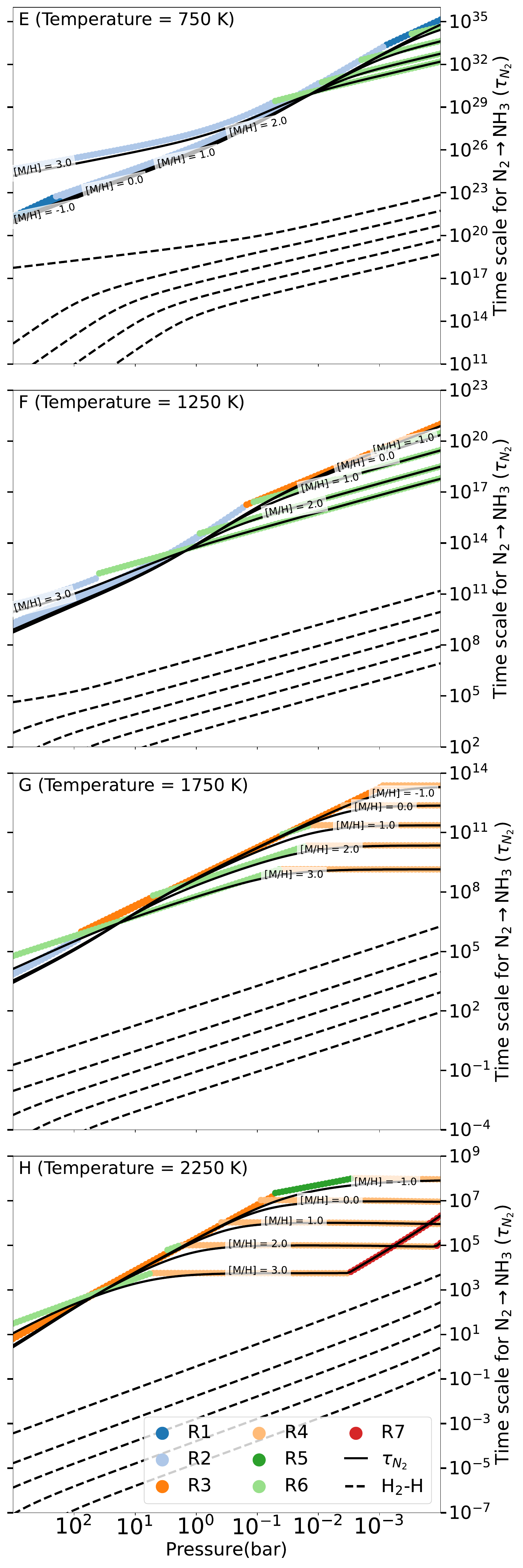}
	\caption{The timescale of conversion of \ch{NH3}$\rightarrow$\ch{N2} ($\tau_{\ch{NH3}}$; left column) 
and \ch{N2}$\rightarrow$\ch{NH3} ($\tau_{\ch{N2}}$; right column) for four different 
		temperatures (750 K, 1250 K, 1750 K, and 2250 K) with five different metallicities (0.1, 1, 10, 
100, and 1000 $\times$ solar metallicity). The colored lines represent the timescale of RLS the corresponding 
temperature-pressure value, and the black dashed lines represent  $\tau_{\ch{H2}} \times \frac{3[\ch{N2}]}{\ch{H2}}$. 
The solid black line is $\tau_{\ch{NH3}}$ (left column) and $\tau_{\ch{N2}}$ (right column), labeled with the respective 
metallicity.} \label{fig:time_scale_NH3_N2}
\end{figure}

\subsection{Timescale of \ch{NH3} and \ch{N2}}
The chemical timescales for the interconversion of \ch{NH3}$\rightleftarrows$\ch{N2} are as follows \citep{Tsai2018}:

\begin{align}
\tau_{\ch{NH3}} &= \frac{1}{2}\Big(\textit{ }\frac{[\ch{NH3}]}{\text{Reaction rate of RLS}} + \tau_{\ch{H2}} 
\times \frac{3[\ch{N2}]}{\ch{H2}}\textit{ }\Big) \label{eq:main_3}\\ 
\tau_{\ch{N2}} &= \frac{[\ch{N2}]}{\text{Reaction rate of RLS}} + \tau_{\ch{H2}} \times \frac{3[\ch{N2}]}{\ch{H2}} 
\label{eq:main_4}.
\end{align}
Here, $\tau_{\ch{NH3}}$ and $\tau_{\ch{N2}}$ are the chemical timescales of conversion of \ch{NH3}$\rightarrow$\ch{N2}
and \ch{N2}$\rightarrow$\ch{NH3} respectively. [\ch{NH3}], [\ch{N2}], and [\ch{H2}] are the number densities of \ch{NH3},
\ch{N2}, and \ch{H2} respectively. The interconversion timescale of \ch{H2}$\rightleftarrows$\ch{H} is $\tau_{\ch{H2}}$ 
and ‘Reaction rate of RLS’ is the rate of the RLS relevant for the desired temperature-pressure and metallicity values. 
The first term in Equations \ref{eq:main_3} and \ref{eq:main_4} is related to the timescale of the RLS. The second 
term is related to the interconversion of \ch{H2}$\rightleftarrows$\ch{H}, which is required because during the 
conversion of \ch{NH3}$\rightleftarrows$\ch{N2}, the \ch{H2}$\rightleftarrows$\ch{H} interconversion also occurs. 
Reconversion of \ch{H2}$\rightarrow$\ch{H} and \ch{H}$\rightarrow$\ch{H2} is also required to achieve the steady-state.

In Figures \ref{fig:time_scale_NH3_N2} , we have respectively plotted $\tau_{\ch{NH3}}$ 
(Equation \ref{eq:main_3}) and $\tau_{\ch{N2}}$ (Equation \ref{eq:main_4}) for four different temperatures 
(750 K, 1250 K, 1750 K, and 2250 K) and five different metallicities (0.1, 1, 10, 100, and 1000 $\times$ solar 
metallicity). The contribution from the first and second terms in Equations \ref{eq:main_3} and \ref{eq:main_4} 
are plotted separately in colored and black dashed lines, respectively. The rate of increase of these two terms is 
a strong function of pressure and temperature. Although the magnitude of the first term is greater than 
the second term at high pressure and low temperature, the rate of increase with increasing temperature and 
decreasing pressure is larger for the second term in $\tau_{ \ch{NH3}}$ (Equation \ref{eq:main_3}). Therefore, 
the relative importance of the \ch{H2}$\rightleftarrows$\ch{H} conversion term changes appreciably over the 
parameter space, especially for the \ch{NH3}$\rightarrow$\ch{N2} conversion.

For 750 K (panel `A and E' in Figures  \ref{fig:time_scale_NH3_N2}), as the metallicity increases 
from 0.1 to 1000 $\times$ solar metallicity, the contribution of the first term in Equation \ref{eq:main_3} ($\tau_{ \ch{NH3}}$)  
is decreased by five orders of magnitude. However for Equation \ref{eq:main_4} ($\tau_{ \ch{N2}}$), in the high pressure 
region, the first term increases by more than three orders of magnitude and then it gradually starts to decrease with 
decreasing pressure, and when the pressure reduces to $\approx$ 0.001 bar, it decreases by nearly three orders of magnitude.  
The second term in Equations \ref{eq:main_3} and \ref{eq:main_4} does not have a contribution at 750 K, although it 
increases with metallicity. The increase is highest at the high-pressure region ($\sim$ ten orders of magnitude), and 
the rate of increase gradually decreases with decreasing pressure (increase is about five orders of 
magnitude at the lowest pressure considered). 

\begin{figure}[h]
	\centering
	\includegraphics[width=0.49\textwidth]{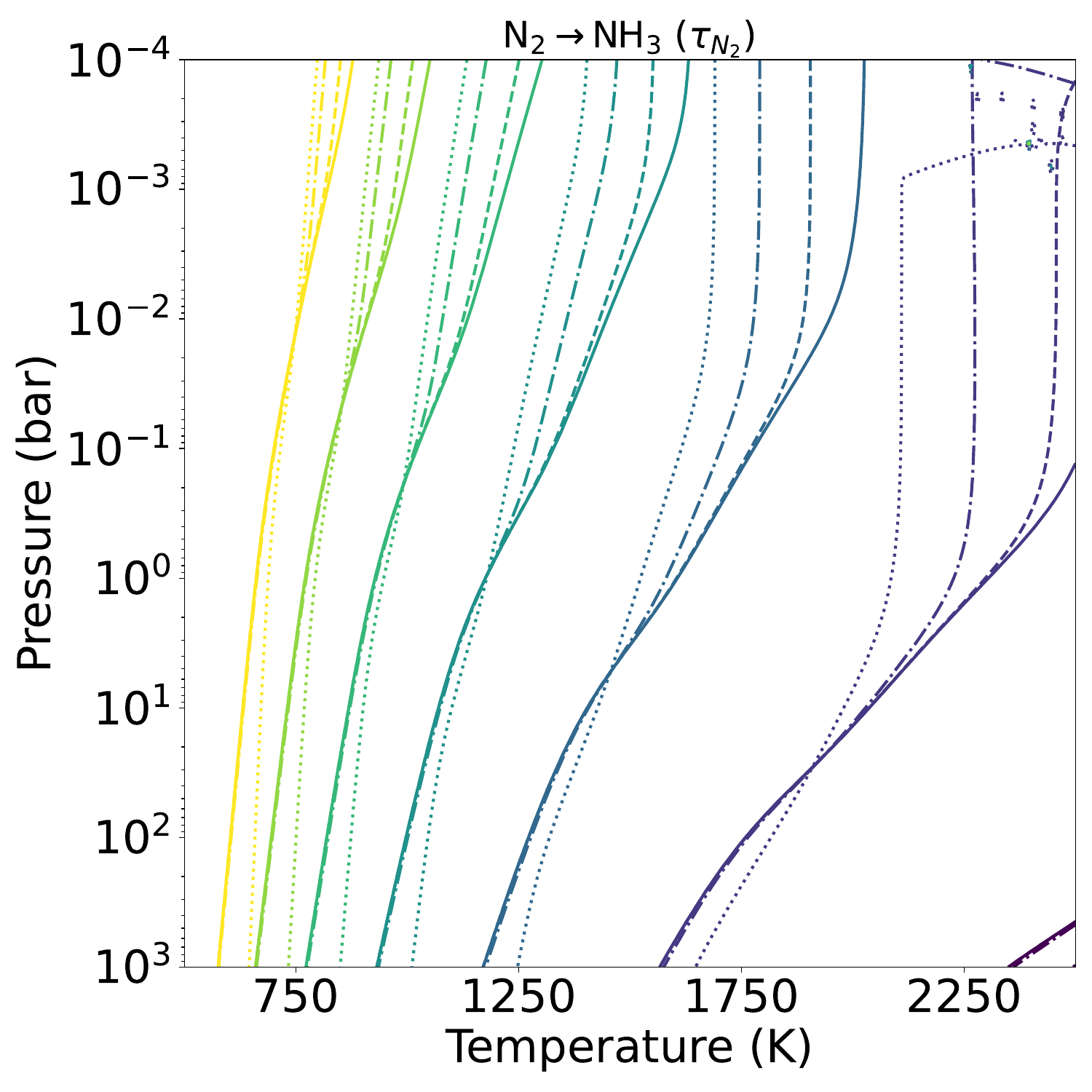}
	\includegraphics[width=0.49\textwidth]{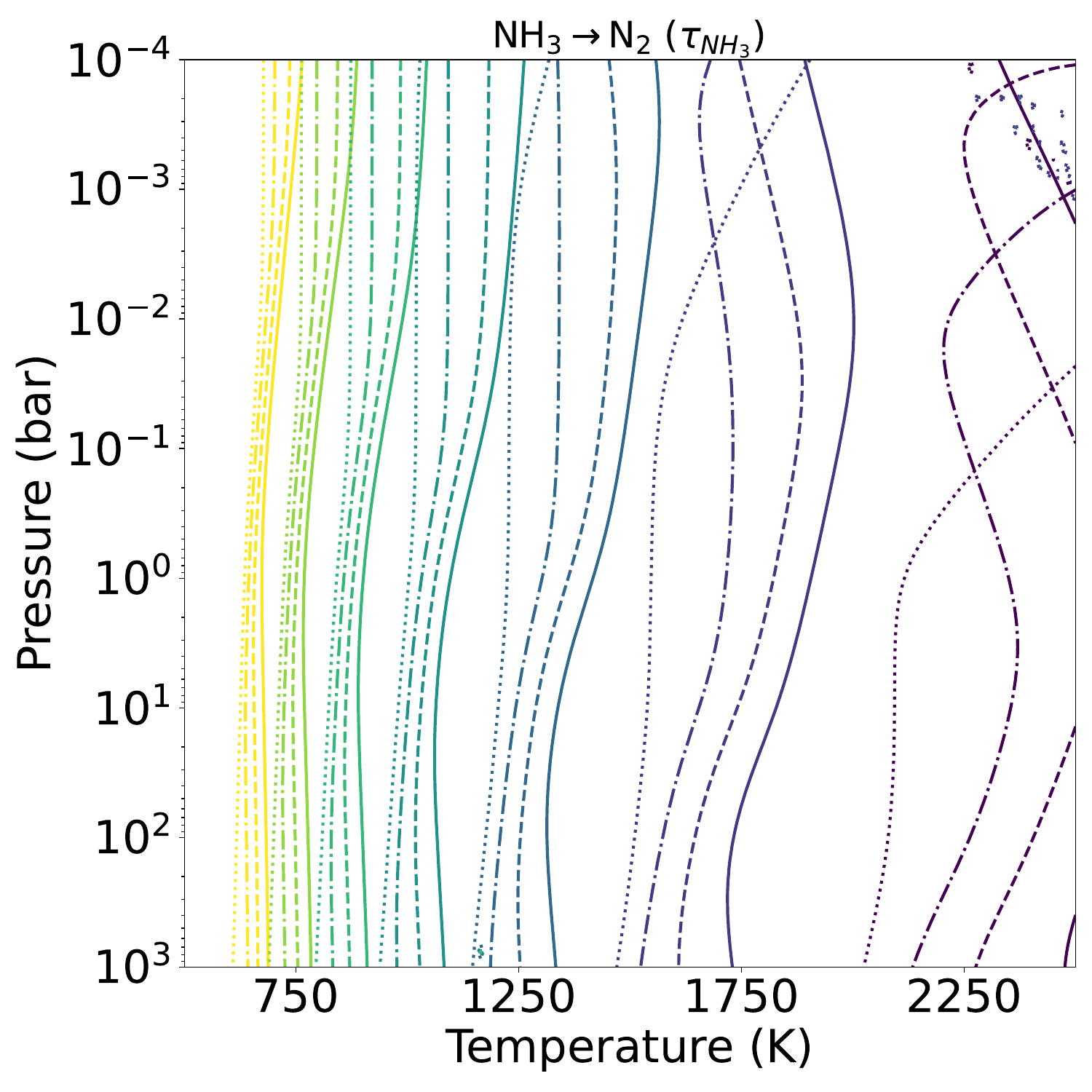}
	\caption{The constant contour line of $\tau_{\ch{N2}}$ (left) and $\tau_{\ch{NH3}}$ (right) are plotted for four 
		different metallicities (0.1, 1, 10, and 1000 $\times$ solar metallicity). The solid, dashed,
		dotted-dashed, and dotted lines show 0.1, 1, 10, and 1000 $\times$ solar metallicity, respectively. The colored lines from blue to yellow represent the constant 
		contours of timescales in log10 scale for 0, 5, 10, 15, 20, 25, and 30.} \label{fig:contour_chemical_time_scale}
\end{figure}

Panels `B and F', `C and G', and `D and H' in Figures \ref{fig:time_scale_NH3_N2} show the pressure variation of 
timescales for the temperatures 1250 K, 1750 K, and 2250 K, respectively. As the temperature increases, the strength of 
these two terms starts to decrease, though at different rates. In the high-pressure region, $\tau_{\ch{NH3}}$ 
decreases with increasing metallicity, whereas in the low-pressure region where the second term dominates, it increases 
with increasing metallicity. As the temperature increases from 1250 K to 2250 K, R4 and R6 become the RLS in 
the high metallicity region, and the RLS term (first term) in $\tau_{\ch{NH3}}$ decreases by more than six 
orders of magnitude. However, for $\tau_{\ch{N2}}$ (Equation \ref{eq:main_4}), for 1250 K: $P>1$ bar, 1750 K: 
$P>10$ bar, and 2250 K: $P>100$ bar, the first term increases with increasing metallicity, and for other pressure regions, 
it decreases with increasing metallicity. Also, the second term increases by around five to seven orders of 
magnitude, but it does not contribute to $\tau_{\ch{N2}}$. However, in $\tau_{\ch{NH3}}$, as the temperature 
increases, the second term becomes comparable to the first term and starts to contribute to $\tau_{\ch{NH3}}$, 
particularly at low-pressure and high-metallicity regions. In this region, $\tau_{\ch{NH3}}$ increases with 
increasing metallicity; otherwise, the RLS term (first term) dominates in $\tau_{\ch{NH3}}$, and $\tau_{\ch{NH3}}$ decreases with metallicity. Clearly, in the region where the second term starts to contribute significantly, the metallicity dependence on the mixing ratio 
of \ch{NH3} becomes complex.

In Figure  \ref{fig:contour_chemical_time_scale},  we have plotted the constant contour lines of $\tau_{\ch{N2}}$ 
and $\tau_{\ch{NH3}}$ in temperature and pressure parameter space for 0.1 (solid), 1 (dashed), 10 (dotted-dashed), 
and 1000 (dotted) $\times$ solar metallicities. The lines from blue to yellow are the constant contour lines of 
$10^{0}$ to $10^{30}$ s. Both the conversion timescales decrease with the increasing temperature and pressure. 
However, the dependence of the timescale on the metallicity is complex and changes with pressure levels. As 
metallicity increases, the constant contour of $\tau_{ \ch{N2}}$ shifts towards the low-temperature region when 
R4 to R6 become RLS and towards the high-temperature region when other reactions become RLS. When the second term 
in Equation \ref{eq:main_4} ($\tau_{ \ch{NH3}}$) is dominant, the increase in metallicity shifts the contour of  
$\tau_{ \ch{NH3}}$ towards the high temperature region, and in the region where first term (RLS term) dominates, 
it shifts towards low temperature with increasing metallicity. 

We have compared the \ch{NH3}, \ch{N2} chemical timescales with the widely used analytical
expressions from \cite{Zahnle2014}. We found that the analytical expressions do not give the correct value for the 
entire parameter space; therefore, they should be used with caution while calculating the quench pressure level (more discussion 
can be found in Appendix A.2).

\begin{figure}[h]
	\centering
	\includegraphics[trim={0cm 0cm 8cm 0cm},clip, width=1\textwidth]{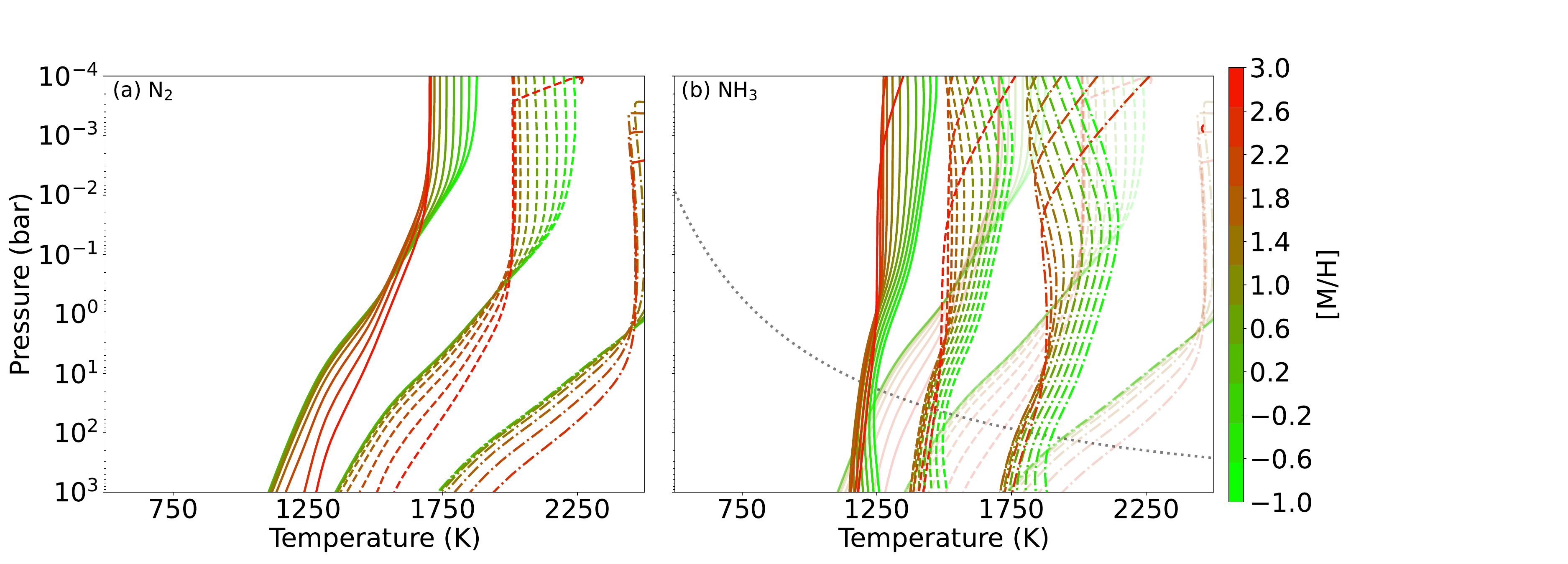}
	\caption{ The contour of (a) $\tau_{\ch{N2}} /\tau_{mix} = 1$ and (b) $\tau_{\ch{NH3}} /\tau_{mix} = 1$ for 
		assorted parameters are shown. The color bar (each value represents one metallicity contour) represents the metallicity 
		value for the plotted quenched curve. The three sets of lines are for different values of $K_{zz}$ coefficients; solid, 
		dashed, and dashed-dotted lines are for  $K_{zz}$ = $10^4$, $10^8$, and $10^{12}$ cm$^2$ s$^{-1}$ respectively. 
		In panel (b), the quenched curves of \ch{N2} are shown in faded lines with the same color and the equal-abundance 
		curves for 0.1 $\times$ solar metallicity ([M/H] = -1) is also shown.}
	\label{quenching_contiur_plot}
\end{figure}
\subsection{Effect of metallicity on the quench level}
 
We compare the previously calculated vertical mixing timescale for the different transport
strengths \citep{Soni2023} with $\tau_{ \ch{N2}}$ and $\tau_{ \ch{NH3}}$ and find the quenched
curve for the same range of metallicity values. Quenched curves are the contour lines in
pressure and temperature space on which the chemical and vertical mixing time scales are
equal. When a thermal profile of the planet is plotted along with a quenched curve of the
relevant $K_{zz}$ and metallicity, then they cross each other at the quench level if it exists.
Figure \ref{quenching_contiur_plot} shows the quenched curves for $K_{zz}$ = $10^4$ cm$^2$
s$^{-1}$ (solid line), $10^8$ cm$^2$ s$^{-1}$ (dashed line), and $10^{12}$ cm$^2$ s$^{-
1}$(dashed-dotted line). Contour lines for eleven metallicities are plotted for every $K_{zz}$
value. This figure shows the general behaviour of how the quench level of \ch{NH3} and \ch{N2}
changes with the $K_{zz}$ and metallicity. For increasing the $K_{zz}$ value (decreasing
vertical mixing time scale), the quenched curve moves towards higher pressure and
temperature region (see different line-style in Figure \ref{quenching_contiur_plot}) because
increasing the pressure and temperature shorter the chemical time scale, which is required to
match with the higher $K_{zz}$ value or smaller vertical mixing time scale. For a fixed $K_{zz}$
value, the increase in metallicity has two effects on the quenched curve of \ch{N2}: for the
region where the chemical time scale increases with metallicity, the quenched curve shifts in
high-temperature region. The region where the chemical time scale decreases, with the
metallicity, the quenched curve shifts toward low-temperature region, and the chemical time scale 
increases which compensates the metallicity effect on chemical time scale. 

The \ch{NH3} quenched 
curves move towards the low-temperature region with increasing metallicity. In the region where 
the second term in Equation \ref{eq:main_3} dominates, it shifts towards the high-temperature and high-pressure regions.

\section{\ch{HCN}}\label{sec:hcn}
\begin{figure}[h]
	\centering
	\includegraphics[width=1\textwidth]{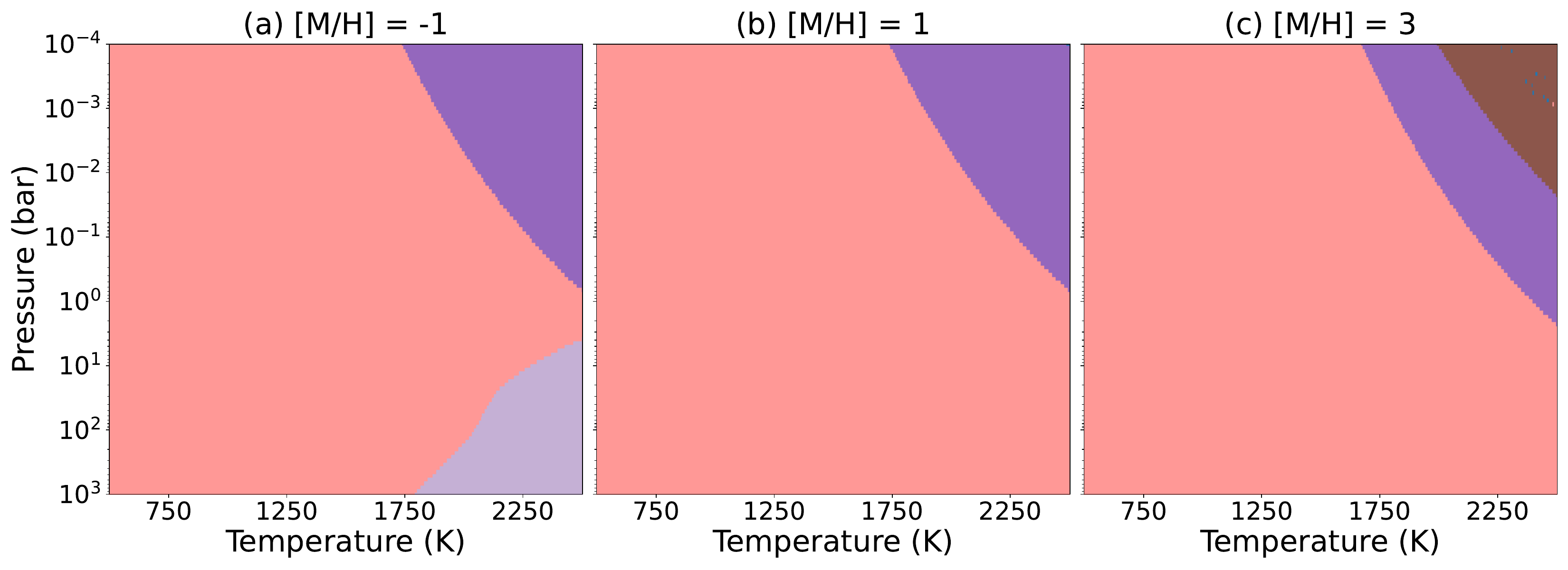}
	\caption{ The pressure-temperature range of different rate-limiting steps is shown for three 
		different metallicities. Each color corresponds to the different RLS in Figure \ref{fig:conversion_N2}.} 
	\label{fig:RLS_area_HCN}	
\end{figure}

The abundance of \ch{HCN} is generally lower compared to \ch{NH3} and \ch{N2} in the majority of the 
pressure-temperature range. However, as the temperature increases, the \ch{NH3} abundance starts to 
decrease; therefore, for certain cases, the \ch{HCN} abundance becomes comparable to or more than the \ch{NH3} 
abundance. In the high metallicity region (log10(HCN)$<-10$ for 100 $\times$ solar metallicity), \ch{HCN} 
can exceed the \ch{NH3} abundance at low-pressure ($P < 10^{-6} $ bar for [M/H]$>$  100 $\times$ solar metallicity) 
and high-temperature regions. The quenching of \ch{HCN} can affect the quenched abundance of \ch{NH3} due to 
its thermal decomposition. In fact, \cite{Zahnle2014} reported that the thermal decomposition of \ch{HCN} can increase the 
\ch{NH3} abundance by 10\%. The conversion scheme for \ch{HCN}$\rightarrow$\ch{NH3} has been studied by several 
authors \citep{Moses2010, Moses2011, Tsai2018, Dash2022}. \cite{Moses2010} find the conversion scheme for \ch{HCN}$\rightarrow$\ch{NH3}, 
in which \ch{CH3NH2} radical is produced via successive hydrogenation of \ch{HCN}, which decomposes into \ch{NH3} 
and \ch{CH4} \citep{Tsai2018}. In another pathway, \ch{HCN} gets converted into \ch{NH3} through HNCO as an intermediate molecule 
\citep{Moses2011, Tsai2018}. Recently, \cite{Dash2022} reported a scheme, in which \ch{HCN} is converted to \ch{H2CN}, 
which reacts with H to produce N and \ch{CH3}, and N is converted into \ch{NH3} via successive hydrogenation. We 
used our network analysis tool to find the conversion schemes for \ch{HCN}$\rightarrow$\ch{NH3} and we found three conversion 
pathways which are listed in Table~\ref{Table1} and their parameter region is given in Figure \ref{fig:RLS_area_HCN}. All three pathways are involved in the conversion but are important in different parameter spaces. The 
pathway involving HNCO (second scheme in Table~\ref{Table1}) as an intermediate molecule remains dominant in 
most of the parameter space. The first pathway i.e., via \ch{H2CN} is important only in a small parameter space 
(low-metallicity, high-temperature and high-pressure). The third conversion path is dominant in the high 
metallicity, high-temperature 
and low-pressure region, where \ch{HCN} is converted into N and subsequently into \ch{NH3}.

\begin{table}[ht]
	\caption{\label{Table1}Different conversion schemes for \ch{HCN}$\rightarrow$\ch{NH3}.}
	\begin{center}
		\begin{tabular}{|l|l|l|}
			\hline
			First conversion path                      & Second conversion path                        & Third conversion path                        \\
			\hline
			\ch{HCN + H + M} $\rightarrow$ \ch{H2CN + M} &\ch{HCN + OH} $\rightarrow$ \ch{HNCO + H} (RLS) &  \ch{HCN + H}   $\rightarrow$ \ch{CN + H2} (RLS)  \\
			\ch{H2CN + H}    $\rightarrow$ \ch{CH3 + N} (RLS)  &\ch{HNCO + H} $\rightarrow$ \ch{CO + NH2} &  \ch{CN + O}    $\rightarrow$ \ch{N + CO} (RLS)   \\
			\ch{N + H2}      $\rightarrow$ \ch{NH + H}   &\ch{NH2 + H2} $\rightarrow$ \ch{NH3 + H}  &  \ch{N + H2}    $\rightarrow$ \ch{NH + H}    \\
			\ch{NH + H2}     $\rightarrow$ \ch{NH2 + H}  &\ch{H2O + H}  $\rightarrow$ \ch{OH + H2}  &  \ch{NH + H2}   $\rightarrow$ \ch{NH2 + H}   \\
			\ch{NH2 + H2}    $\rightarrow$ \ch{NH3 + H}  &---------------------------------------   &  \ch{NH2 + H2}  $\rightarrow$ \ch{NH3 + H}   \\
			\ch{CH3 + H2}    $\rightarrow$ \ch{CH4 + H}  &\ch{HCN + H2O}$\rightarrow$ \ch{NH3 + CO} &  \ch{OH + M}    $\rightarrow$ \ch{O + H + M} \\
			\ch{H + H + M}   $\rightarrow$ \ch{H2 + M}   &                                          &  \ch{H2O + H}   $\rightarrow$ \ch{OH + H2}   \\
			-----------------------------------------    &                                          & \ch{H + H + M} $\rightarrow$ \ch{H2 + M}    \\
			\ch{HCN + 3 H2}  $\rightarrow$ \ch{NH3 + CH4}&                                          &  ------------------------------------------  \\
			&                                          &  \ch{HCN + H2O} $\rightarrow$ \ch{NH3 + CO}  \\
			\hline
		\end{tabular}
	\end{center}
\end{table}




\subsection{Timescale of HCN}\label{S5P1}

\begin{figure}[b!]
	\centering
	\includegraphics[width=1\textwidth]{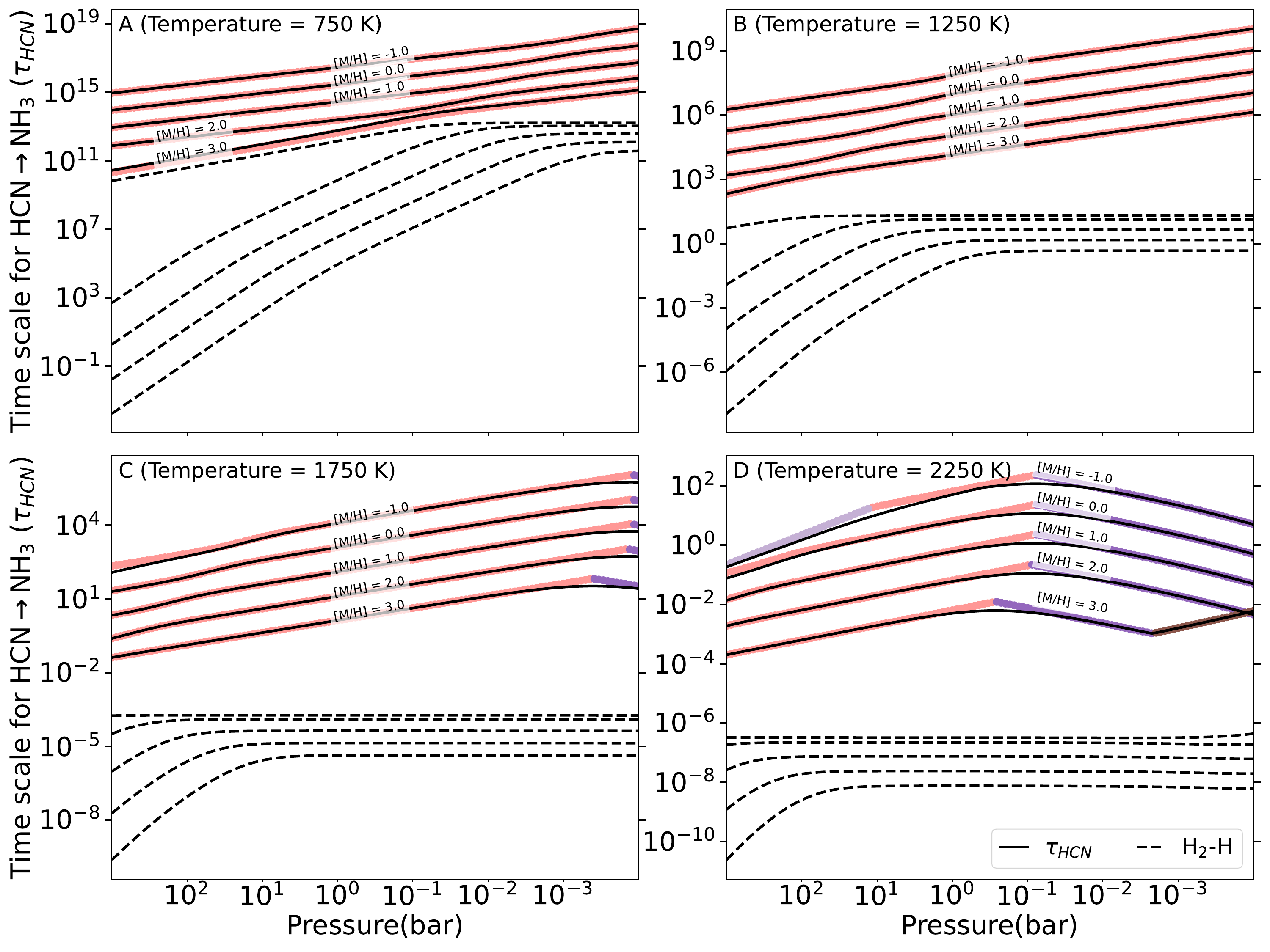}
	\caption{The timescale of conversion of \ch{HCN}$\rightarrow$\ch{NH3} ($\tau_{\ch{HCN}}$) for four different 
		temperatures (750 K, 1250 K, 1750 K, and 2250 K) with five different metallicities 
(0.1, 1, 10, 100, and 1000 $\times$ solar metallicity). The colored lines represent the timescale of RLS in the 
corresponding temperature-pressure
		value, and the black dashed lines represent  $\tau_{\ch{H2}} \times \frac{3[\ch{HCN}]}{\ch{H2}}$. The solid black line is 
		$\tau_{\ch{HCN}}$, labeled with the respective metallicity.} \label{fig:time_scale_HCN-NH3}
\end{figure}

The chemical timescale of \ch{HCN} follows the same convention as \ch{NH3} and \ch{N2} and is given by 
the following equation:
\begin{align}
\tau_{\ch{HCN}} &= \Big(\textit{ }\frac{[\ch{HCN}]}{\text{Reaction rate of RLS}} + \tau_{\ch{H2}} 
\times \frac{3[\ch{HCN}]}{\ch{H2}}\textit{ }\Big) \label{eq:main_5}.
\end{align}
Here the first term is related to the RLS, and the second term is related to the conversion of 
$\ch{H2}\rightleftarrows\ch{H}$. For the second and third conversion schemes, the second term 
does not apply; the first term is used to calculate $\tau_{\ch{HCN}}$. The second term will 
only come when the \ch{HCN}$\rightarrow$\ch{NH3} conversion takes place through \ch{H2CN} 
(first scheme, Table~\ref{Table1}) (This is assuming that $\ch{H2O}+\ch{H}\rightarrow\ch{OH}+\ch{H2}$ is 
fast enough and does not affect the $\ch{HCN}\rightarrow\ch{NH3}$ conversion). However, its 
strength is always significantly lower than the RLS term and hence it does not contribute to $\tau_{\ch{HCN}}$. 
It can be seen from Figure \ref{fig:time_scale_HCN-NH3}, in which the conversion timescale of 
$\ch{HCN}\rightarrow\ch{NH3}$ is plotted for assorted temperatures and metallicities, that the second 
term is unimportant for $\tau_{\ch{HCN}}$.

\begin{figure}[h]
	\centering
	\includegraphics[width=0.45\textwidth]{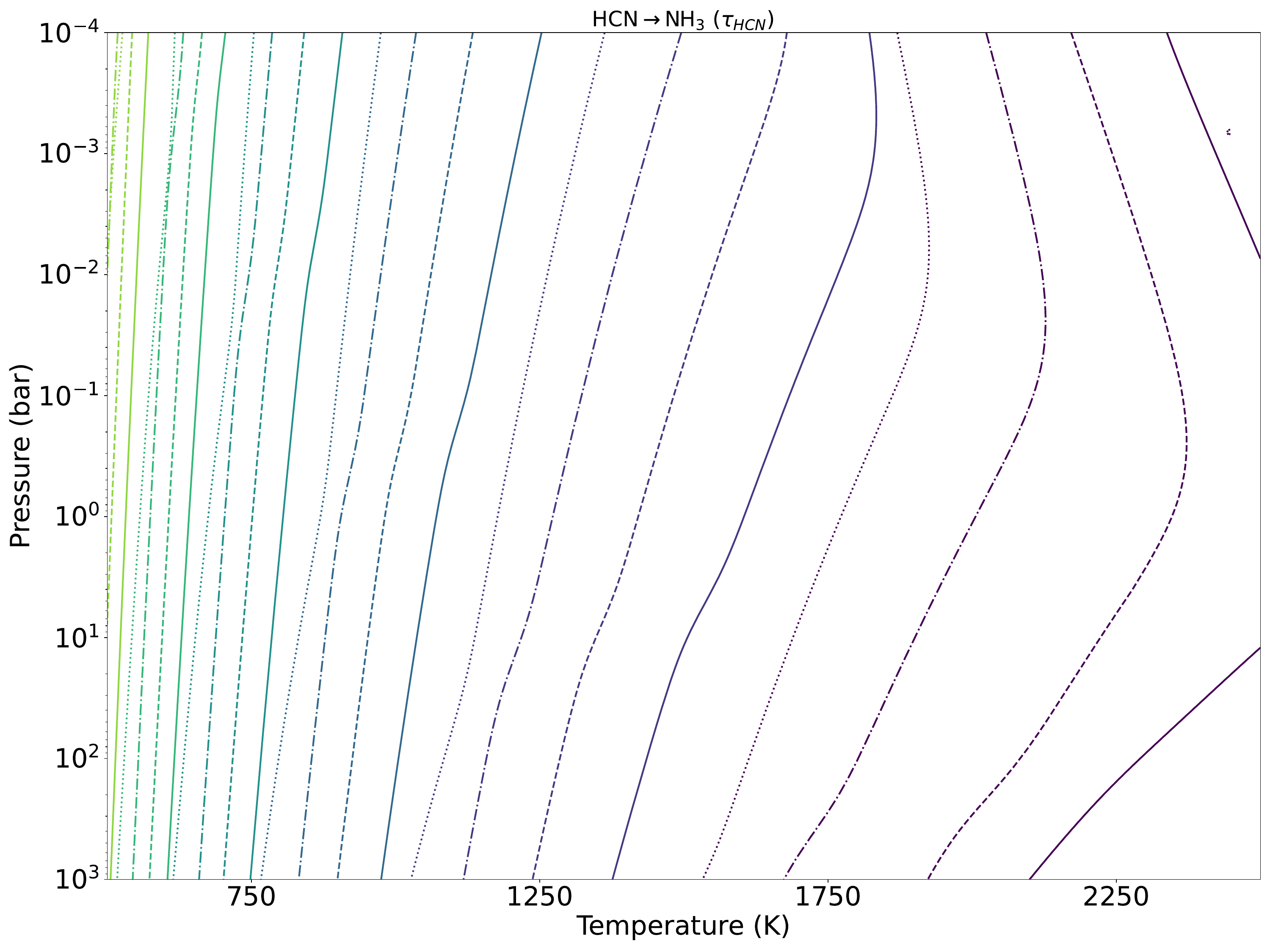}
	\includegraphics[width=0.45\textwidth]{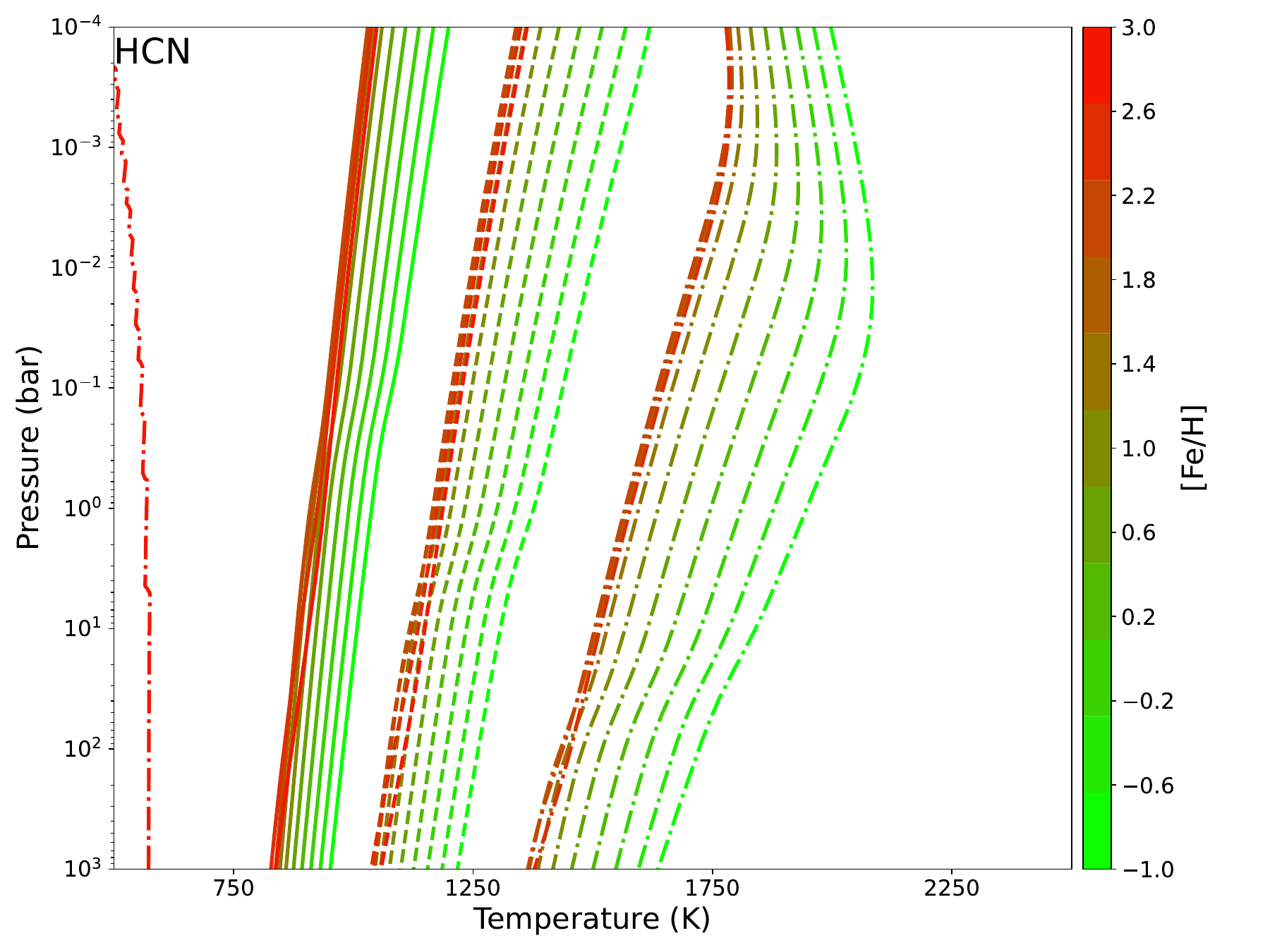}
	\caption{The left panel shows the contour of the chemical timescale of \ch{HCN} (following the same convention as Figure \ref{fig:contour_chemical_time_scale}. The right panel shows the quenched curve of \ch{HCN} (following the same convention as Figure \ref{quenching_contiur_plot}).} 
	\label{fig:quench_HCN}	
\end{figure}

In most of the parameter region, $\ch{HCN + OH} \rightarrow \ch{HNCO + H}$ is the RLS, which makes 
$\tau_{\ch{HCN}}$ decrease linearly with metallicity. Also, increasing the temperature and pressure 
decrease $\tau_{\ch{HCN}}$. In the region where $\ch{HCN + H }\rightarrow \ch{CN + H2}$ becomes 
the RLS,  $\tau_{\ch{HCN}}$ increases slowly with increasing metallicity. The chemical timescale of 
\ch{HCN} is many orders of magnitude less than the \ch{N2} and \ch{NH3} chemical timescales. At 
low-temperature, this difference is around ten orders of magnitude; however, this gap decreases to 
a few orders as the temperature increases from 500 K to 2500 K. Therefore, \ch{HCN} quenches well above the 
quench level of \ch{NH3} and \ch{N2} in the hot atmosphere. In contrast, in the cold atmosphere, 
HCN is quenched along with \ch{NH3} and \ch{N2}. Temperature and pressure also play a crucial 
role in defining the quench level. $\tau_{\ch{HCN}}$ increases around three orders of magnitude 
with decreasing pressure from 10$^3$ to 10$^{-4}$ bar, whereas $\tau_{ \ch{N2}}$ increases by 
more than ten orders of magnitude, and $\tau_{ \ch{NH3}}$ is a comparatively weak function of pressure for $T<$ 1250 K 
and it decreases with pressure for $T>$ 1250 K where R4 dominates. In Figure  \ref{fig:quench_HCN} (left panel), we have plotted the 
constant contour lines of $\tau_{\ch{HCN}}$ with the same convention that we used for Figure 
\ref{fig:contour_chemical_time_scale}. $\tau_{\ch{HCN}}$ decreases with increasing temperature 
and pressure for the region of the parameter space where the first scheme is dominant. In the 
region where $\ch{HCN} + \ch{H} \rightarrow \ch{CN} + \ch{H2}$ becomes the RLS, $\tau_{\ch{HCN}}$ 
decreases with increasing pressure, and decreasing metallicity.

In Figure  \ref{fig:quench_HCN} (right panel), we have plotted the contour line on which the dynamical and chemical 
conversion timescales of \ch{HCN} are equal. We follow the same convention as Figure \ref{quenching_contiur_plot}. 
Only one RLS is dominant in most of the parameter range resulting in a simpler behavior of the \ch{HCN} quenched 
curve on the temperature-pressure and metallicity space. The quenched curve of \ch{HCN} shifts towards low-temperature 
and low-pressure regions with increasing metallicity for most of the parameter space. 

We have compared the \ch{HCN} chemical timescales with the widely used analytical expressions from \cite{Zahnle2014}, 
similar to \ch{NH3} and \ch{N2}. We found that the analytical expressions for HCN also do not give the correct value for the 
entire parameter space (more discussion can be found in Appendix A.2).
 
\subsection{Quenched abundance of HCN}\label{S5P2}

Quenching approximation is a simple and computationally efficient method to constrain the transport 
abundance of dominant molecules in the atmosphere. However, this method also possesses some limitations 
\citep{Tsai2017}. The chemical timescale is computed using the chemical equilibrium, and the true 
chemical timescale can deviate if the reactant of the RLS deviates from the equilibrium abundance. 
The other limitation is that molecule's abundance can deviate from its thermochemical equilibrium abundance 
well below its actual quench level if it remains in equilibrium with molecules that 
have already quenched. Some multi-dimensional studies have suggested that horizontal mixing (zonal and 
meridional wind) can also affect the \ch{NH3} abundance \citep{Agundez2014B, Drummond2020, Baeyens2021, Zamyatina2023}. 
The effect of horizontal mixing dominates over vertical quenching in the high-pressure region (P$>$1 bar) where 
the vertical quench level of \ch{NH3} lies. It changes the \ch{NH3} abundance from its thermochemical equilibrium 
abundance at the vertical \ch{NH3} quench level. This effect cannot be explored in the 1D model, and in this study, 
we did not incorporate horizontal mixing and only considered vertical mixing.   However, the previous limitation 
can be lifted, if we know the quenched abundance of the molecules. As 
discussed in Section~\ref{S5P1}, the chemical timescale of \ch{HCN} is shorter or comparable to the \ch{NH3} 
and CO timescales. As a result, \ch{HCN} is quenched above the quench level of CO and \ch{NH3}. \ch{HCN} remains 
in equilibrium with CO, \ch{NH3}, OH, H, and \ch{H2} in the region where the second chemical scheme is dominant. 
The mixing ratio of \ch{HCN} is given by the following equation:

\begin{equation}
[\ch{HCN}] = k\frac{[\ch{NH3}][\ch{CO}][\ch{H}]}{[\ch{OH}][\ch{H2}]},
\end{equation}
where $k$ is the equilibrium constant. The RLS in the second scheme is $\ch{HCN + OH}\rightarrow \ch{HNCO + H}$, 
and OH remains in equilibrium with \ch{H2O} and \ch{H2}, and \ch{H2}$\leftrightarrows$\ch{H} conversion timescale is faster 
than the chemical timescale of HCN. As a result, $\tau_{\ch{HCN}}$ remains close to its thermochemical equilibrium 
with CO, \ch{H2O}, and \ch{NH3}. We can safely write the quench abundance of \ch{HCN} by the following:
\begin{equation}
[\ch{HCN}_q] = k\frac{[\ch{NH3}_{, q}][\ch{CO}_q][\ch{H}_{eq}]}{[\ch{OH}_{eq}][\ch{H2_{,eq}}]} \label{Eq:HCN}
\end{equation}
where $[\ch{NH3}_{, q}]$ and $[\ch{CO}_q]$ are respectively the non-equilibrium abundance of \ch{NH3} and CO at the \ch{HCN} quench level.

\section{Applying on the Test Exoplanets}\label{sec:apply}
We compare our quenched abundance calculated from quenching approximation with the chemical kinetics model with 
photochemistry switched off. 
We use two test exoplanets, HD 189733 b and GJ 1214 b, the same as in our previous work \citep{Soni2023}. The thermal 
profiles of these exoplanets cross the \ch{NH3-N2} boundary; therefore, transport can play a crucial role in altering the 
atmospheric composition from the thermochemical equilibrium composition. HD 189733 b is a gas giant with an orbital 
period of 2.22 days,  equilibrium temperature $T_{\text{equi}}\approx $ 1200 K, and surface gravity $g_{\text{surface}}\approx 
$ 21.5 $\text{m s}^{-2}$ \citep{Moutou2006}. GJ 1214 b is a Neptune-sized planet with an orbital period of 1.58 days, 
$T_{\text{equi}}\approx $ 600 K, and $g_{\text{surface}}\approx$ 8.9 $\text{m s}^{-2}$ \citep{Charbonneau2009}. To find 
the quenched abundance, we have used the method given in \cite{Soni2023} in which we plot the quenched curve of \ch{NH3}, 
\ch{N2}, and \ch{HCN} on the thermal profile of these exoplanets. The pressure level where the quenched curve intersects 
with the thermal profile gives the quench level. The thermochemical equilibrium mixing ratio at the quench level is compared 
with the chemical kinetics model (with only transport) and found good agreement. 
We have used mixing length of 0.1 and 1 $\times$ pressure scale heights as well as the value calculated using the method described by 
\cite{Smith1998}. A discussion on effect of different mixing lengths is given in Appendix A.1.

\begin{figure}[h]
	\centering
	\includegraphics[width=0.3\textwidth]{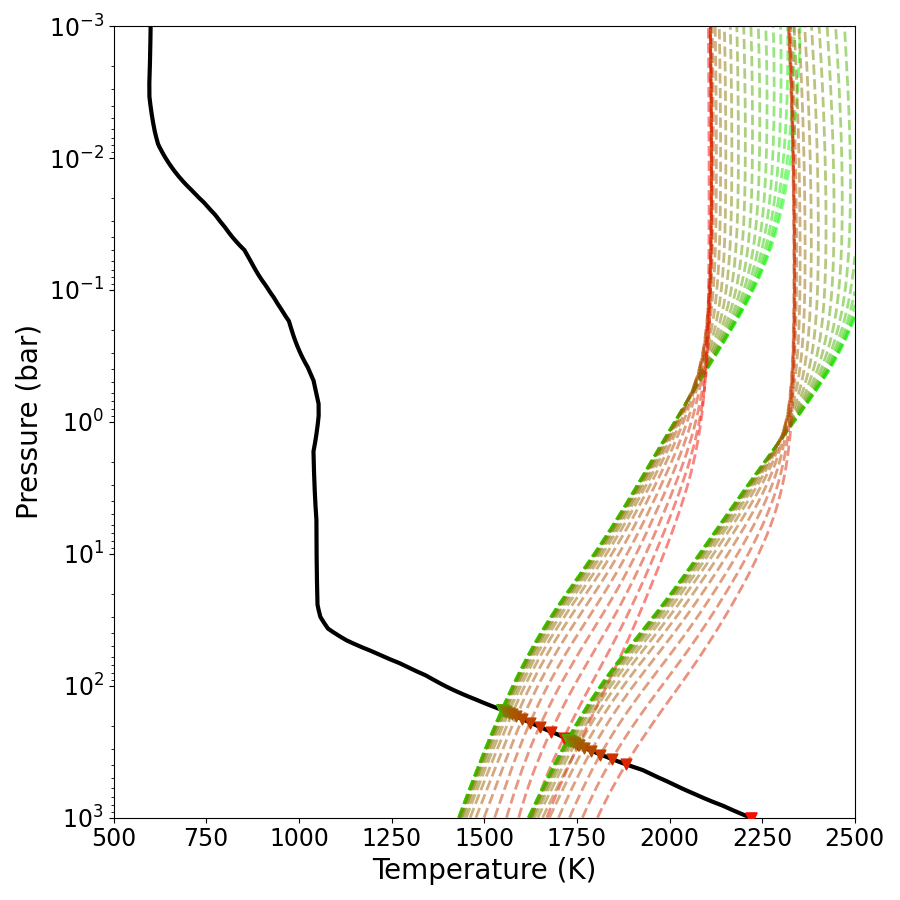}
	\includegraphics[width=0.3\textwidth]{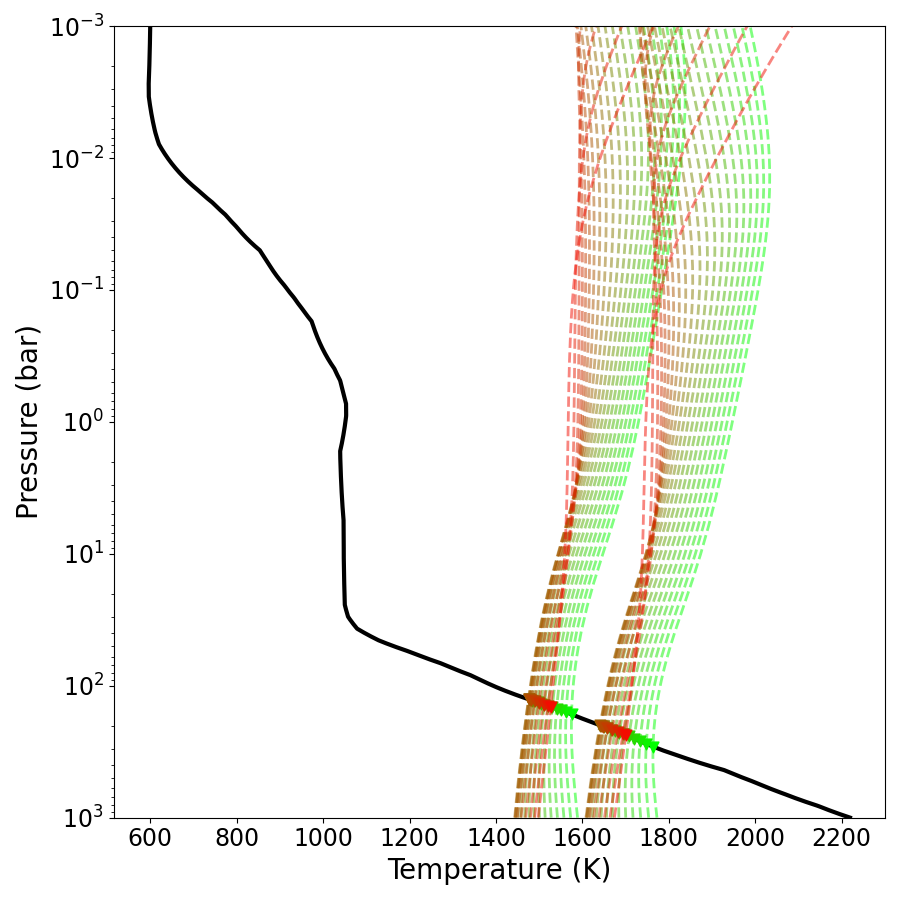}
	\includegraphics[width=0.3\textwidth]{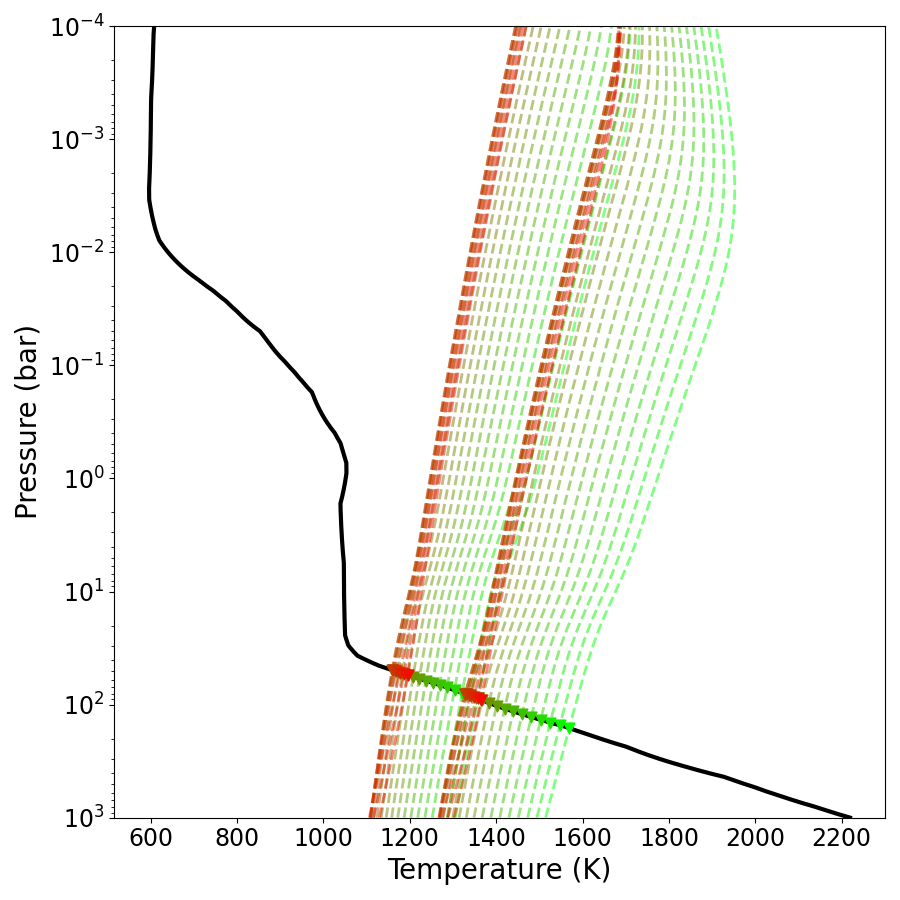}
	\includegraphics[width=0.3\textwidth]{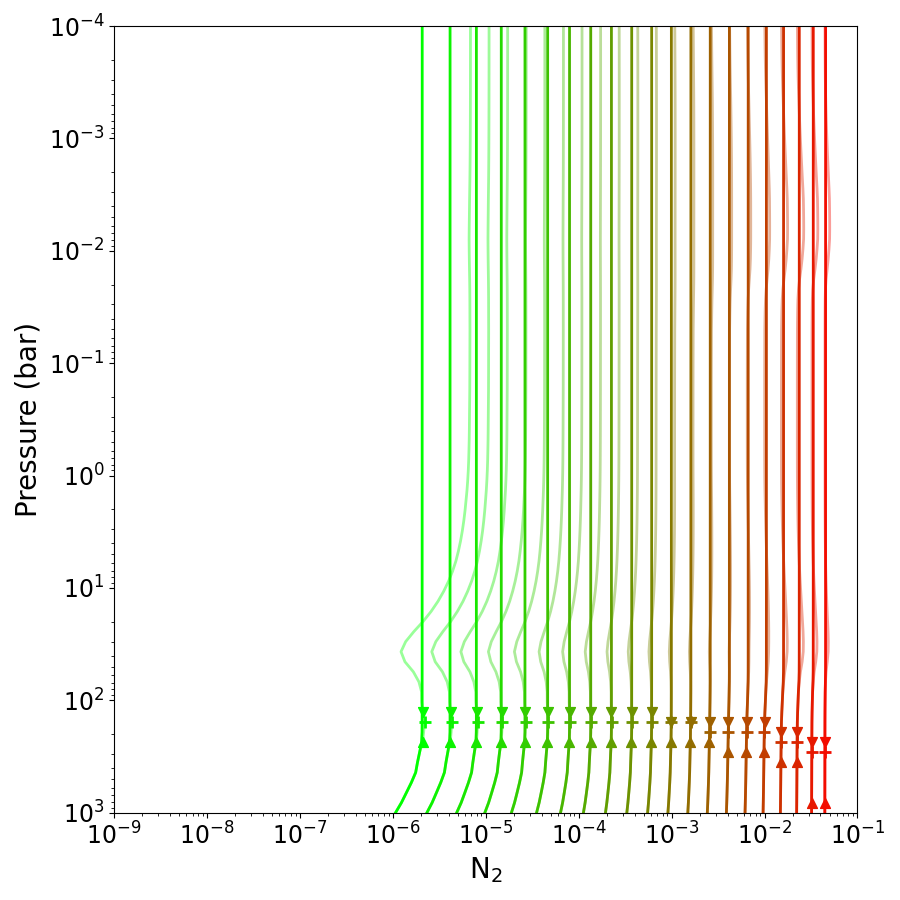}
	\includegraphics[width=0.3\textwidth]{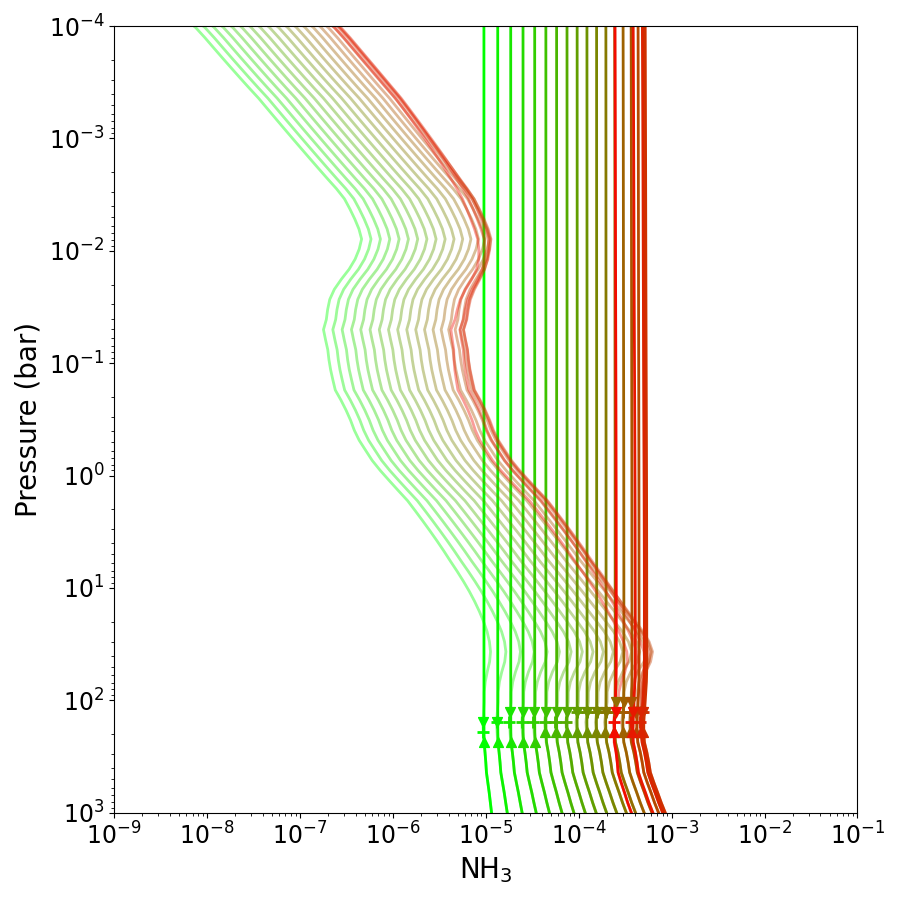}
	\includegraphics[width=0.3\textwidth]{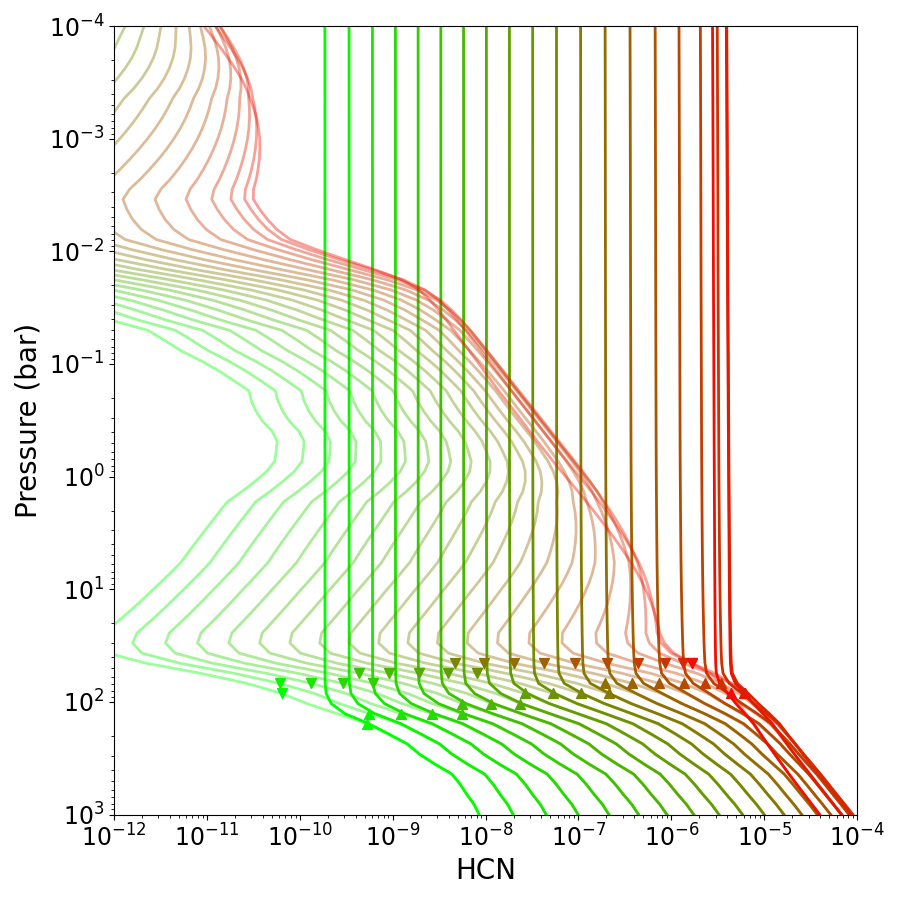}
	
	\caption{The top panel shows the T-P profile overplotted with the quenched curve of \ch{N2} (left), \ch{NH3} (middle) 
and \ch{HCN} (right) for 20 different metallicities (green to red lines are from 0.1 to 1000 $\times$ solar 
		metallicity). The quenched curve is calculated for $K_{zz}$ = 10$^9$ cm$^2\text{ s}^{-1}$ 
		and assuming the mixing length is equal to 0.1 $\times$ (upper triangles) and 1 (lower triangles) atmospheric scale height. The 
		bottom panel shows the mixing ratio of \ch{N2} (left), \ch{NH3} (middle) and \ch{HCN} (right), for the 
		same set of metallicities. The colored lines are the output of the chemical kinetics model and the 
		corresponding faded colored lines are the equilibrium abundances. The `+' symbol is the quench level 
		calculated using the Smith method \citep{Smith1998}. }\label{fig:SN}
\end{figure}

\subsection{GJ 1214 b}
In Figure \ref{fig:SN}, we have over-plotted the quenched curve of \ch{N2} (left), \ch{NH3} (middle) and \ch{HCN} (right) with 
the thermal profile of  GJ 1214 b, which is adopted from \cite{Charnay2015}. The quench level lies on the pressure 
level where the temperature falls sharply with decreasing pressure. As a result, the quench level for different
metallicity remains near the same pressure level (See figure \ref{quenching_contiur_plot}). \ch{NH3} and \ch{N2} 
quench at the same pressure level, and \ch{HCN} quenches at a slightly lower pressure level. As shown in 
Figures \ref{fig:NH3/N2} and \ref{fig:N2-NH3}, at the quench level (10$^2$ bar), the equal-abundance curve spans 
from 2000 K ([M/H] = -1) to 500 K ([M/H] = 3). The quench temperature of \ch{NH3} and \ch{N2} is around 1500 K 
for GJ 1214 b; as a result, increasing metallicity changes the dominant species from \ch{NH3} to \ch{N2} and a shift 
from \ch{NH3} dominant to \ch{N2} dominant atmosphere happens around [M/H] = 1 for the infrared photosphere 
($P\approx100$ mbar). In the case of thermochemical equilibrium, \ch{N2} is the dominant species at all the 
metallicities at the infrared photosphere (100 mbar and 1 mbar). 

\begin{figure}[h]
	\centering
	\includegraphics[width=0.3\textwidth]{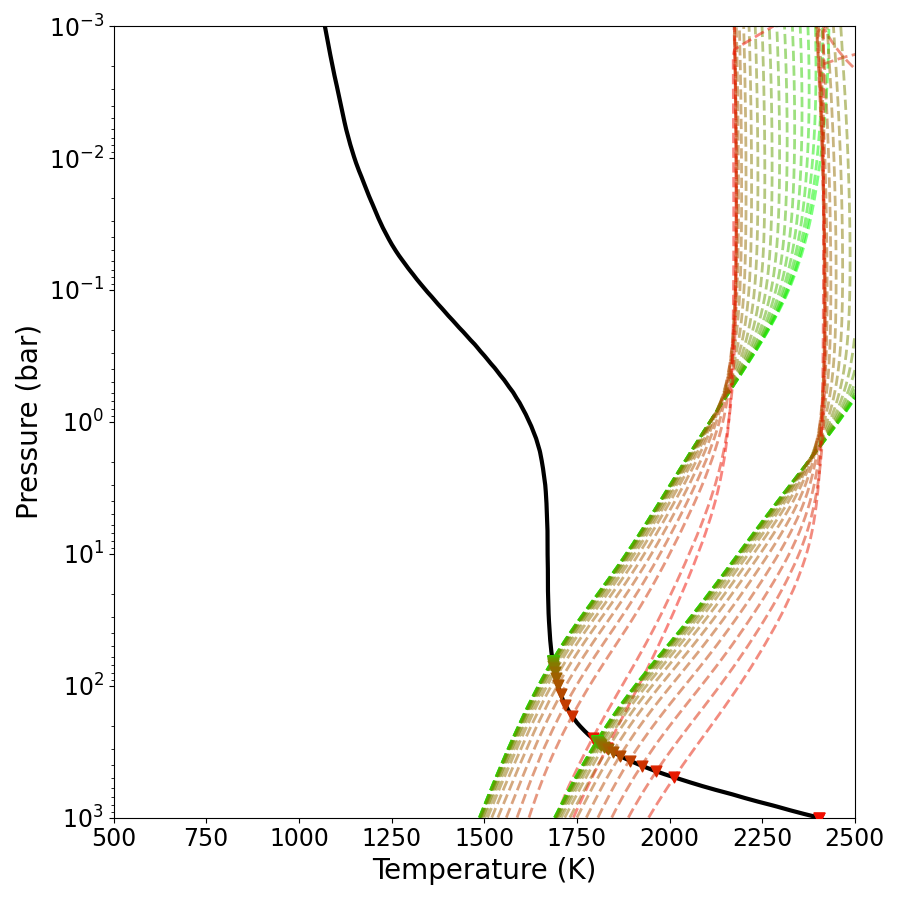}
	\includegraphics[width=0.3\textwidth]{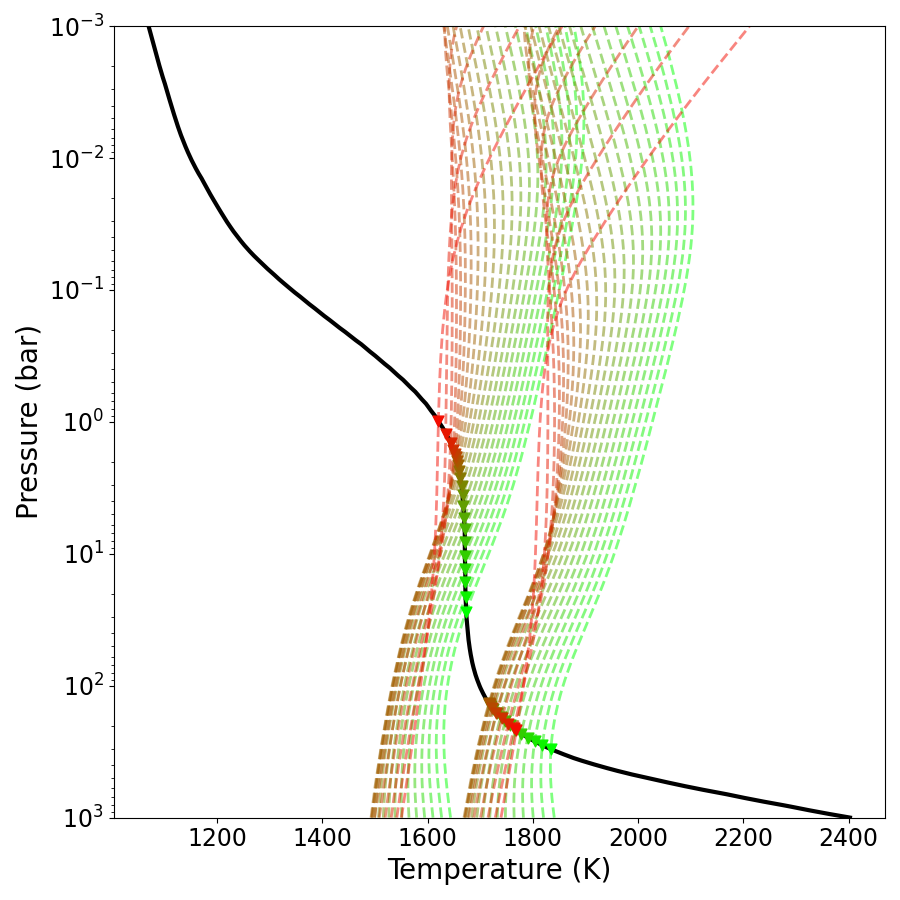}
	\includegraphics[width=0.3\textwidth]{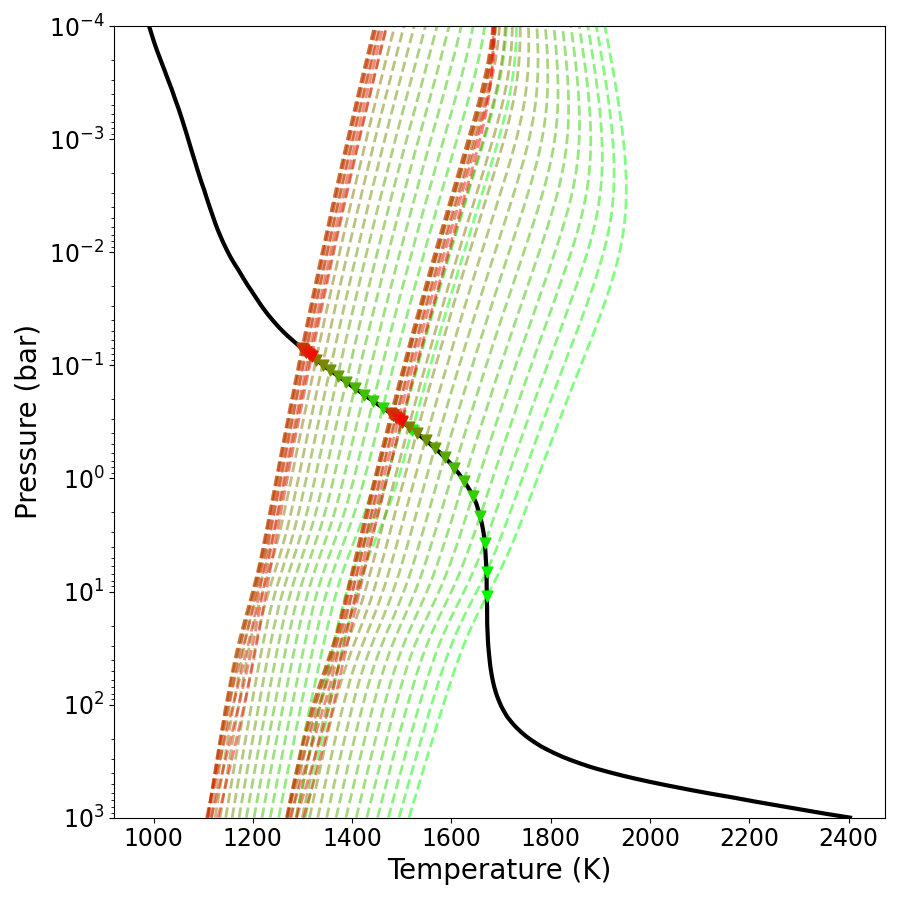}
	
	\includegraphics[width=0.3\textwidth]{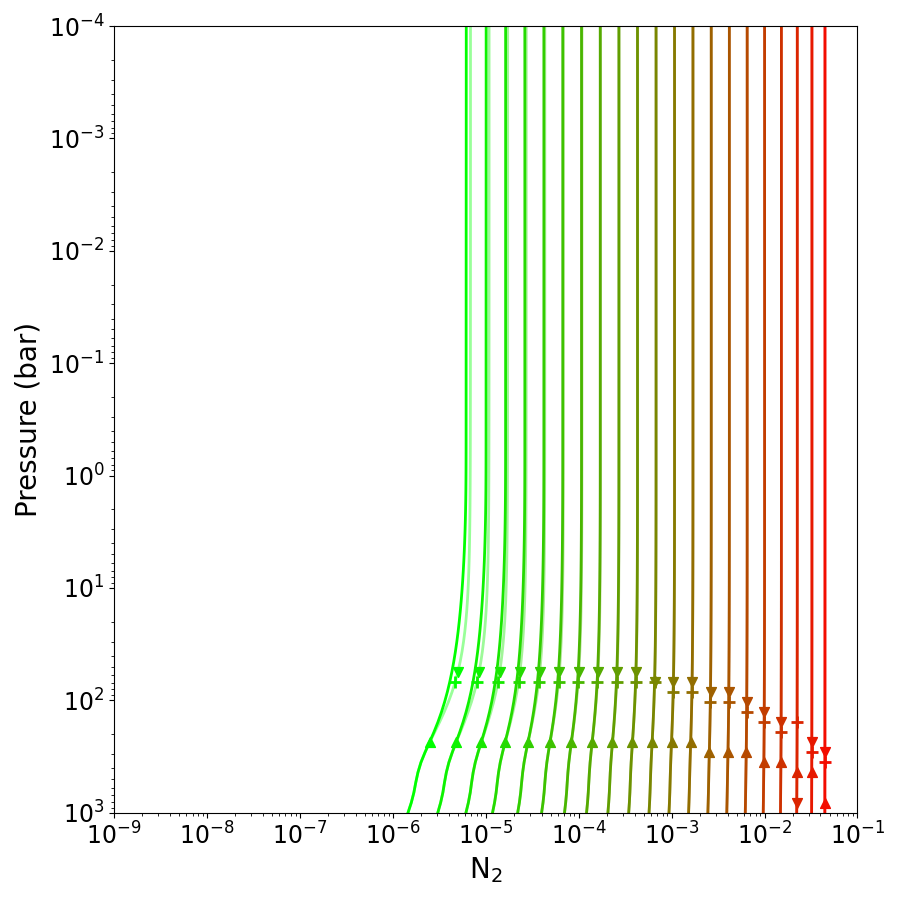}
	\includegraphics[width=0.3\textwidth]{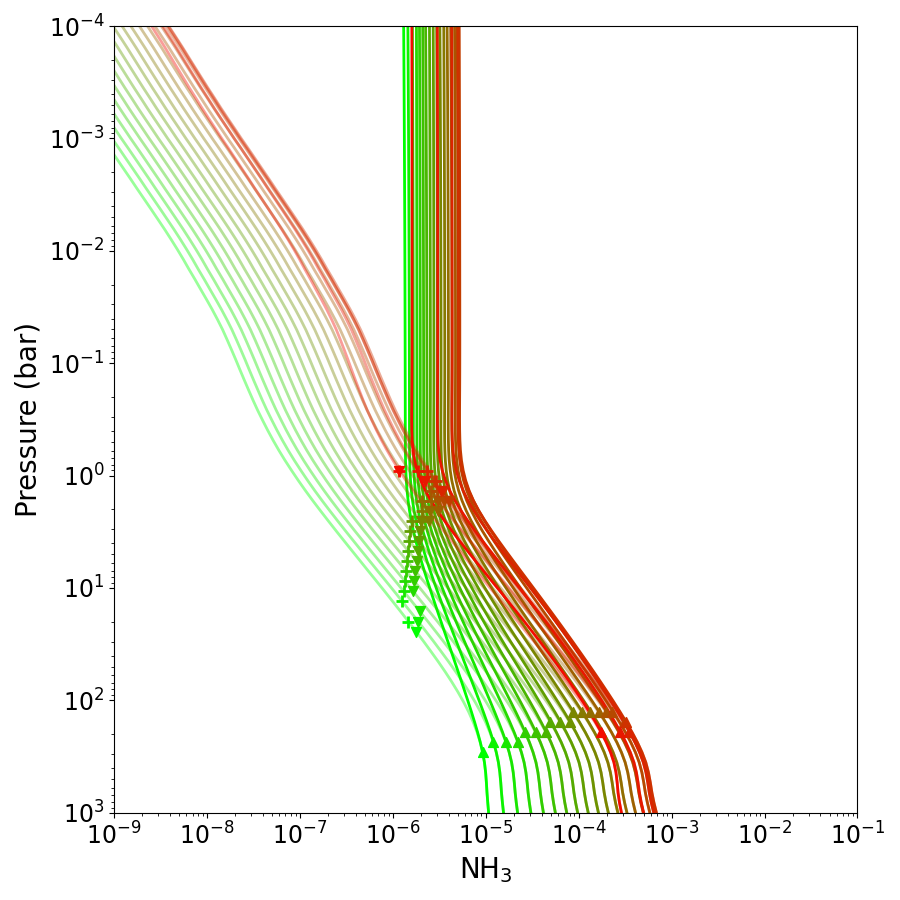}
	\includegraphics[width=0.3\textwidth]{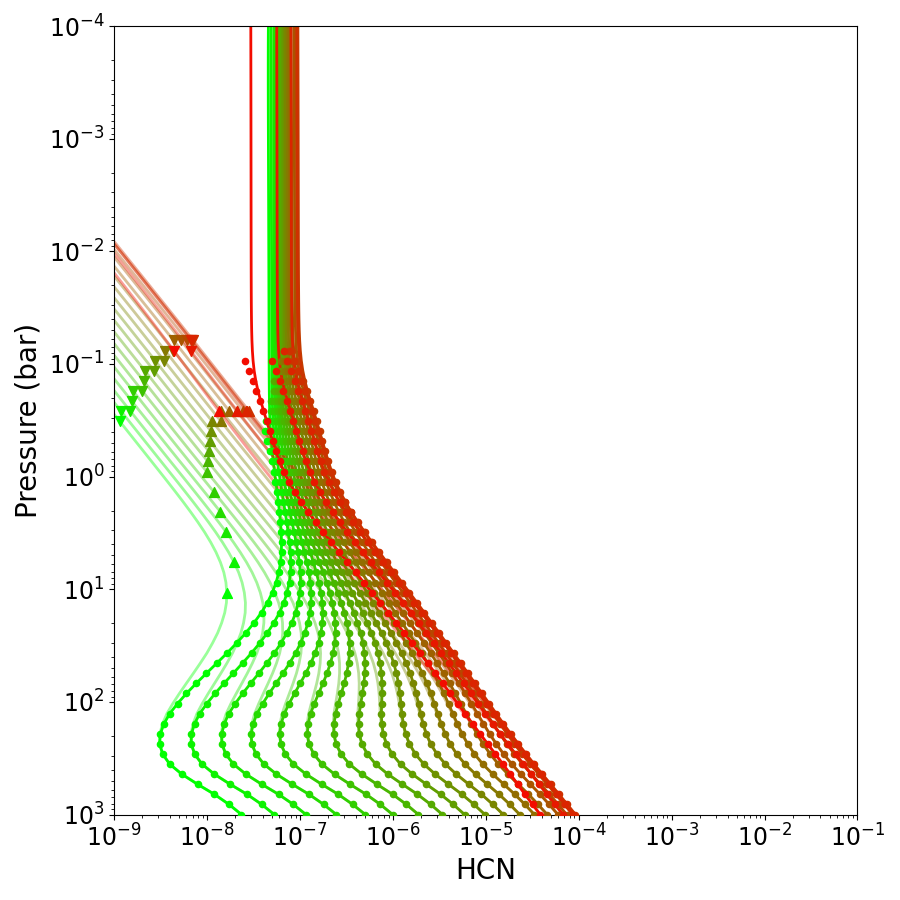}
	\caption{Same as Figure \ref{fig:SN} but for HD 189733 b. The solid circles (bottom right panel) represent the \ch{HCN} 
		abundance calculated from Equation \ref{Eq:HCN}, where [\ch{NH3}$_q$] and [\ch{CO}$_q$] is taken from 
		photochemistry-transport. The solid circles are plotted till the quench level (for mixing length = $0.1 \times H$) 
		of HCN.}\label{fig:HG}
\end{figure}

The thermochemical equilibrium abundance profile of \ch{HCN} mostly follows the \ch{NH3} and CO 
abundance profile. The quench curve of \ch{HCN} intersects at the high-temperature region of 
the atmosphere (T $\approx$ 1200-1600 K) of GJ 1214 b, and at this temperature-pressure, $\tau_{\ch{HCN}}$ is four 
to six orders of magnitude smaller than $\tau_{ \ch{NH3}}$ and $\tau_{ \ch{N2}}$. The temperature 
falls sharply at the quench pressure level and leads to a steep decrease of \ch{CO} \citep{Soni2023}. 
However, the \ch{NH3} abundance does not fall sharply. The collective effect of \ch{CO} and \ch{NH3} 
on \ch{HCN} leads to a decrease in \ch{HCN} sharply at its quench level. As discussed in 
Section \ref{S5P2}, the quenched abundance of \ch{HCN} is affected by CO and \ch{NH3} quenched abundance.

\subsection{HD 189733 b}
In Figure \ref{fig:HG}, we have over-plotted the quenched curve of \ch{N2} (left), \ch{NH3} (middle) and \ch{HCN} (right) with 
the thermal profile of  HD 189733 b.
The thermal profile is adopted from \cite{Moses2011}, and it remains nearly isothermal at the quench pressure 
level of \ch{NH3} and \ch{N2}. The thermal profile of HD 189733 b is such that \ch{N2} is the most dominant 
nitrogen-bearing species for most of the metallicities except for [M/H] $<$ 0 and $P <$ 10 bar. Thus, in 
thermochemical equilibrium, \ch{N2} is the dominant N species at the infrared photosphere ($P\approx100$ mbar) 
for all the parameter ranges. The presence of transport does not favor \ch{NH3} over \ch{N2}. The \ch{NH3} mixing ratio 
remains around 10$^{-5}$ and slightly increases with increasing metallicity, whereas the \ch{N2} abundance increases 
linearly with metallicity. \ch{NH3} and \ch{N2} quench at the same pressure levels when $L = 0.1 \times H$, 
and the \ch{NH3} quench level lies at a slightly lower pressure than the \ch{N2} quench level for $L = 1 \times H$. The \ch{HCN} quench 
level lies at one order of magnitude lower pressure than the \ch{NH3} quench level. As \ch{HCN} remains in equilibrium 
with \ch{NH3}, the \ch{HCN} abundance deviates from its thermochemical equilibrium abundance well below its quench level. 
The transport abundance of \ch{HCN} starts to deviate from its thermochemical equilibrium at 100 bar (at 100 bar, 
\ch{NH3} starts to deviate from its thermochemical equilibrium). However, above the quench level of \ch{HCN} ( $P \approx$ 
100 mbar), it freezes at its quenched abundance. The metallicity dependence of the quenched \ch{HCN} abundance is directly 
related to the metallicity dependence of the quenched abundance of \ch{NH3}, CO, and \ch{H2O}. As the effect of 
metallicity on \ch{HCN} due to CO and \ch{H2O} is canceled out, it mainly follows the quenched \ch{NH3}. 
As a result, it changes by a small factor as metallicity increases by four orders of magnitude.  

\section{Constraint on Metallicity and Transport strength}\label{sec:constraint}
\begin{figure}[h]
	\centering
	\includegraphics[trim={0cm 0cm 3cm 0cm},clip,width=1\textwidth]{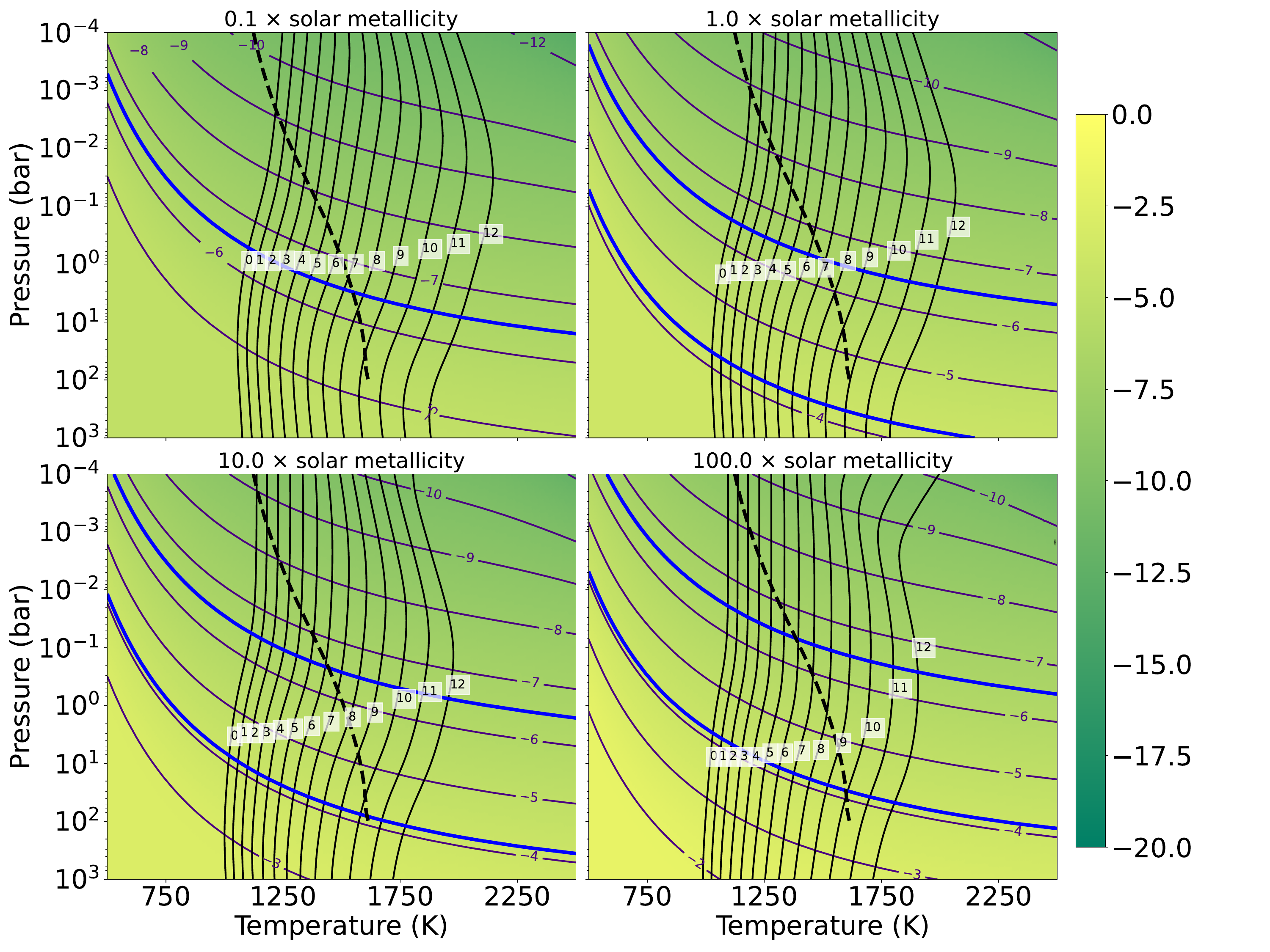}
	\caption{The color bar shows the mole fraction of \ch{NH3} in the log10 scale as a function of pressure 
		and temperature. The region between the solid blue lines are the retrieval constrain of HD 209458 b taken from 
		\cite{MacDonald2017}, the black dashed line is the 	T-P profile adopted for effective temperature $T_{\text{eff}}$ = 1000 K, 
		$g_{\text{surface}} = 3000$ cm s$^{-2}$, which is adopted from \cite{MacDonald2017}. The solid black lines are the quench lines for different 
		$K_{zz}$ values (labeled in the plots). The four plots represent the different values of atmospheric metallicity. }
	\label{fig:NH3_HD20945b}
\end{figure}

In \cite{Soni2023}, we have shown that the disequilibrium mixing ratios 
derived using quenching approximation can be used to constrain the transport strength and metallicity 
of the atmosphere for a given observed abundance of \ch{CO} and \ch{CH4}. In this work, we 
examined if N-bearing species can also be used to constrain the transport strength. We used abundance of \ch{NH3} 
to constrain the transport strength for HD 209458 b. We 
overplotted the retrieved \ch{NH3} (10$^{-6.5}$ $<$ \ch{NH3}$_{\text{,mix}}$ $<$ 10$^{-4.15}$) 
abundance with the quenched curve in the equilibrium abundance data in Figure \ref{fig:NH3_HD20945b}, in which 
the retrieved abundance is adopted from \cite{MacDonald2017}. It can be seen that all four values of 
metallicity, along with 6 $<$ log10($K_{zz}$) $<$ 12, can explain the observational mixing ratio 
of \ch{NH3}. However, low water abundance indicates the subsolar metallicity or high C/O ratio; here, 
we consider the subsolar metallicity case, for which the observational mixing ratio of \ch{NH3} can be 
well constrained by the high transport strength ($K_{zz}$ $>$ 10$^7$ cm$^2$ s$^{-1}$). We also find that similar transport 
strength is required to constrain the \ch{CH4} abundance (\ch{CH4}$_{\text{,mix}} \approx 10^{-8}$). The thermal 
profile lies in the CO dominant region, and for subsolar metallicity, $10^{-5}<\ch{CO}_{\text{mix}}<10^{-4}$. 
As discussed in the previous section, the quenched abundance of \ch{NH3} and CO can constrain the quenched 
abundance of HCN. We use Equation \ref{Eq:HCN} at the quench level of \ch{HCN} for $K_{zz}$  $>$ 10$^7$ cm$^2$ s$^{-1}$) along with the 
quenched $\ch{CO} [10^{-5} - 10^{-4}]$ and quenched $\ch{NH3} [10^{-6.5} - 10^{-4.15}]$ and found that the 
range of quenched abundance of \ch{HCN} is $[10^{-9} - 10^{-7}]$, which overlaps with the observed abundance of 
HD 209458 b. The observational signature of \ch{NH3} is low, and \ch{NH3} is less sensitive to pressure-temperature and transport strength in the \ch{N2} dominant region as compared to \ch{CH4} in the \ch{CO} dominant region. 
This makes \ch{CH4} better potential molecules compared to \ch{NH3} to constrain the transport strength. 

\section{Observability of N bearing species}\label{sec:observability}
Detection of \ch{NH3} and \ch{HCN} in the exoplanetary atmosphere is challenging due to their 
low photospheric abundance and due to the presence of contribution of \ch{H2O} in the total 
transmission spectrum, which is substantially larger than that of other species. The strength of 
their spectral signature in the planet spectrum increases with their abundance, and shows 
the 100 to 300 ppm transit-signature if their abundance exceed  10$^{-2}$ $\times$ \ch{H2O} mixing 
ratio \citep{MacDonald2017}. Supporting figures for this section (Figure~\ref{fig:T_int_vs_T_equi_HCN}, 
\ref{fig:T_int_vs_T_equi_HCN_mat_1}, 
\ref{fig:T_int_vs_T_equi_HCN_mat_10}, \ref{fig:T_int_vs_T_equi_NH3}) are in Appendix B.

\subsection{HCN}
To find the optimal parameter for the thermal profile, we have used the petitRADTRANS code \citep{Molliere2019} 
to generate 1D thermal profiles (petitRADTRANS uses the \cite{Guillot2010} method to generate the thermal profile). 
We have generated 2500 thermal profiles for different combination of $T_{\text{int}}$ (150 - 400 K), $T_{\text{equi}}$ 
(800-1600 K), kappa-ir = 0.01, and gamma = 0.4. Subsequently, we calculated the quenched abundance of \ch{HCN} for 
different gravity and $K_{zz}$ values. 


The quenched 
\ch{HCN} abundance increases with increasing temperature 
for $T_{\text{equi}}$ $<$ T$_{\text{q, HCN}}$ ($T_{\text{{q, HCN}}}$ is the $T_{\text{equi}}$ for the maximum 
quenched \ch{HCN} abundance), and then it decreases rapidly with increasing temperature for $T_{\text{equi}}$ $>$ 
$T_{\text{q, HCN}}$. As $g_{\text{surface}}$ and $K_{zz}$ increase, the $T_{\text{q, HCN}}$ shifts towards high 
$T_{\text{equi}}$ temperature. The quenched \ch{HCN} abundance decreases with $T_{\text{int}}$ for 
$T_{\text{equi}} \approx $  $T_{\text{q, HCN}}$ and it becomes independent or increases with $T_{\text{int}}$ as 
$T_{\text{equi}}$ deviates from $T_{\text{q, HCN}}$ (Figure \ref{fig:T_int_vs_T_equi_HCN} in Appendix B). 
This behavior can be attributed to its dependence on the 
quenched \ch{NH3} and \ch{CO}. 
The increase of $T_{\text{equi}}$ shifts the quench level of \ch{CO} and \ch{NH3} towards 
the high-pressure and high-temperature region. As a consequence, the \ch{CO} quenched abundance increases with $T_{\text{equi}}$, 
and \ch{NH3} quenched abundance decreases. 
The increase of the quenched \ch{CO} stops when the \ch{CO} quench level enters 
the \ch{CO} dominant region. However, the \ch{NH3} quench abundance continues to decrease. This results in an optimal 
$T_{\text{equi}}$ ($T_{\text{q, HCN}}$) for which \ch{HCN} quench abundance attains the maximum value at $T_{\text{equi}}$ 
= $T_{\text{q, HCN}}$. The quenched \ch{HCN} abundance is maximum around T$_{\text{equi}}$ = 1100 - 1300 K 
for T$_{\text{int}}$ = 150 K (see Figure \ref{fig:T_int_vs_T_equi_HCN} panel (d) in Appendix B).  
Recently \cite{Ohno2022} also found that the \ch{HCN} abundance has non-monotonic dependence on $K_{zz}$, 
and there can be a sweet spot of $K_{zz}$ for which the \ch{HCN} abundance is maximum. They found that the \ch{HCN} 
observational signature peaks at T$_{\text{equi}}$ = 1000 K (for $K_{zz}$  = 10$^{8}$ cm$^{2}$ s$^{-1}$, 
$g_{\text{surface}}$ = 19.96 m s$^{-2}$ and T$_{\text{int}}$ = 157 K). 
A slight difference may arise from the photochemistry and use of the quenching approximation; 
nevertheless, the estimate from quenched abundance is reasonably close and demonstrates its effectiveness in finding 
solutions without a full chemical kinetics model.

We have studied the 
variation of quenched \ch{HCN} abundance with T$_{\text{int}}$ and T$_{\text{equi}}$ for sub-solar 
(0.1 $\times$ solar) and super-solar (10 $\times$ solar) metallicity (Figures \ref{fig:T_int_vs_T_equi_HCN_mat_1} 
and \ref{fig:T_int_vs_T_equi_HCN_mat_10} in Appendix B). 
We found that when the quench point lie in the \ch{N2} dominated region, the \ch{HCN} abundance increases 
with power of $\sim$ 0.5 with metallicity, while the quench point lie in the \ch{NH3} dominated region, \ch{HCN}
abundance increases linearly with the metallicity. It is to be noted that in CO-dominant region, 
\ch{CO} thermochemical abundance increases linearly with metallicity; however, 
its effect is nullified since the denominator in equation \ref{Eq:HCN} also has a linear dependence 
on metallicity. Thus the metallicity dependence of \ch{NH3} determines the HCN metallicity 
dependence. Since \ch{NH3} increases as a power of 0.5 in the \ch{N2} dominated region and linearly in the \ch{NH3}
dominated region, HCN show the same behaviour.
The increase of \ch{CO}-dominated region with metallicity increases the parameter space 
of the sweet spot for the HCN$_{\text{max}}$ and shifts the $T_{\text{q, HCN}}$ towards lower temperature. 
As deeper \ch{CH4}/\ch{CO} boundary will allow lower $T_{\text{equi}}$ to have their \ch{CO} 
quench level in the \ch{CO} dominant region, and the upper limit of $T_{\text{equi}}$ comes from 
the \ch{NH3} quench level.	

\subsection{\ch{NH3}}

The maximum \ch{NH3} ($\ch{NH3}_{\text{,max}}$), lies in the parameter range for which the \ch{NH3} 
quench level lies in the \ch{NH3} dominant region, $\ch{NH3}_{\text{,max}}$ can be achieved for 
$T_{\text{int}}$ $<$ 150 K, $T_{\text{equi}}$ $<$ 1200 K and $K_{zz}$ $>$ 10$^{8}$ cm$^{2}$ s$^{-1}$. 
For a higher value of $K_{zz}$, the sweet spot expands towards a higher value of $T_{\text{equi}}$ 
(Figure~\ref{fig:T_int_vs_T_equi_NH3} in Appendix B). 
Our results are similar to \cite{Ohno2022}. For the higher $T_{\text{equi}}$, the photochemistry efficiently 
depletes the \ch{NH3}, and the effect of this is out of the scope of this study, and we refer the reader 
to see \cite{Ohno2022}. We found that at higher pressure levels, the thermal profile (adiabatic) mostly 
follows the \ch{NH3}/\ch{N2} contour lines, which move in the higher pressure region as the metallicity 
increases. However, the contour of \ch{NH3} abundance shifts towards lower pressure (see Figures 
\ref{fig:NH3/N2} and \ref{fig:NH3_HD20945b}). For a fixed thermal profile, the $\ch{NH3}_{\text{,max}}$ 
increase with metallicity with a proportionality of $\sim$ 0.5 in the region where the quench level 
lies in the \ch{N2} dominant region and linearly where the quench level lies in \ch{NH3} dominant region.

\subsection{Effect of photochemistry}
The photochemistry can efficiently remove the \ch{NH3} in the upper part of the atmosphere ($P<10^{-3}$ bar), 
and this photochemical depletion region shifts in the lower pressure region with increasing the strength of 
vertical mixing \citep{Hu2021}. The \ch{NH3} dissociation cross-section is large and comparable to the other 
photoactive molecules (\ch{CO}, \ch{H2O}, \ch{CH4}) in the longer wavelength ($225 nm > \lambda > 190 nm$) and this 
can efficiently produce HCN in the presence of \ch{CH4} \citep{Hu2021, Ohno2023a}. The chemical time scale of 
HCN is sufficiently large (see Figure \ref{fig:time_scale_HCN-NH3}) to allow the vertical mixing to supply the 
photochemically produced HCN in the deeper part of the atmosphere and increase the HCN abundance in the transmission 
and emission infrared spectrum. For a moderate vertical mixing strength (Kzz $\sim10^{8} cm^{2}s^{-1}$), the 
photochemically produced HCN can be transported to the 100 bar pressure level for warm exoplanets (Temperature 
at 100 mbar $\sim$ 1200 K: Figure \ref{fig:quench_HCN}). The lower $T_{\text{equi}}$ resulted in a higher 
chemical time scale for HCN and lower availability of photon flux. The higher chemical time scale can enhance 
transportation of the photochemically produced HCN into the higher pressure region; on the other hand, the lower photon flux can 
limit the photochemically produced HCN. There should be a sweet spot for the maximum photochemically produced 
HCN at the infrared photosphere, which can be studied in future work.

\subsection{Effect of Clouds and Hazes}
The presence of clouds and hazes, which we neglected in the present work, can affect the \ch{NH3} and 
\ch{HCN} abundances along with their spectral signatures. The clouds and hazes can provide extra opacity 
and obscure the spectral feature \citep{Molaverdikhani2020}. Fortunately, the effect of opacity is lesser 
in the longer wavelengths (($\lambda \approx 11 \mu m $)),  where the spectral feature of HCN and \ch{NH3} 
are more pronounced \citep{Kawashima2019, Ohno2020, Ohno2022}. In the shorter wavelengths, 
opacity due to clouds and hazes can affect the atmosphere's thermal structure, leading to the changing of 
the position of the quench level of the species. A thick cloud can increase the temperature in the 
high-pressure region \citep{Molaverdikhani2020} and can affect the quench level of \ch{NH3} in two ways.
Depending upon the verticle mixing strength, it can increase the temperature around the \ch{NH3} quench 
level, as well as it can shift the quench level in the low-pressure region, which depends on the shape of the thermal 
profile. The thermochemical equilibrium abundance of \ch{NH3} decreases in both cases. 

The presence of haze formed in the photochemical region can increase the opacity in the upper part, 
thereby increasing the temperature. On the other hand, it blocks the stellar flux from the lower part 
of the atmosphere, thus decreasing the temperature, which leads to the opposite effect from the clouds 
(increase the \ch{NH3} abundance) \citep{Ohno2022}. The dependency of \ch{HCN} abundance on \ch{CO} and 
\ch{NH3} abundance can lead to a complex effect. An increase in temperature due to clouds can increase 
the \ch{CO} abundance when the thermal profile lies inside the \ch{CH4} dominant region and leads to an 
increase in \ch{HCN} chemical equilibrium abundance. In the case where the thermal profile lies 
in the \ch{CO} dominant region, the \ch{CO} abundance remains constant with the increasing temperature 
resulting in the reduction of \ch{HCN} abundance (see Figure \ref{fig:N2-NH3}).

\section{Potential exoplanets for \ch{HCN} search}\label{sec:candidates}
\begin{figure}[h]
	\centering
figure/	\includegraphics[trim={0cm 0cm 0cm 0cm},clip,width=0.7\textwidth]{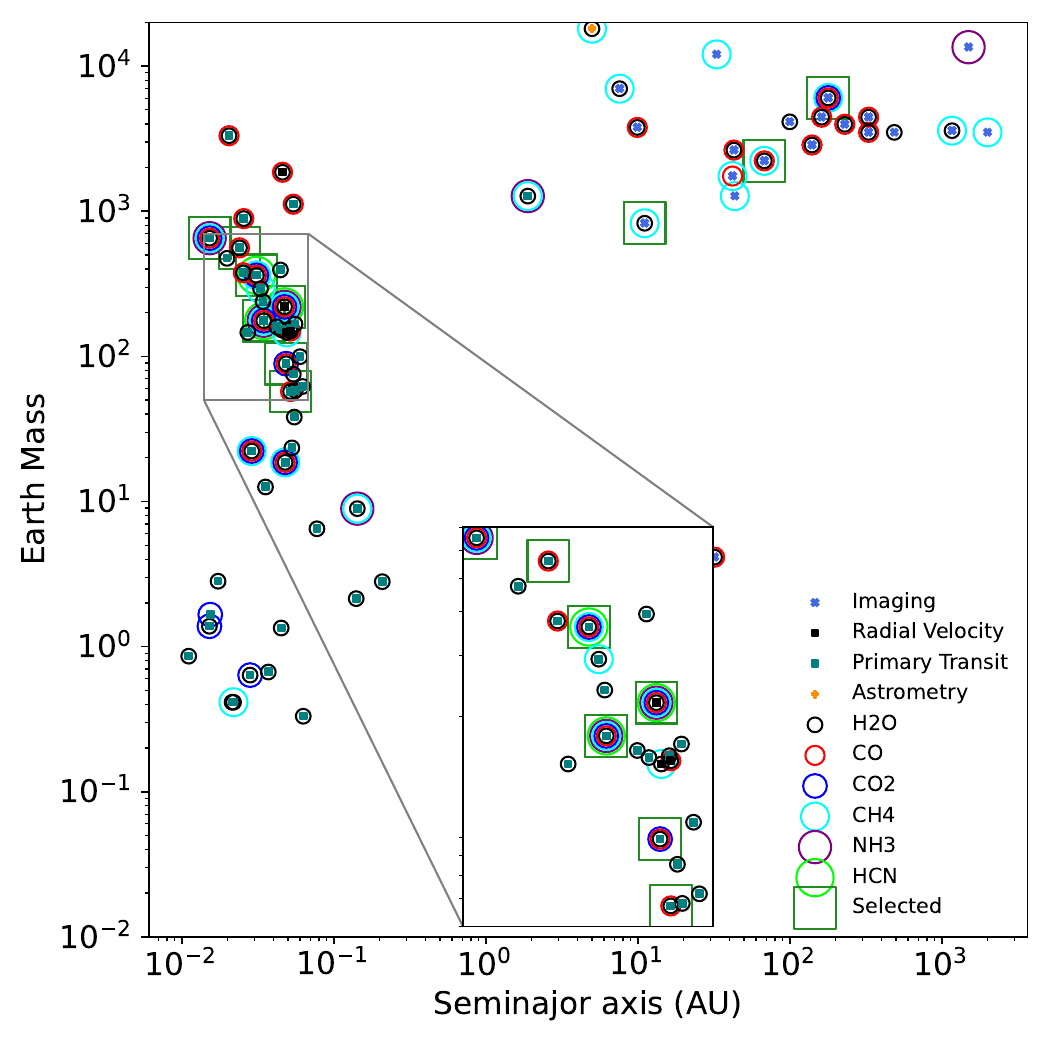}
	\caption{the exoplanets whose atmosphere are probed and at least one H-C-N-O bearing molecule is found are plotted in their mass and orbital separation parameter space. Each colored circle denotes one H-C-N-O bearing molecule.  
		The different marker refer to their detection method, given in the legend (\url{exoplanet.eu}). }
	\label{fig:mole}
\end{figure}

We used quenched dataset to find the \ch{HCN} abundance to identify the potential candidate exoplanet 
for observation using observatories like JWST. We selected the exoplanets with already observed C-O 
species from  \url{exoplanet.eu}, which is about fifty (shown in the Figure \ref{fig:mole}). 
Then we excluded very large and small planets since 
the lower mass planets are hard to observe due to their large star-to-planet radius ratio ($R_s/R_p$), and 
higher mass exoplanets have large gravity, resulting in smaller scale heights in the atmosphere (smaller 
atmospheric thickness). Therefore we chose planets in the mass range between 0.01  and 20 M$_J$. 
Chosen exoplanets are marked with green box. Then 
we chose exoplanets with $T_{\text{equi}}$ between 700 and 1700 K; lower $T_{\text{equi}}$ results in a low CO mixing ratio, 
and higher $T_{\text{equi}}$ favors \ch{N2} over \ch{NH3}. Thus, a moderate $T_{\text{equi}}$  should be targeted to 
find the HCN signature. After selecting the mass and $T_{\text{equi}}$ ranges, we collected $K_{zz}$ values from the 
literature, finding these values for seven exoplanets. We further constrained $K_{zz}$ by using 
\ch{CH4} and \ch{NH3} abundance and we have a total eleven exoplanets (Table~\ref{HCN}).
For calculating HCN abundance, we need to generate thermal profiles, for which we used petitRADTRANS \citep{Molliere2019}. 
petitRADTRANS require $T_{\text{equi}}$, $T_{\text{int}}$ and surface gravity of the planet. $T_{\text{int}}$ is 
constrained by the theoretical model of the evolution of the planet's interior and is a function of planet mass, age, 
bulk elemental abundance, and star-planet interaction\citep{Burrows2001}. However, in this study, we did not use the 
complex theoretical model; instead, we have used $T_{\text{int}}$ = 150 K, 200 K, and 300 K for age $>$ 1 Gyr, 
1 Gyr $>$ age $>$ 0.2 Gyr, and age $<$ 0.2 Gyr respectively. $T_{\text{equi}}$  and surface gravity is collected 
from the literature (See Table~\ref{HCN}).

Table~\ref{HCN} shows the list of promising exoplanets for HCN detection. Columns 6 
contains abundances found by using HCN abundance by quenching approximation for the mixing 
length calculated using the Smith method. Columns 7 and 8 show HCN abundance calculated for 100 mbar
and 1 mbar pressure using the chemical kinetics code, which has both the transport and photochemistry.

First, we studied the exoplanets for which HCN is already detected, i.e.,  HD 209458 b \citep{MacDonald2017},  
HD 189733 b \citep{Cabot2019}, and WASP-80 b \citep{Carleo2022}. We found that the HD 209458 b has the highest 
quenched HCN mixing ratio ($\approx -5.7$), making it one of the best candidates for detection. We also calculated 
the \ch{HCN} abundance with the full chemical kinetics model and found that photochemistry can further increase its 
abundance. For  WASP-80 b, we found that the observed $T_{\text{equi}}$ of 817 K cannot produce HCN adequately for 
any $K_{zz}$ value \citep{Carleo2022}. However, including the photochemistry in the model, we found that high vertical 
mixing (log(Kzz) $\sim$ 10) and lower temperature (817 K) can increase the HCN abundance by more than four to six orders 
of magnitude in the infrared photosphere. Finally, HCN is also detected in HD 189733 b; we found a value 
of $\approx$ -7.1 quenched mixing ratio, and photochemically produced HCN increases the mixing ratio by 
more than one order of magnitude in the emission photosphere and by two orders of magnitude in the transmission photosphere.

Other potential targets for HCN are given in Table \ref{HCN} (fourth row onwards). The most promising candidates are  
WASP-43 b, WASP-77 A b, and WASP-39 b, for which the quenched \ch{HCN} mixing ratio is -6.2, -6.3, and -6.7, 
respectively using the mixing length obtained by the Smith method. 
The $K_{zz}$ on WASP-77 A b is 10$^{12}$ cm$^2$ s$^{-1}$ constrained by the \ch{CH4} abundance.  The presence of \ch{HCN} 
in WASP-77 A b can be evidence of strong disequilibrium chemistry as the \ch{HCN} abundance in WASP-77 A b
decreases rapidly with decreasing $K_{zz}$, for $K_{zz}$ $<$ 10$^{11}$ cm$^2$ s$^{-1}$. WASP-77 A is a G8 spectral type, 
and the small orbital distance (0.024 AU) can increase the photochemical HCN abundance, which can also degenerate 
the HCN abundance for lower $K_{zz}$ values. Besides, WASP-69 b and WASP-127 b can also be reasonable targets for HCN detection. 
When using the mixing length calculated using the Smith method, the quenched HCN abundance is generally closer to the chemical 
kinetics model with photochemistry and transport.

\begin{table}[ht]
	\caption{\label{HCN}Potential observational candidates for HCN.}
	\begin{center}
		\begin{tabular}{|l|l|l|l|l|l|l|l|l|}
			\hline
			Planet           & Age        & $T_{\text{equi}}$ & g   & $K_{zz}$             & Quenched                            &  Photochemistry  &Photochemistry   \\
			& (Gyr)      &   (K)             &   m s$^-2$     &(cm$^2$ s$^{-1}$)     & HCN                                 & -transport       &-transport   \\
			&            &                   &        &                      & $L_{\text{mix}}$=$L_{\text{smith}}$ &  model 100 mbar  &model 1 mbar           \\
			\hline                                                                                  
			HD 209458 b      & 4          &  1450[1]          & 9[2]   & $3\times10^{11}$[3]  & -5.7                                & -5.7             &  -5.3   \\
			WASP-80 b        & 1          &  817[4]           & 13[4]  & $10^{10}$            & - 10.8                              & -6           &  -4.3      \\
			HD 189733 b      & 0.6        &  1200[1]          & 20[1]  & $10^{10}$[3]         & -7.1                                & -6               & -4.5    \\
			\hline
			WASP-43 b        & 0.4        &  1440[7]          & 40[7]  & $10^{9}$ [8]         & -6.2                                & -6.2             &  -5.7      \\
			WASP-77 A b      & 1          &  1715[5]          & 25[5]  & $10^{12}$            & -6.3                                & -6.4             &  -6.4     \\
			WASP-39 b        & 7          &  1150[1]          & 6[1]   & $5\times 10^{8}$ [11]& -6.7                                & -6.3             & -4.7    \\
			VHS 1256-1257 b  & 0.2        &  1100[6]          & 316[6] & $10^{8}$ [6]         & -7                                  & -7               & -7      \\
			WASP-127 b       & 11         &  1400[12]         & 2.3[12]& $10^{10}$            & -7.1                                & -7.6             & -8      \\
			HR 8799 b        & 0.06       &  1000             & 31     & $10^{8}-10^{9}$ [10] & -8.8                                &-8.5              & -8.5    \\
			WASP-69 b        & 2[9]       &  963[9]           & 5[9]   & $10^{10}$            & -9.4                                & -7             &  -5      \\
			51 Eri b         & 0.02       &  700[10]          & 32[10] & $10^{7}$ [10]        & -10.5                               & -10              & -10       \\  
			\hline
		\end{tabular}
	\end{center}
	{Reference: [1] \cite{Kawashima2021}, [2] \cite{MacDonald2017}, [3] \cite{Moses2011}, [4] \cite{Dymont2022}, [5] \cite{Cortes-zuleta2020}, [6] \cite{Miles2023}, [7] \cite{Blecic2014}, [8] \cite{Helling2020}, [9] \cite{Anderson2014}, [10] \cite{Moses2016}, [11] \cite{Tsai2022}, [12] \cite{Boucher2023}.}
\end{table}

\section{Conclusion}\label{sec:conclusion}

In this work, we have studied the effect of metallicity on the thermochemical equilibrium abundance
and the quenched abundance of the N-bearing species \ch{N2}, 
\ch{NH3}, and \ch{HCN}.  We calculated the chemical timescale of \ch{NH3}, \ch{N2}, and \ch{HCN} in the 3D grid 
of temperature (500 to 2500 K), pressure (0.01 mbar to 1 kbar), and metallicity  (0.1-1000 $\times$ solar 
metallicity). We compared the chemical timescale with the vertical mixing timescale and found the quenched curve. 
We used this quenched curve to study the effect of metallicity on the quenched abundance of the molecules. 
Our conclusions are as follows:

\begin{itemize}
	\item 
As metallicity is increased, \ch{N2} equilibrium abundance increases linearly in the \ch{N2} dominated region, 
while in the \ch{NH3} dominant region, it increases more rapidly with increasing metallicity. Whereas, 
\ch{NH3} equilibrium abundance increases linearly with metallicity in \ch{NH3} dominant region, and in \ch{N2} dominant 
region, it increases slowly with metallicity. In the high metallicity region ([M/H] $ > 2.5$), the \ch{NH3} equilibrium 
abundance starts to decrease with increasing metallicity as the bulk H decreases. The metallicity dependence 
of the equilibrium abundance of HCN changes in different regions. In the \ch{NH3} dominant region, 
it rapidly increases with metallicity; in contrast, in the \ch{N2} dominant region, the rate of increase decreases, 
and in the CO dominant region, its abundance almost remains constant with metallicity. HCN remains in 
equilibrium with CO, \ch{H2O} and \ch{NH3}.

\item We studied the metallicity dependence of the two main conversion schemes for \ch{NH3-N2} for the equilibrium composition. 
In the first 
scheme, the conversion occurs through \ch{N2H} and is important in low-temperature regions. In the second, 
conversion occurs through NO or N and is important in the high-temperature region. The effect of metallicity 
on the rate of RLS of the second scheme is more prominent than the RLS of the first scheme. As the metallicity 
increases, the second scheme dominates over the first scheme, covering almost the entire parameter space in high 
metallicity. The conversion of \ch{HCN-NH3} for the equilibrium composition
takes place through \ch{HNCO} in which the HCN loses its C to CO. This scheme is dominant in most of the parameter 
range; as a result, HCN remains in equilibrium with \ch{NH3} and CO. 

\item The vertical mixing timescale is decreased by two orders of magnitude as the metallicity increases 
by four orders of magnitude. $\tau_{ \ch{N2}}$ remains 
constant for the first conversion scheme and as a result the quenched curve of \ch{N2} shifts towards the 
high-temperature and the high-pressure region as the metallicity increases. However, for the region where the 
second scheme is dominant, it shifts towards 
low temperature as the metallicity increase. The quenched curve of \ch{NH3} shifts towards the low-temperature region with 
increasing metallicity for most of the parameter space. In the region where R7 or the second term dominates in 
$\tau_{ \ch{NH3}}$, the quenched curve shift towards a high-temperature region with increasing metallicity. The 
quenched curve of HCN shifts towards the low-temperature region with increasing metallicity for all the parameter 
ranges. 
	
\item We have used two test exoplanets (HD 18973 b and GJ 1214 b) and compared the result of the 
quenching approximation with the photochemistry-transport model. We use the Smith method to improve the error 
in the quenching approximation. For  GJ 1214 b, the quenched abundance of \ch{NH3} and \ch{N2} are accurate 
within $\approx$ a factor of 0.9. For HD 189733 b, the quenched \ch{NH3} abundance is accurate within $\approx$ 
a factor of 0.5, and \ch{N2} is accurate within $\approx$ a factor of 0.9. The quenched abundance of HCN depends 
upon the quenched abundance of \ch{NH3} and CO, and as a result, the error in the \ch{NH3} and CO quenched abundance is 
propagated in HCN. For GJ 1214 b the quenched HCN abundance is accurate within $\approx$ a factor of 0.1 (this main 
deviation comes from the error in the CO quenched abundance). In the case of HD 189733 b, the HCN abundance is accurate
 within $\approx$ a factor of 0.5 (this main deviation comes from the error in the \ch{NH3} quenched abundance). As 
the metallicity increases, the error in the quenched CO decreases, and as a result, the quenched HCN is accurate within 
$\approx$ a factor of 0.5 for high metallicity. 

\item For a given $T_{\text{int}}$ and $T_{\text{equi}}$, there is a sweet spot in the $K_{zz}$ parameter space 
for which the quenched \ch{HCN} or \ch{NH3} abundance is maximum. The \ch{NH3} quenched abundance increases with 
increasing $K_{zz}$ and becomes independent after a certain value of $K_{zz}$ at which the \ch{NH3} quench level 
lies on the adiabatic part of the thermal profile. In this parameter space, decreasing $T_{\text{int}}$ increases the 
quenched \ch{NH3}. For a given thermal profile, the \ch{HCN} quenched abundance first increases with increasing $K_{zz}$ 
untill it reaches its maximum value, and further increasing the $K_{zz}$ decreases the HCN quenched abundance. 
Lower $T_{\text{equi}}$ favors \ch{NH3} over \ch{N2} and \ch{CH4} over \ch{CO}, and a higher value of $T_{\text{equi}}$ 
favors \ch{CO} over \ch{CH4} and \ch{N2} over \ch{NH3}. This results in an optimal value of $T_{\text{equi}}$ to achieve 
maximum quenched HCN. We also found that as the metallicity is increased, the parameter space moves towards the lower 
temperature, and HCN abundance increases.

\item We searched potential candidates for HCN detection using the data set for quenched HCN abundance 
and generating thermal profiles using petitRADTRANS.  Along with the exoplanets for which HCN is 
already detected (HD 209458 b, HD 189733 b, and WASP-80 b), we found that the most promising candidates are
WASP-43 b, WASP-77 A b, and WASP-39 b.

\end{itemize}

\section*{Acknowledgements}
The authors thank the anonymous referee for constructive comments which strengthened the paper. We thank Sana Ahmed 
for suggestions that improved the overall readability of the manuscript.
The work done at the Physical Research Laboratory is supported by the Department of Space, Government of India.

\newpage
\bibliographystyle{astron}
\bibliography{references.bib}

\appendix
\renewcommand\thefigure{\thesection.\arabic{figure}}
\setcounter{figure}{0}
\section{Model and parameters} \label{sec:model_A}
A detailed description of the model is provided in \citet{Soni2023}, and here we briefly 
describe the main aspects. The 1D chemical kinetics model solves the mass continuity equation for each species as follows:
\begin{equation}
\frac{\partial n_i}{\partial t} = P_i - n_i L_i - \frac{\partial \phi_i}{\partial z}, \label{Eq:1}
\end{equation}
where $n_i$ is the number density of the $i$-th species, $P_i$ and $n_iL_i$ are the production and 
loss rates due to thermochemical and photochemical reactions, and $\phi_i$ represents the transport 
flux. Equation \ref{Eq:1} is numerically solved for each layer and each species in the network until the 
convergence criteria are fulfilled. The transport processes include eddy diffusion and molecular diffusion; due 
to the large uncertainty in the eddy diffusion coefficient, it is taken as a parameter, while the molecular 
diffusion coefficient is calculated by the description given in \cite{Chapman1991}. To find the flux in each 
atmospheric layer, we use the two-stream approximation of radiative transfer following \cite{Heng2014}. For 
the present work, we have used a reduced network of 52 species involving H-C-N-O which are connected by 600 
chemical reactions. Although, our chemical network contains all the important species up to six hydrogen, 
two carbon, two nitrogen, and three oxygen atoms, and single atoms for He, Na, Mg, Si, Cl, Ar, K, Ti, and Fe 
following \cite{Tsai2017, Tsai2018} (for H, C, N, and O) and \cite{Rimmer2016} (for other species).
Details of the network and benchmarking can be found in \cite{Soni2023}.

\subsection{Smith method}
\begin{figure}[h]
	\centering
	\includegraphics[width=1\textwidth]{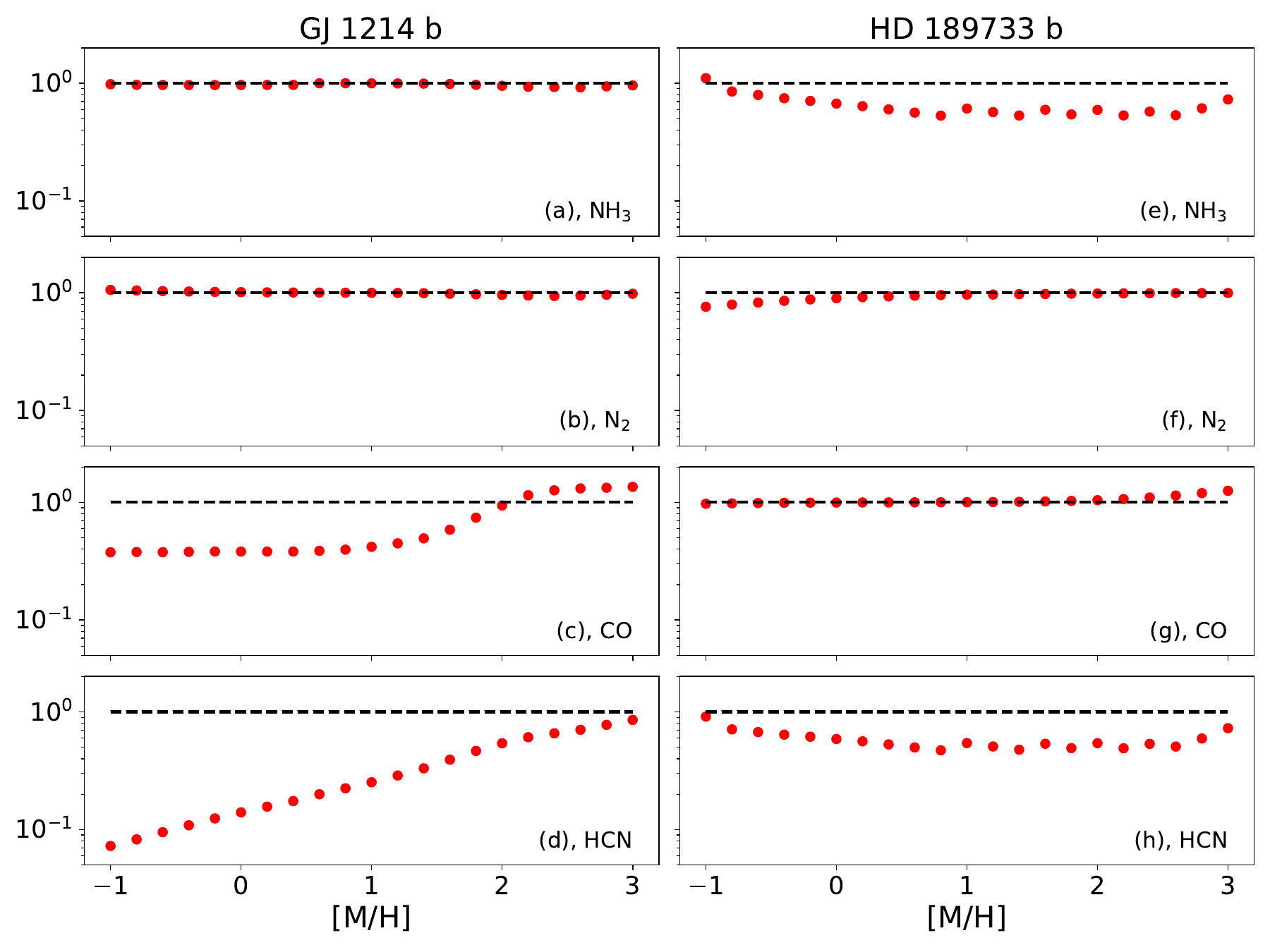}
	\caption{The ratio of chemical kinetics model (with only transport) output mixing ratio at 100 mbar with 
the output of quenching approximation 
mixing ratio is shown. The vertical mixing timescale is calculated using the mixing length, which is calculated using the 
Smith method.}\label{fig:error_quench}
\end{figure}

The mixing length theory calculates the dynamical time scale in the quenching approximation. 
Generally, the mixing length is assumed to be one pressure scale height \citep{Smith1998, Madhusudhan2011, Fortney2020}. 
\cite{Smith1998} found that the mixing length lies between 0.1 and 1 $\times$ pressure 
scale height. We used both limits in our previous work \citep{Soni2023} and calculated the error (e$_{q}$) in 
the quenching approximation by measuring the difference between the quenched value of $L = 0.1 \times H$ 
and $1\times H$. It yields an error of a few orders of magnitude in \ch{NH3} in HD 189733 b, and the error for 
\ch{N2} and \ch{NH3} in GJ 1214 b is around a factor of 2. The value of e$_{q}$ in \ch{N2} for 
HD 189733 b is minimal, as the thermochemical equilibrium abundance of \ch{N2} does not change with pressure at the 
quench level (is not affected by transport). However, a more accurate mixing length can be calculated using the 
method described in \cite{Smith1998}, which can increase the accuracy. In Figure \ref{fig:error_quench}, we have 
plotted the ratio of chemical kinetics modeled (with only transport) mixing ratio at 100 mbar with the quenched 
mixing ratio using the Smith method. As shown in Figure \ref{fig:error_quench}, the error of \ch{NH3} 
for HD 189733 b is reduced from two orders of magnitude to 0.5, and the error in GJ 1214 b \ch{N2} 
quenched abundance is also reduced from 2 to 0.7. As discussed in Section \ref{S5P1}, the quenched abundance of \ch{HCN} 
depends upon the quenched \ch{NH3} and CO abundance; as a consequence, any error in the \ch{NH3} or CO quenched abundance 
also propagates in the quenched HCN. In HD 189733 b, the combined error of \ch{HCN} is reduced from two orders of magnitude 
to a factor of 0.5. For the case of GJ 1214 b, this error is one order of magnitude for low metallicity 
and is reduced to 0.9 as the metallicity is increased. The chemical equilibrium abundance of CO is very 
sensitive to pressure at its quench pressure \citep{Soni2023}; as a result, a small error in the quench pressure can lead 
to an order-of-magnitude error in the quenched abundance of CO. 

\subsection{Comparison with the analytical expression}
We have compared the \ch{NH3}, \ch{N2}, and \ch{HCN} chemical timescales with the widely used 
analytical expressions from \cite{Zahnle2014}. \cite{Zahnle2014} ran several chemical kinetics models and 
found the quench level by finding the highest pressure level where the non-equilibrium abundance 
deviates from its thermochemical equilibrium abundance. At the quench level, they fit the vertical 
mixing timescale by an analytical expression and use it as the chemical timescale (at the quench 
level, the chemical timescale is comparable to the vertical mixing timescale). We have seen in \cite{Soni2023} 
that chemical timescales for \ch{CO}, \ch{CH4}, and \ch{CO2} from analytical expressions are in reasonably good 
agreement in the temperature range between 1000 and 2500 K. However, for the case of \ch{NH3-N2}, the adiabatic 
thermal profile remains close to the contour of \ch{NH3/N2} \citep{Fortney2020, Ohno2022}; as a 
result, the transport abundance does not deviate from its thermochemical equilibrium abundance after the quench 
level. It creates a greater obstacle  to finding the quench level in the study by \cite{Zahnle2014}. The chemical 
timescales from the analytical 
expression for \ch{NH3} and \ch{N2} are degenerate, as any set of Arrhenius coefficients which gives 
the same value at 10 bar and 1750 K can fit the quench level data.

\begin{figure}[h]
	\centering
	\includegraphics[width=0.4\textwidth]{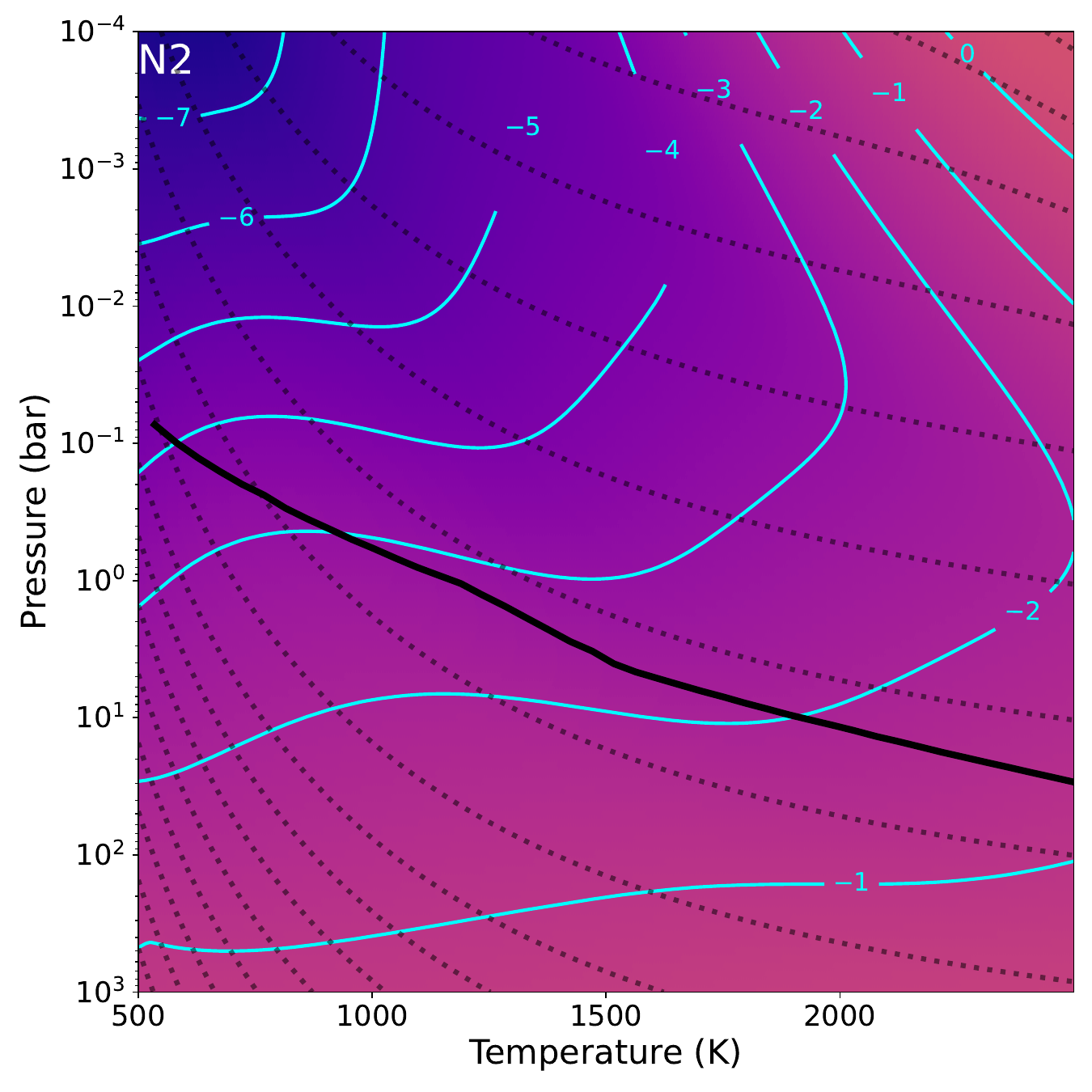}
	\includegraphics[width=0.4\textwidth]{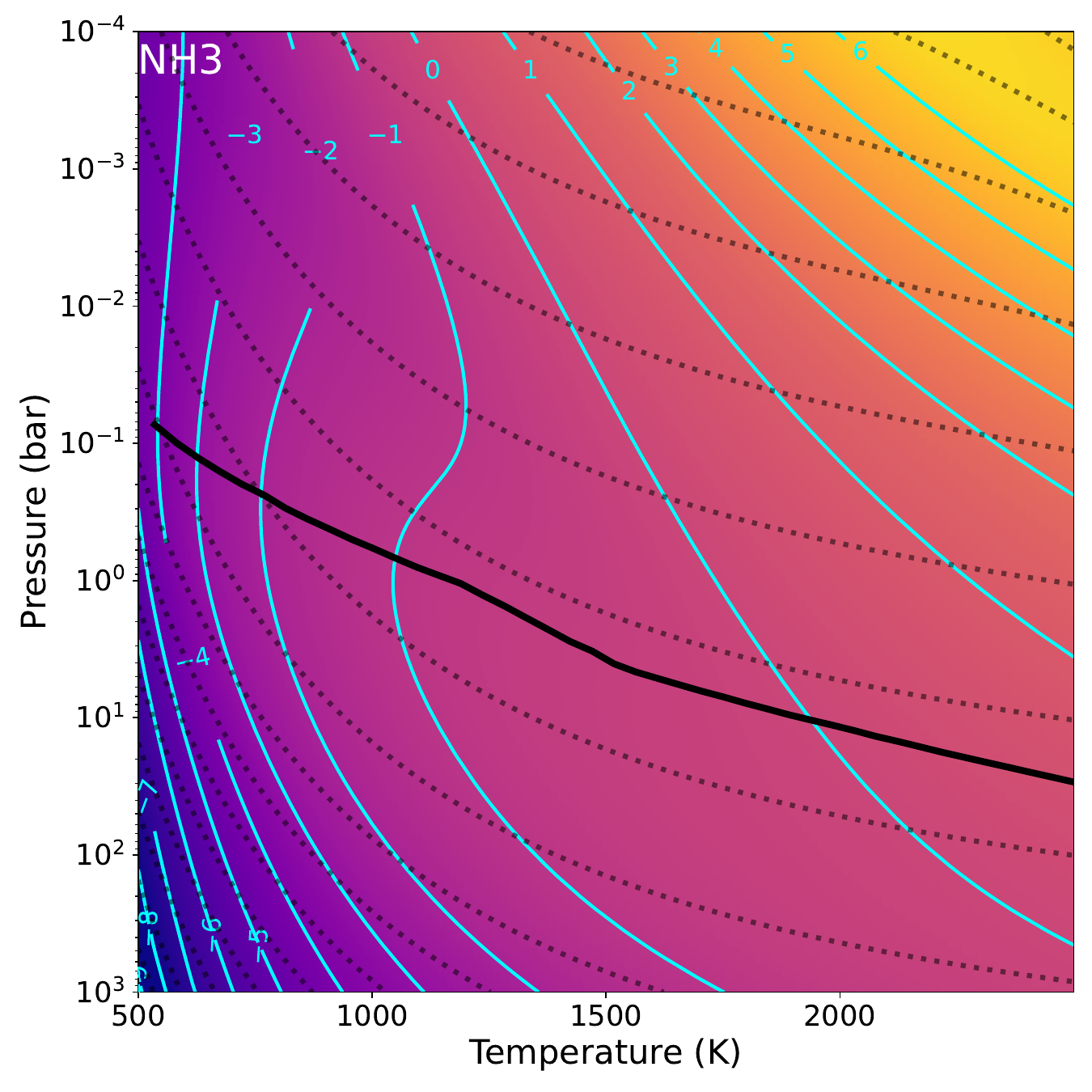}
	\includegraphics[width=0.4\textwidth]{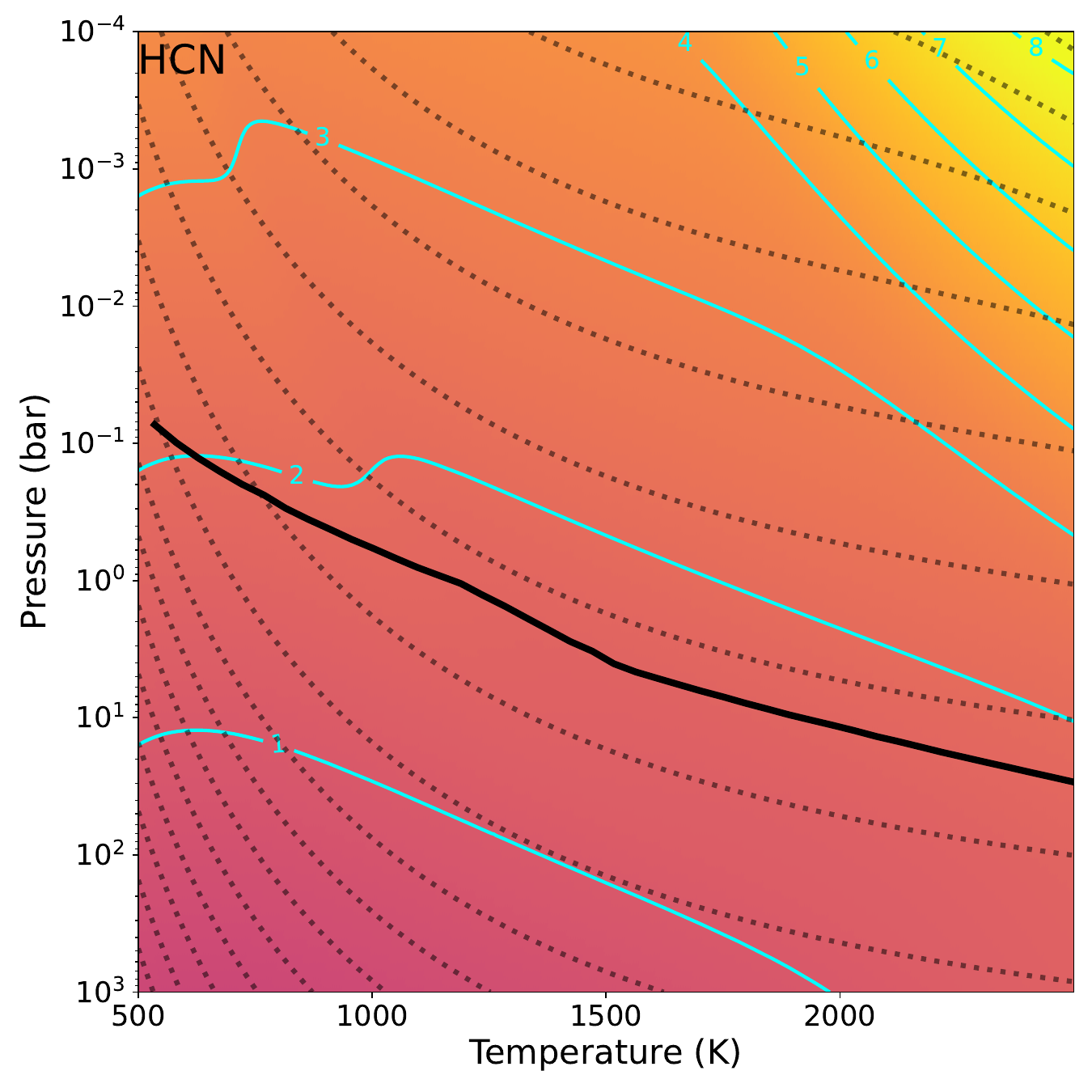}
	\includegraphics[width=0.4\textwidth]{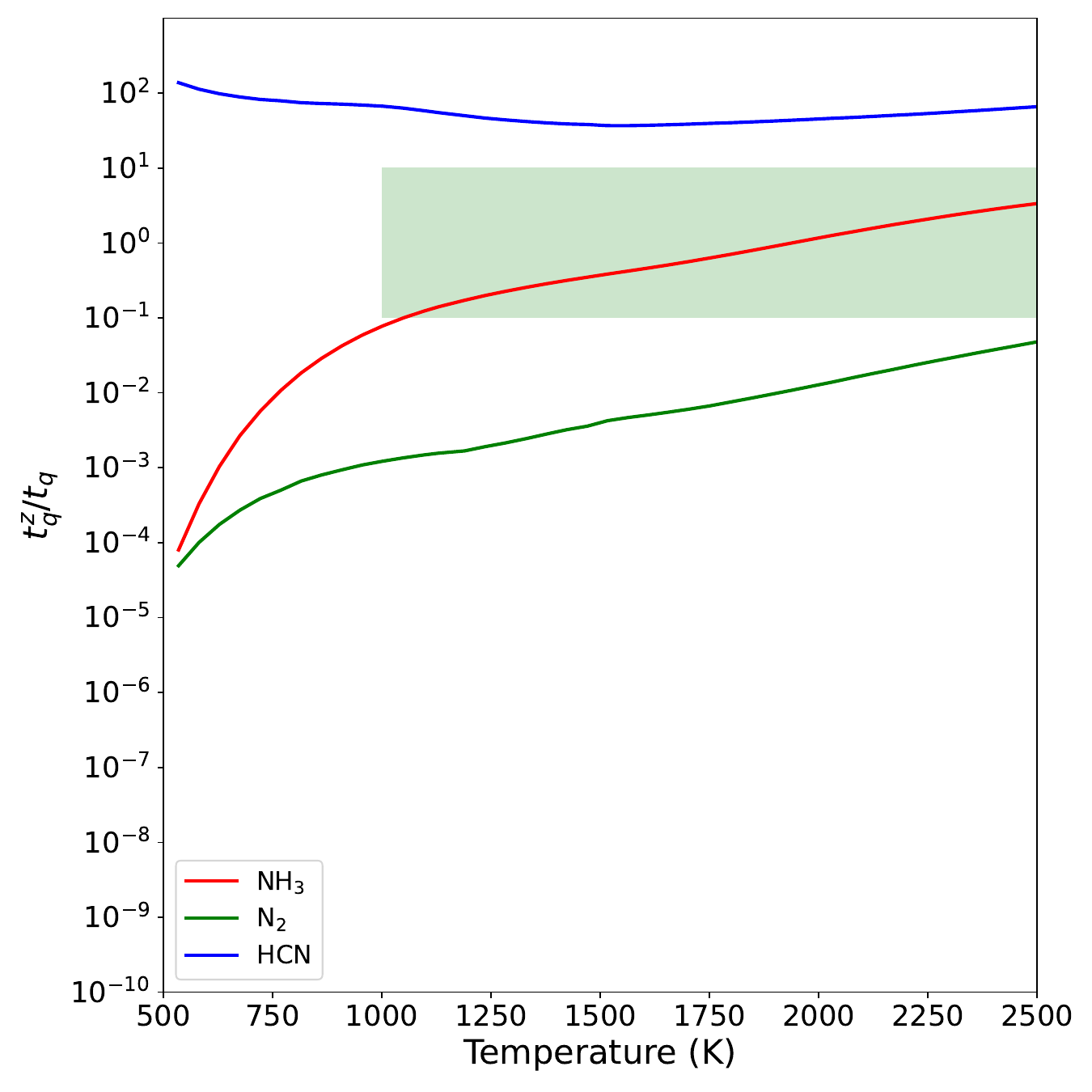}
	
	\caption{The contours of the ratio of our chemical timescales with an analytical timescale 
		from \cite{Zahnle2014} for \ch{N2}, \ch{NH3}, and \ch{HCN} are shown in the 
		log10 scale. The black solid line is the T-P profile taken from \cite{Zahnle2014} for effective temperature $T_{\text{eff}}$ = 600 K 
		and $g_{\text{surface}} = 10^3$ cm s$^{-2}$. Black dotted lines are the contours of \ch{NH3/N2}.
		The bottom right panel shows the ratio of the chemical 
		timescale obtained in our study ($t_{q}$) and from \cite{Zahnle2014} ($t_q^z$), for the thermal 
		profile.}\label{fig:comparision}
\end{figure}

In Figure \ref{fig:comparision}, we have compared the chemical timescale from both studies. For \ch{NH3}, 
chemical timescales from analytical expressions deviate significantly at low-pressure and high-temperature 
($P<10^{-1}$ bar, $T>1500$ K) and $T<1250$ K. However, along the thermal profile, the \ch{NH3} timescale 
remains close to the analytical expression for $T>1500$ K. For \ch{N2}, the analytical timescales provide 
reasonably correct chemical timescales in the high-pressure region ($P>10^2$ bar) and high temperature 
and low-pressure region ($P<10^{-2}$ bar, $T>2000$ K). However, along the thermal profile, the deviation 
is one order of magnitude at T = 2500 K and more than four orders of 
magnitude at T = 500 K. The deviation of the chemical timescale of \ch{HCN} is one to three orders of magnitude 
in the region where $\ch{HCN + OH} \rightleftarrows \ch{HNCO + H}$ becomes the RLS. In the region where \ch{HCN}$\rightarrow$\ch{NH3}
conversion takes place through \ch{CN}, the deviation is four to eight orders of magnitude. Along the thermal 
profile, the deviation remains around two orders of magnitude. The analytical expression for \ch{N2} and \ch{HCN} 
deviate more than two orders of magnitude, and \ch{NH3} deviation is one order of magnitude 
for T $<$ 1000 K. Thus, analytical expressions do not always give the correct value; therefore they should be used with caution 
while calculating the quench pressure level.

\clearpage
\section{Supporting Figures for \S~8}

\setcounter{figure}{0}
\begin{figure}[h]
	\centering
	\includegraphics[trim={0cm 0cm 3cm 0cm},clip,width=1\textwidth]{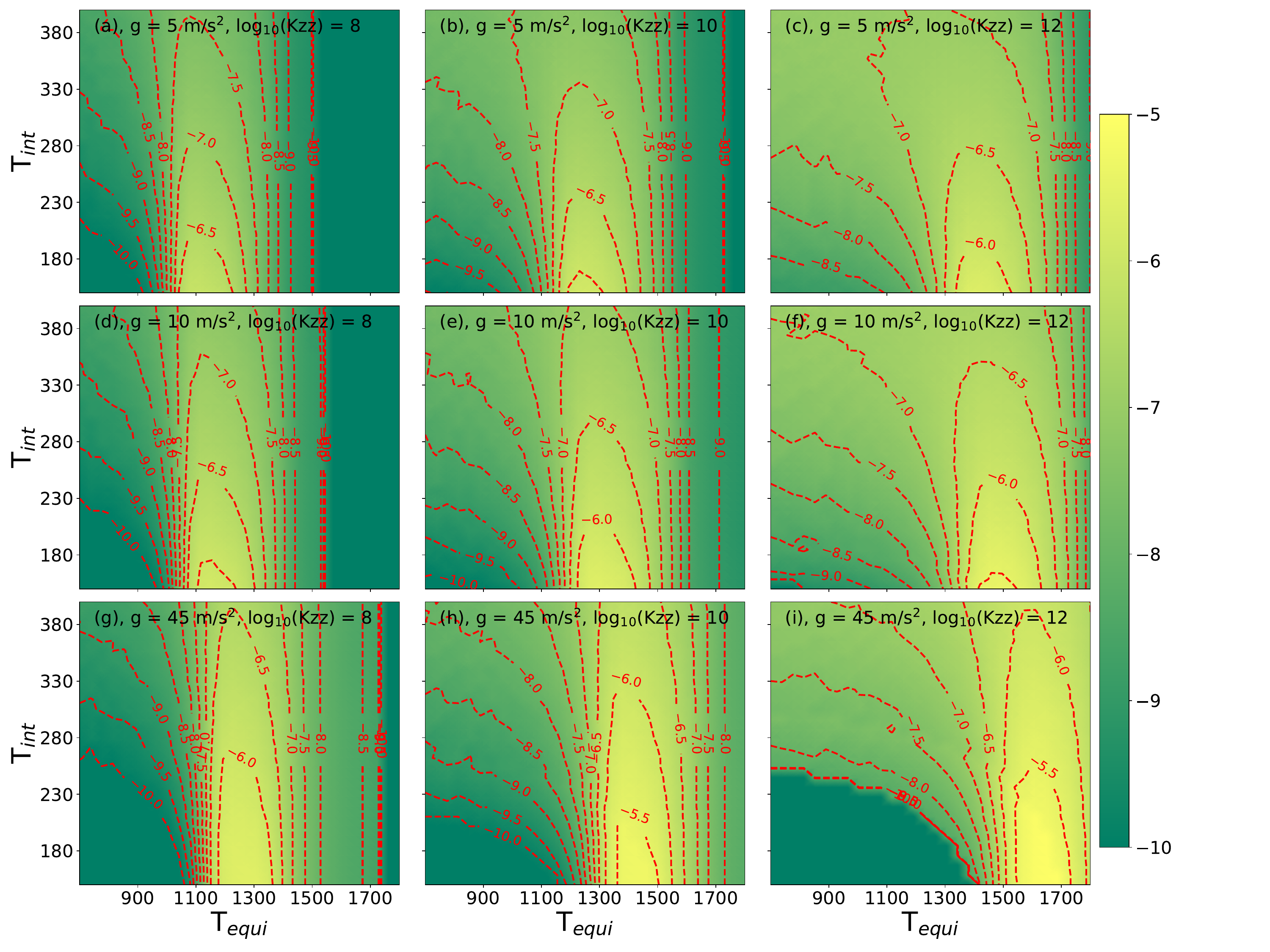}
	\caption{The color bar shows the mole fraction of quenched \ch{HCN} in the log10 
		scale as a function of pressure and temperature. The different panel represents the different 
		values of $g_{\text{surface}}$ (5, 15, 45 m s$^{-2}$) and $K_{zz}$ (10$^{8}$, 10$^{10}$, 10$^{12}$  cm$^{2}$ s$^{-1}$) 
		which are labeled in the panel.}
	\label{fig:T_int_vs_T_equi_HCN}
\end{figure}

\begin{figure}[h]
	\centering
	\includegraphics[width=1\textwidth,page=1,trim= 1cm 0cm 1.5cm 0cm]{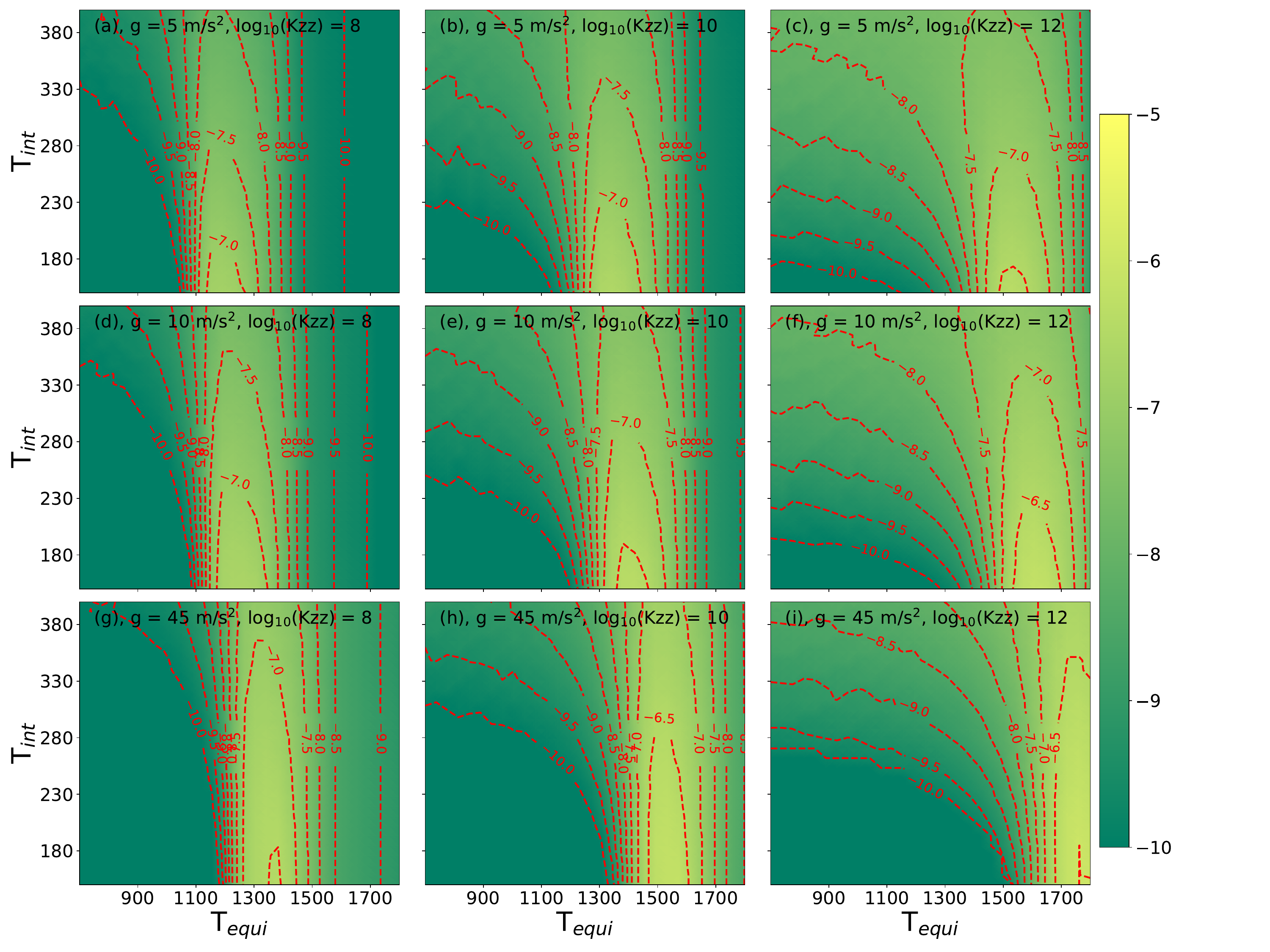}
	\caption{Same as Figure \ref{fig:T_int_vs_T_equi_HCN}, but for 0.1 $\times$ solar metallicity.}
	\label{fig:T_int_vs_T_equi_HCN_mat_1}
\end{figure}

\begin{figure}[h]
	\centering
	\includegraphics[width=1\textwidth,page=1,trim= 1cm 0cm 1.5cm 0cm]{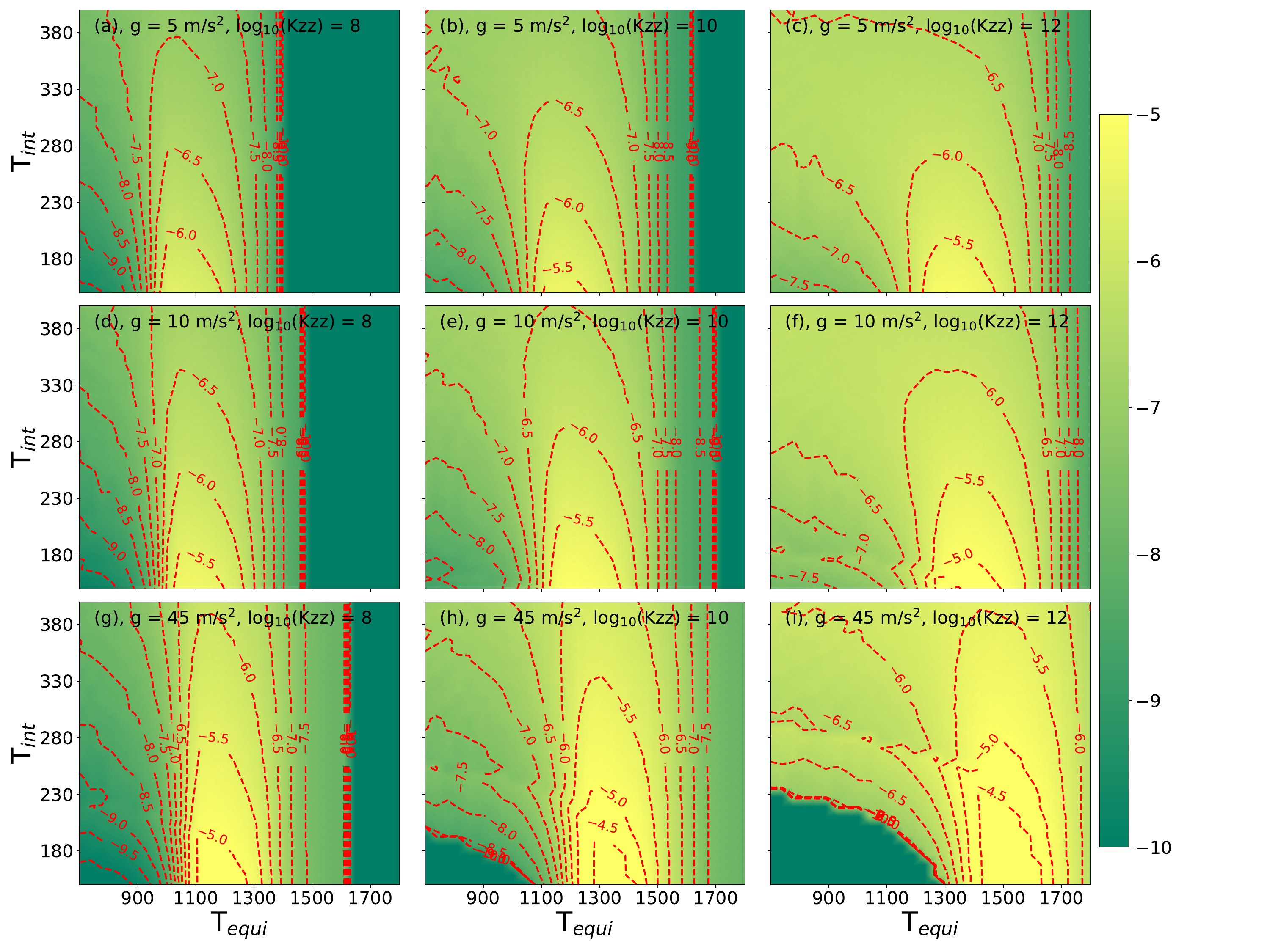}
	\caption{Same as Figure \ref{fig:T_int_vs_T_equi_HCN}, but for 10 $\times$ solar metallicity.}
	\label{fig:T_int_vs_T_equi_HCN_mat_10}
\end{figure}

\begin{figure}[h]
	\centering
	\includegraphics[trim={0cm 0cm 3cm 0cm},clip,width=1\textwidth]{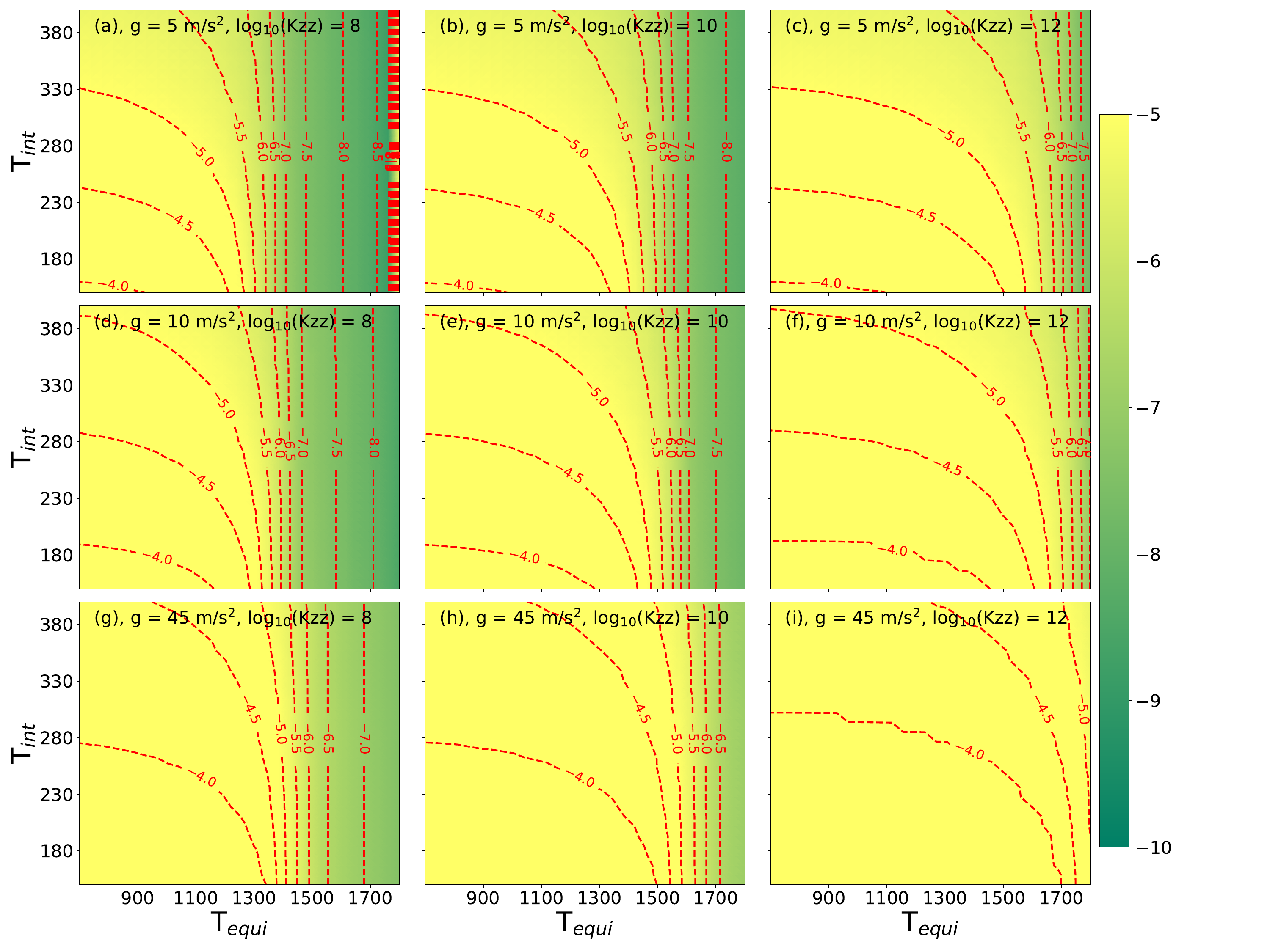}
	\caption{Same as Figure \ref{fig:T_int_vs_T_equi_HCN}, but for \ch{NH3}}
	\label{fig:T_int_vs_T_equi_NH3}
\end{figure}

\end{document}